\documentclass[amsmath,amssymb,prd,aps,showpacs,twocolumn,superscriptaddress,floatfix,nofootinbib]{revtex4-1}

\usepackage{graphicx}
\usepackage{dcolumn}
\usepackage{bm}
\usepackage{hyperref}
\usepackage[mathlines]{lineno}

\usepackage[usenames, dvipsnames]{xcolor}
\usepackage{bbold}
\usepackage{float}
\usepackage{color}
\usepackage[caption=false]{subfig}
\usepackage{euscript} 
\usepackage{placeins}
\usepackage{diagbox}
\usepackage[thinlines]{easytable}
\usepackage{appendix}

\newcommand{\p}{{\bm{p}}}

\newcommand{\q}{{\bm{q}}}

\def\mD{m_{\mathrm{D}}}

\def\gs{g}
\def\d{\mathrm{d}}
\def\e{\mathrm{e}}
\def\CR{C_\mathrm{R}}
\def\CA{C_\mathrm{A}}
\def\bb{{\boldsymbol b}}
\def\bp{{\bb_\perp}}
\def\bbp{b_\perp}

\def\qp{q_\perp}
\def\bqp{\boldsymbol q_\perp}

\def\Cbp{C(\bbp)}
\def\Cqp{C(\qp)}

\newcommand{\nn}{\nonumber}
\def\tp{\tilde{p}}
\def\tq{\tilde{q}}
\def\ttp{\tilde{\p}}
\def\ttq{\tilde{\q}}
\def\tpsi{\tilde{\psi}}
\def\qhn{\hat{q}_0}

\def\CR{C_\mathrm{R}}
\def\CA{C_\mathrm{A}}

\def\nf{N_{\mathrm{f}}}
\def\Tf{T_{\mathrm{f}}}

\newcommand{\qt}{q_\perp}

\def\mDsq{m_{\mathrm{D}}^2}

\begin{document}

\title{Splitting rates in QCD plasmas from a non-perturbative determination of the momentum broadening kernel $C(\qt)$}

\author{S\"oren Schlichting}
\affiliation{Fakult\"at f\"ur Physik, Universit\"at Bielefeld\\ 
D-33615 Bielefeld, Germany.}
\author{Ismail Soudi}
\affiliation{Fakult\"at f\"ur Physik, Universit\"at Bielefeld\\ 
D-33615 Bielefeld, Germany.}
\affiliation{Department of Physics and Astronomy, Wayne State University\\
Detroit, MI 48201.}
\date{\today}

\begin{abstract}
We exploit a recent non-perturbative determination of the momentum broadening kernel $C(b_{\bot})$ in impact parameter space~\cite{Moore:2021jwe}, to determine the momentum space broadening kernel $C(q_{\bot})$ in high-temperature QCD plasmas.
We show how to use the non-pertubatively determined kernel $C(q_{\bot})$ to compute the medium-induced splitting rates in a QCD plasma of finite size.
We compare the resulting in-medium splitting rates to the results obtained with leading-order and next-to-leading order perturbative determinations of $C(q_{\bot})$, as well as with various approximations of the splitting employed in the literature. Generally, we find that the differences in the splitting rates due to the momentum broadening kernel are larger than the errors associated with approximations of the splitting rate.
\end{abstract}

\maketitle

\section{Introduction}
One of the clearest signals for the formation of a Quark Gluon Plasma (QGP) in heavy ion collisions is the suppression of the yields of highly energetic particles.
When highly energetic partons or jets traverse the medium, they interact with the constituents of the QGP leading to a loss of energy, commonly referred to as jet quenching \cite{Bjorken:1982tu}. 
While for highly energetic partons only a small fraction of the energy is lost due to elastic interactions with the medium, the interactions of hard partons with the medium constituents also induce additional ``medium-induced`` radiation~\cite{Gyulassy:1993hr,Baier:1994bd,Zakharov:1996fv}, which provides the dominant energy loss mechanism for highly energetic particles. Studies of medium-induced radiation in QCD plasmas date back to the early determinations of the Landau-Pomeranchuk-Migdal effect \cite{Landau:1953um,Migdal:1956tc} in QCD \cite{Baier:1996kr,Baier:1996sk,Zakharov:1996fv,Zakharov:1997uu}, and include determinations of the radiation rate in QCD plasmas of infinite \cite{Arnold:2002ja} and finite spatial extent \cite{CaronHuot:2010bp,Andres:2020vxs}. Beyond the development of different theoretical formalism to determine the in-medium splitting rates~\cite{Gyulassy:1993hr,Zakharov:1996fv,Zakharov:1997uu,Zakharov:1998sv,Baier:1994bd,Baier:1996kr,Baier:1996sk,Wiedemann:1999fq,Gyulassy:1999zd,Salgado:2003gb,Blaizot:2012fh,Apolinario:2014csa,Sievert:2019cwq,Andres:2020vxs}, there have also been ongoing efforts to construct suitable approximations of the medium-induced splitting rates in various different limits \cite{Gyulassy:1999zd,Wiedemann:1999fq,Gyulassy:2000fs,Mehtar-Tani:2019tvy,Mehtar-Tani:2019ygg,Barata:2020sav,Andres:2020kfg}.

In all the different formalisms to obtain splittings rates, the interaction of hard partons with the medium are described using the rate of transverse momentum broadening 
\begin{align}
    \Cqp & \equiv (2\pi)^2\frac{\d \Gamma}{
    \d^2 \bqp}
    \;,
\end{align}
which defines the rate to exchange transverse momentum $\bqp$ with the QCD medium. While perturbative broadening kernels have been used to successfully predict observables (see e.g. \cite{Mehtar-Tani:2013pia,Cao:2020wlm} for recent reviews) it is known that due to the infrared (IR) problem, the perturbative expansion breaks down in the IR regime even at small coupling \cite{Linde:1980ts}. Nevertheless, effective field theories coupled with lattice calculations can be used to evade the IR problem \cite{Braaten:1995cm}. Specifically, for the momentum broadening kernel $C(\qt)$, one can define the zero-subtracted Fourier transform 
\begin{align}
    \Cbp & \equiv \int \! \frac{\d^2 \bqp}{(2\pi)^2}
    \left( 1 - e^{i\bqp \cdot \bp} \right) \Cqp \label{eq:subtraction_FT} \;.
\end{align}
which can be defined non-perturbatively in terms of certain light-like Wilson loops \cite{CasalderreySolana:2007qw}. For temperatures well above the critical temperature $T_c$ these light-like Wilson loops can be recast in the dimensionally reduced long-distance effective theory for QCD, 3D Electrostatic QCD (EQCD) \cite{CaronHuot:2008ni}. 
In an earlier study \cite{Moore:2021jwe} we showed how the short distance behavior of the broadening kernel $C_\mathrm{EQCD}(\bp)$ determined from non-perturbative lattice simulations of EQCD \cite{Panero:2013pla,DOnofrio:2014mld,Moore:2019lua,Moore:2019lgw}, can be matched to obtain a non-perturbative determination of $C_\mathrm{QCD}(\bp)$ in QCD at all scales. In this study the broadening kernel was computed in impact parameter $(\bp)$ space, which is favorable for the calculation of medium-induced radiation in an infinite medium \cite{Arnold:2002zm}. However, in order to extend the framework to a QCD medium of finite size, 
it is highly favorable to work in momentum $(\bqp)$ space \cite{CaronHuot:2010bp}.

Central objective of this paper is to employ the non-perturbative determination of $C_\mathrm{QCD}(\bp)$ in ~\cite{Moore:2021jwe}, to determine the medium-induced radiation rates in the physically relevant situation of a finite medium. First, we Fourier transform the non-perturbative kernel to momentum space in Sec.~\ref{sec:Broadening}. 
Subsequently in Sec.~\ref{sec:Formalism}, we recapitulate the formalism of~\cite{CaronHuot:2010bp} to obtain the splitting rates in a finite medium and introduce the opacity expansion \cite{Wiedemann:2000za,Gyulassy:1999zd} together with an expansion around the multiple soft scattering limit \cite{Mehtar-Tani:2019tvy,Mehtar-Tani:2019ygg,Barata:2020sav} and a resummed opacity expansion method \cite{Andres:2020kfg}. Numerical results for the splitting rates are presented in Sec.~\ref{sec:Results}, where we compare our calculations with the results obtained using leading order (LO) and next-to-leading order (NLO) perturbative determinations of the broadening kernels, and investigate the quality of various approximations. We conclude with a summary of our important findings in Sec.~\ref{sec:Conclusion}.

\section{Non-perturbative broadening kernel}\label{sec:Broadening}
Building on earlier works that established the procedure~\cite{Panero:2013pla,DOnofrio:2014mld,Moore:2019lua}, non-perturbative contributions to the momentum broadening kernel were extracted from an EQCD lattice calculation in~\cite{Moore:2020wvy}. However, since EQCD is an IR effective theory of QCD, the EQCD broadening kernel from \cite{Moore:2020wvy} is only valid in the IR regime.
In \cite{Moore:2021jwe} we demonstrated how one can perform a matching to the QCD broadening kernel in UV regime to obtain a broadening kernel valid over the entire range of momenta/impact parameters~\cite{Moore:2021jwe}. By following the arguments~\cite{Ghiglieri:2018ltw,Moore:2021jwe}, the non-perturbative broadening kernel is determined as  
\begin{align} \label{eq:match2}
    C_{\mathrm{QCD}}(\bbp) \approx \left( C_{\mathrm{QCD}}^{\mathrm{pert}}(\bbp) - C^\mathrm{pert}_\mathrm{subtr}(\bbp) \right) + C_{\mathrm{EQCD}}^{\mathrm{latt}}(\bbp) \,,
\end{align}
where $C_{\mathrm{QCD}}^{\mathrm{pert}}(\bbp)$ is the UV limit $(\qp \gg m_D)$ of the QCD kernel, which is known analytically in momentum space~\cite{Arnold:2008vd,CaronHuot:2008ni}\footnote{See Eq.~(\ref{eq:CLO}) for the full leading order QCD kernel, without the limit $(\qp \gg m_D)$}
\begin{align}	\label{Cqp_hard}
    &C^{\rm pert}_\mathrm{QCD}(\qp) = \frac{\gs^4 \CR}{\qp^4} \!\int \! \frac{\d^3 p}{(2 \pi)^3} \frac{p - p_z}{p} \nonumber\\
    &\left[ 2 \CA n_\mathrm{B}(p) \left( 1 {+} n_\mathrm{B}(p') \right) + 4 \, \nf \Tf \, n_\mathrm{F}(p) \left( 1 {-} n_\mathrm{F}(p') \right) \right],
\end{align}
where $p' = p + \frac{\bqp^2 + 2 \bqp \hspace{-1pt} \cdot \, \p}{2 (p - p_z)}$ , the equilibrium distributions are given in terms of the Bose-Einstein distribution $n_B(p) = \frac{1}{e^{p/T}-1}$ and the Fermi-Dirac distribution $n_F(p) = \frac{1}{e^{p/T}+1}$. Throughout this manuscript, we will take three quark flavors $N_f=3$ and the color algebra constant are $C_A=3$ and $T_f = \frac{1}{2}$.
The subtraction term $C^\mathrm{pert}_\mathrm{subtr}(\qp)$ is given by~\cite{Ghiglieri:2018ltw}
\begin{align}	\label{Cqp_subtr}
    C^\mathrm{pert}_\mathrm{subtr}(\qp) = \frac{\CR \gs^2 T \mDsq}{\qp^4} - \frac{\CR \CA \gs^4 T^2}{16 \, \qp^3} \, ,
\end{align}
where, as discussed in detail in ~\cite{Moore:2021jwe}, the first term cancels against the (unphysical) IR limit of $C^{\rm pert}_\mathrm{QCD}(\qp)$, while the second term cancels out the (unphysical) UV behavior of the EQCD kernel. 
\subsection{Broadening kernel in impact parameter space}
\begin{table}[t!] 
    \centering {\small
    \begin{tabular}{c|c|c}
    & $\left. \frac{\Cbp }{ g^2} \right\vert^{N_\mathrm f = 3}_{250~\mathrm{MeV}}$ & $\left. \frac{\Cbp }{ g^2} \right\vert^{N_\mathrm f = 3}_{500~\mathrm{MeV}}$ \\
    & & \\
    \hline
    $\gs^2$ & $3.725027$ & $2.763516$ \\
    \hline
    $\hat{q}_{0} / {\gs^6 T^3}$ & $0.1465(78)$ & $0.185(10)$  \\
    \hline
    $A$ & $-0.6717$ & $-0.4885$  \\
    \hline
    $\xi$ & $0.1780$ & $0.1702$  \\
    \hline
    $\sigma_{\rm EQCD}/g^4T^2$ & $0.2836(10)$ & $0.2867(10)$  \\
\end{tabular} }
\caption{Strong coupling constant $\gs$ and various constants that determine the limiting behavior of the non-perturbative momentum broadening kernel (c.f. Eqns.~ (\ref{IR_limit},\ref{UV_limit},\ref{eq:Separation}). Numerical values are reproduced from \cite{Moore:2019lgw,Moore:2021jwe}. }
\label{tab:full_Cbp}
\end{table}
Before we proceed to Fourier transform the resulting kernel to momentum space, we briefly recall the limiting behaviors of the kernel in impact parameter space. At long-distances the Wilson loop follows an area-law behavior \cite{Laine:2012ht} with asymptotic corrections which are important for smoothening the transition to the numerical data values
\begin{align}	\label{IR_limit}
    &\frac{C_\mathrm{QCD} (\bbp)}{\gs^2T} \xrightarrow{\bbp \gg \; 1/g^2T} A + \frac{\sigma_\mathrm{EQCD}}{\gs^4 T^2} \gs^2 T \bbp \nonumber\\
    &+  \frac{\gs^2 \CR}{\pi} \left[ \frac{\mD^2}{4\gs^2T^2} \left( \frac{1}{6} - \frac{1}{\pi^2} \right) + \frac{\CA}{8 \pi^2} \right] \log (\gs^2 T \bbp) \, .
\end{align}
Here $\sigma_\mathrm{EQCD}$ is the string tension of EQCD~ \cite{Laine:2005ai} and $A$ is a constant fitted to the EQCD lattice data~\cite{Moore:2021jwe}.

Conversely, at short-distances, the broadening kernel follows a similar behavior to the leading order QCD behavior
\begin{align}	\label{UV_limit}
    \frac{C_\mathrm{QCD} (\bbp)}{\gs^2T} \xrightarrow{\bbp \ll \; 1/\mD}& \frac{1}{4} \frac{\qhn}{\gs^6T^3} (\gs^2 T \bbp)^2   \nonumber\\
    & - \frac{\CR\mathcal{N}}{8 \pi }  (\gs T \bbp)^2 \log(\gs^2 T \bbp)  \, ,
\end{align}
where $\mathcal{N} = \frac{\zeta (3)}{\zeta(2)} \left( 1 + \tfrac{\nf}{4} \right)$ and we provide the numerically extracted value of $\hat{q}_0$ in Tab.~\ref{tab:full_Cbp} which is reproduced from \cite{Moore:2021jwe} for the sake of completeness of the presentation.

\subsection{Broadening kernel in momentum space}
\begin{figure}
    \includegraphics[width=0.42\textwidth]{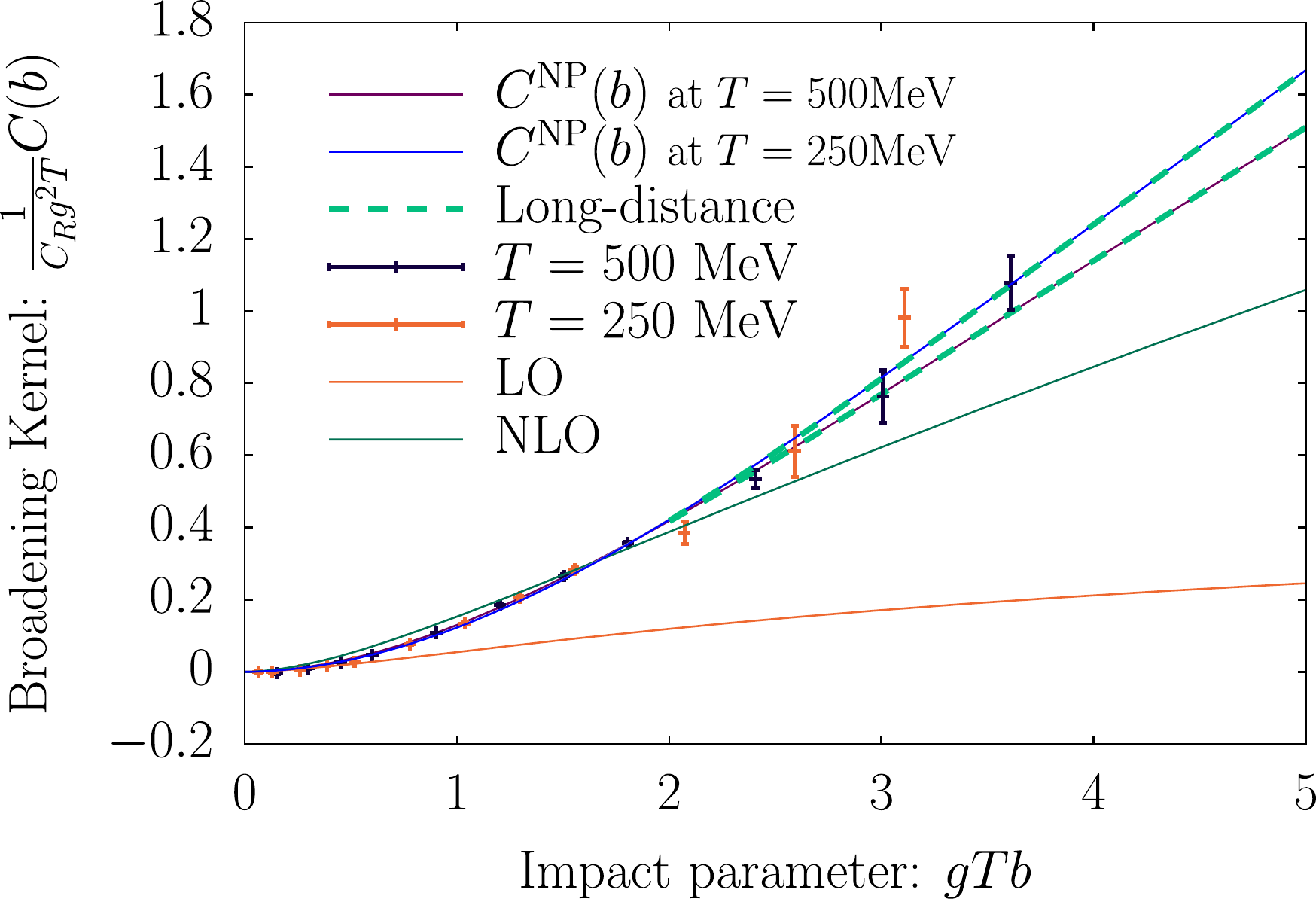}
    \includegraphics[width=0.45\textwidth]{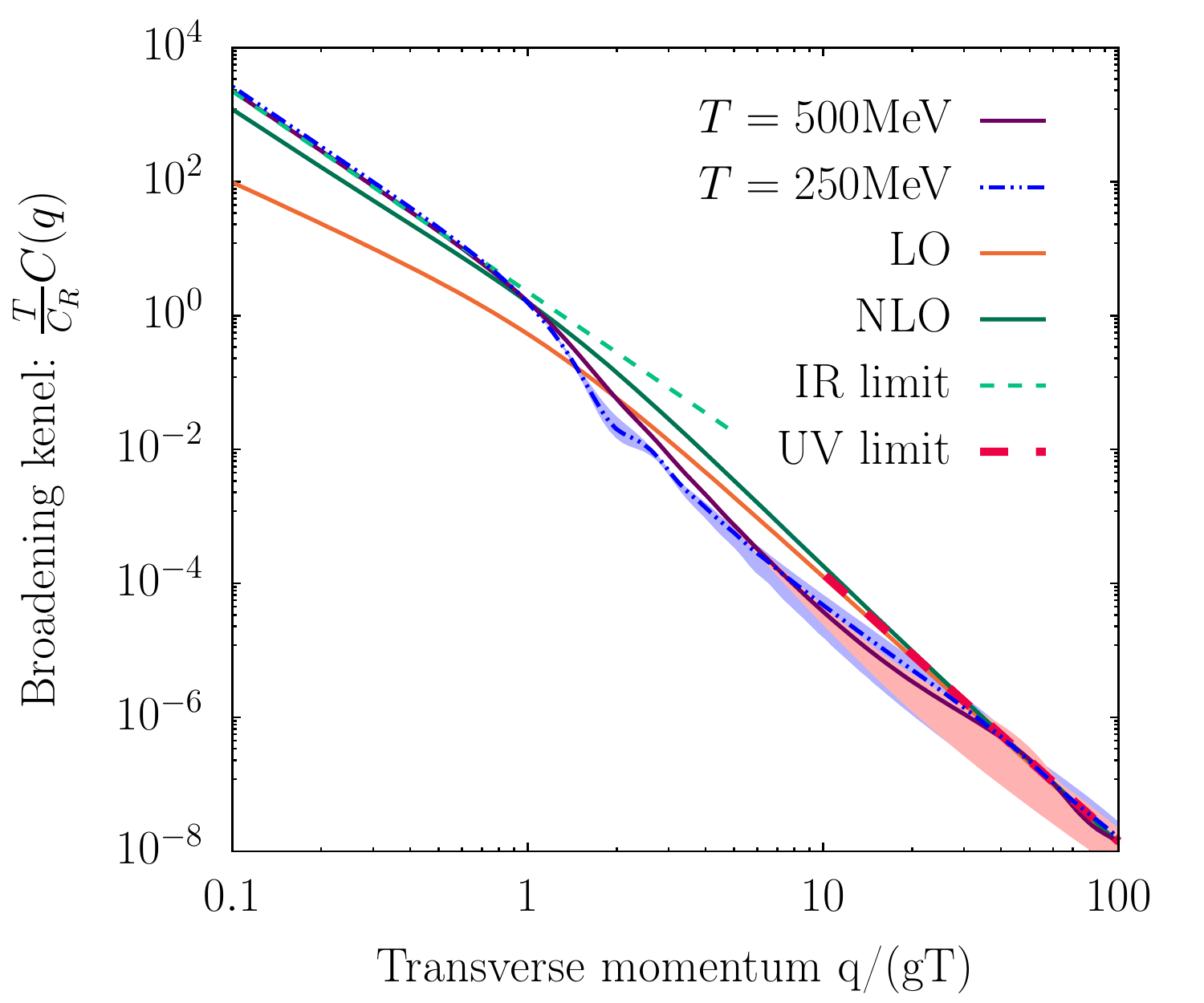}
    \caption{(top) Non-perturbative elastic broadening kernel $C_{\rm QCD}(\bp)$ in impact parameter space.  Data points for two different temperatures $T=250,500 {\rm MeV}$ are shown alongside the interpolating splines. We also compare to the short-distance limit in Eq.~(\ref{UV_limit}) and the long-distance limit in Eq.~(\ref{IR_limit}). (from \cite{Moore:2021jwe}). (bottom) Elastic broadening kernel $C_{\rm QCD}(\bqp)$ in momentum space for $T=250,500$MeV. Blue and purple bands represent uncertainties of the spline interpolation for $250$MeV and $500$MeV respectively. We also compare the kernel to leading-order (LO) Eq.~(\ref{eq:CLO}) and next-to-leading order (NLO) Eq.~(\ref{eq:CEQCDbp_NLO}) determinations at $T=500$MeV, as well as to the UV limit in Eq.~(\ref{eq:UV_limitqp}) and the IR limit in Eq.~(\ref{eq:IR_limitqp}). }
    \label{fig:InverseKernel}
\end{figure}

We now proceed to perform the Fourier transform of the non-perturbative broadening kernel \cite{Moore:2021jwe} back to momentum space. 
In principle, the inverse Fourier transform is standard and should be straightforward to compute. However, due to the spareness of the data points, the divergent behavior of the kernel at large impact parameter and the highly oscillatory nature of the integrals involved, performing the numerical integral is actually rather challenging. To avoid these difficulties, we found that it is best to Fourier transform the coordinate space derivative $\frac{\d C(\bp)}{\d b_\perp}$ of the momentum broadening kernel. Using Eq.~(\ref{eq:subtraction_FT}), one can write
\begin{align}
    \frac{\d C(\bp)}{\d \bbp} =& \int \! \frac{\d^2 \qp}{(2 \pi)^2}  \e^{-i \bqp \hspace{-1pt} \cdot \, \bp} \left[  \frac{i\bqp\hspace{-1pt} \cdot \, \bp}{b_\perp} C(\qp) \right] \, .
\end{align}
Exploiting the fact that $C(\bqp)$ does not depend on the direction $\bp/|\bbp|$ then leads to the following Hankel transform,
\begin{align}
    \frac{\d C(\bp)}{\d \bbp} =&\int_0^{\infty} \! \frac{\d q_\perp}{(2 \pi)} ~\qp J_1(b_\perp\, \qp) \left[ \qp C(\qp) \right] \;,
\end{align}
where $J_1(x)$ is the Bessel function of the first kind of order $1$. By use of the inverse Hankel transform, one then obtains the broadening kernel in momentum space as
\begin{align} \label{eq:HankelTransform}
    C(\qp)  =& \frac{2\pi}{\qp}\int_0^{\infty} \! \d b_\perp~  b_\perp J_1(b_\perp\,\qp) \frac{\d C(b_\perp)}{\d b_\perp}\, .
\end{align}
While the integral in Eq.~(\ref{eq:HankelTransform}) is still highly oscillatory, it can be computed numerically as long as the integrand is sufficiently well behaved at the integration boundaries. In order to ensure numerical convergence, we therefore subtract the leading asymptotic behavior at large distances 
\begin{equation}
    \frac{\d C^{\rm IR}(\bp)}{\d \bbp} = \sigma_\mathrm{EQCD} \;.
\end{equation}
and only perform a numerical Hankel transform of the remainder
\begin{align}
    \frac{\d }{\d \bbp}\Delta C_{\rm QCD}(\bp) = \frac{\d C_{\rm QCD}(\bp)}{\d \bbp}-\frac{\d C^{\rm IR}_{\rm QCD}(\bp)}{\d \bbp} \;,
\end{align}
which by construction vanishes for large impact parameters. By numerically performing the Hankel transform 
\begin{align}\label{eq:HankelTransformDeriv}
    \Delta C_{\rm QCD}(\qp) =& \frac{2\pi}{\qp}\int_0^{\infty} \! \d b_\perp~  b_\perp J_1(b_\perp\,\qp) \frac{d}{db_\perp}\Delta C_{\rm QCD}(b_\perp)\,,
\end{align}
and supplying it with the analytic result for the Hankel transform of $C^{\rm IR}(\bp)$, given by (c.f.~Appendix.~\ref{ap:Hankel})
\begin{equation}
    C^{\rm IR}_{\rm QCD}(\qp) =\frac{2\pi}{\qp^3} \sigma_\mathrm{EQCD}\;.
\end{equation}
we obtain the full momentum broadening kernel as
\begin{align}
    C_{\rm QCD}(\qp) =  \Delta C_{\rm QCD}(\qp) + C^{\rm IR}_{\rm QCD}(\qp)\;,
\end{align}
We note that, due to the fact that the Bessel function is highly oscillatory for large momenta $\qt$, sufficient care should be taken in performing the integral, and we describe the procedure we employ in Appendix.~\ref{ap:Hankel}.

Next, in order to construct the momentum broadening kernel $C(\qt)$ at all scales we proceed to transform the limiting behaviors of the kernel, which can be used to extrapolate the results beyond the tabulated range of $\qt$ values. In the deep infrared regime, the momentum broadening kernel is determined by the string tension such that 
\begin{align}
    C_{\rm QCD}(\qp) \xrightarrow{\qp \ll \; g^2T}&2\pi\frac{\sigma_\mathrm{EQCD}}{\qp^3}  \nonumber\\ 
    \label{eq:IR_limitqp}
\end{align}
In the UV limit the momentum broadening kernel follows the same behavior as the perturbative QCD kernel in Eq.~(\ref{Cqp_hard}), and one obtains~\cite{Arnold:2008vd}
\begin{align}
    C_{\rm QCD}(\qp) \xrightarrow{\qp \gg \; \mD} \frac{\CR \gs^4T^3 \mathcal{N}}{\qp^4} \,.\label{eq:UV_limitqp}
\end{align}

\subsection{Perturbative kernel in EQCD}
Before we present results for the non-perturbative determination of $C(\qt)$, we briefly recall the results of perturbative calculations, following ~\cite{Ghiglieri:2018ltw,Moore:2021jwe},  which we will use as a reference for comparison. At leading order (LO) $\mathcal{O}(\gs^4)$, the QCD collisional broadening kernel can be expressed in momentum space \cite{Arnold:2008vd} as
\begin{align}\label{eq:CLO}
    &C^{\rm LO}_{\rm QCD}(\qp) = \frac{ \gs^4 \CR}{\qt^2(\qt^2+ \mDsq)} \int \! \frac{\d ^3p}{(2\pi)^3}
    \frac{p-p_z}{p} \nonumber \\ 
    &  
    \left[ 2 \CA n_\mathrm{B}(p)(1 + n_\mathrm{B}(p')) + 4 \nf \Tf n_\mathrm{F}(p)(1 - n_\mathrm{F}(p'))\right]\;,
\end{align}
with $p' = p + \frac{\bqp^2 + 2 \bqp \hspace{-1pt} \cdot \, \p}{2 (p - p_z)}$ and displays the following asymptotic behaviors
\begin{align}
C^{\rm LO}_{\rm QCD}(\qp) =& \gs^2 T \CR
    \begin{cases}
        \frac{ \mDsq - \gs^2 T^2 \CA \frac{\qp}{16 T}}{\qp^2 (\qp^2 + \mDsq)} \;, & \qp \ll \gs T \;,\\\\
        \frac{ \gs^2 T^2}{\qp^4} \frac{\zeta (3)}{\zeta(2)} \left( 1 + \tfrac{\nf}{4} \right)\;, & \qp \gg \gs T \;.\\
    \end{cases}
\end{align}

Next-to-leading order (NLO) corrections are of order $\gs^5$ and arise from infrared corrections that are suppressed by an additional factor of $\mD \sim \gs$, which can be calculated in EQCD~\cite{CaronHuot:2008ni}. Similar to the treatment of the non-perturbative kernel, the NLO broadening kernel is computed using perturbative results for the soft contributions from EQCD and supplying the hard contribution by a matching \cite{CaronHuot:2008ni}. Specifically,
\begin{align} \label{eq:CEQCDbp_NLO}
    C^{\rm NLO}_{\rm QCD}(\qp) =&C^{\rm LO }_{\rm EQCD}(\qp)+C^{\rm NLO }_{\rm EQCD}(\qp) + C_{\rm QCD}^{\rm pert}(\qp)\nonumber\\
    &- C_{\rm subtr}^{\rm pert}(\qp)\;,
\end{align}
where the leading and next-to-leading order contributions from soft modes are given by~\cite{Aurenche:2002wq,Arnold:2008vd,CaronHuot:2008ni} 
\begin{align}	
    &C^{\rm LO }_{\rm EQCD}(\qp) = \CR \gs^2 T \frac{\mDsq}{\qt^2(\qt^2 + \mDsq)}\;,\\
    &\frac{C^{\rm NLO }_{\rm EQCD}(\qp)}{\gs^4T^2 \CR \CA} =
    \frac{7}{32 \qt^3} +
    \frac{  {-} \mD -2\frac{\qt^2{-} \mDsq}{\qt}
    \tan^{-1} \left(\frac{\qt}{ \mD}\right) }
    {4\pi(\qt^2{+} \mDsq)^2}
    \nonumber\\
    &
    +\frac{ \mD - \frac{\qt^2{+}4 \mDsq}{2\qt}
    \tan^{-1}\left(\frac{\qt}{2 \mD}\right)}
    {8\pi \qt^4}
    \nonumber\\
    &
    - \frac{\tan^{-1} \left(\frac{\qt}{ \mD}\right) }
    {2\pi \qt(\qt^2 + \mDsq)}
    +\frac{\tan^{-1} \left(\frac{\qt}{2 \mD}\right) }{2\pi \qt^3}
    \nonumber\\
    & +
    \frac{ \mD}{4\pi(\qt^2{+} \mDsq)}\left[
    \frac{3}{\qt^2{+}4 \mDsq}
    -\frac{2}{(\qt^2{+} \mDsq)}
    -\frac{1}{\qt^2} \right]\;,
\end{align}
and if not stated otherwise, we employ the leading order perturbative expressions for $\mD$ (see Eq.~(\ref{eq:mDSqr})) when evaluating the LO and NLO kernels.

\subsection{Numerical results}

We display the perturbative and non-perturbative (NP) determinations of the momentum broadening kernels $C_\mathrm{QCD} (\bbp)$ in impact parameter and $C_\mathrm{QCD} (\qp)$ in momentum space in Fig.~\ref{fig:InverseKernel} ,
where the LO and NLO perturbative kernels are computed for $T=500$MeV (see Tab~\ref{tab:full_Cbp} for the coupling $\gs$ employed) .
The top panel presents the kernel $C_\mathrm{QCD} (\bbp)$ in impact parameter space, where bands represent the uncertainty in the spline definition as discussed in \cite{Moore:2021jwe}.
We use the same color coding when presenting the broadening kernel $C_\mathrm{QCD} (\qp)$ in momentum space in the bottom panel of Fig.~\ref{fig:InverseKernel}, where the bands represent the transformation of the different splines in the band from the top panel. 
We also show the limiting behaviors in the infrared in Eq.~(\ref{eq:IR_limitqp}) and ultra-violet in Eq.~(\ref{eq:UV_limitqp}), as well as the LO and NLO kernels in Eqns.~(\ref{eq:CLO}) and (\ref{eq:CEQCDbp_NLO}). Strikingly, when expressing $T C_\mathrm{QCD} (\qp)/C_R $ as a function of $\qt/\gs T$ in momentum space, or $C(\bbp)/\gs^2TC_R$ as a function of $\gs T \bbp$ in impact parameter space, both data sets at $T=250,500$MeV display very similar behavior. We find that, as expected from Eq.~(\ref{eq:IR_limitqp}), the broadening kernel in the IR follows a $\sim1/\qp^3$ and respectively $\sim\bbp$. While this feature is missing at LO, the infrared behavior of the NP kernel is qualitatively similar to the NLO kernel; however on a quantitative level the slopes differ by an order one prefactor due to the difference in the string tension $\sigma_{\rm EQCD}$. Interestingly, for intermediate values of $\qt/(gT)$, the momentum broadening kernel determined from EQCD lattice data falls below the LO and NLO results.  In the UV limit, all kernels display the same $ \sim 1/\qp^4$ and respectively $ \sim \bbp^2\log(\bbp)$ behavior, associated with the contribution from hard scatterings in Eq.~(\ref{Cqp_hard}).

\section{Medium-induced splitting rates} \label{sec:Formalism}
Equipped with the broadening kernel $C(\qt)$ in momentum space, we now proceed to compute the rate of medium-induced radiation. 
Starting point of the rate calculation is the formal expression \cite{Baier:1996sk,Zakharov:1997uu,Zakharov:2004vm,CaronHuot:2010bp} 
\begin{align}\label{eq:Formal}
    \frac{dP^a_{bc}}{dz}= &\frac{g^2P^a_{bc}(z)}{4\pi P^2 z^2(1-z)^2} {\rm Re}\int^\infty_{0} dt_1~\int_{t_1}^{\infty} dt~ \int_{\p,\q} ~ \q.\p \nonumber\\
    &[G(t,\q;t_1,\p) - (\text{vac})]  \;,
\end{align}
which describe the total probability of (nearly) collinear in-medium splitting of the particle $a$ with momentum $P$ into particles $b$ and $c$ with momentum $zP$ and $(1-z)P$ respectively.
The propagator $G(t,\q;t_1,\p)$ satisfies the evolution equation 
\begin{align}\label{eq:Schrodinger}
    (\partial_t + i \delta E(\q) + \Gamma_3(t)) G(t,\q;t_1,\p) &=0\;, 
\end{align}
with initial condition 
\begin{equation}\label{eq:InitialProp}
    G(t_1,\q;t_1,\p) = \frac{1}{4P^2z^2(1-z)^2}(2\pi)^2\delta^{(2)}(\p-\q)\;.
\end{equation}
The energy is given by
\begin{align}
    \delta E(\p) \equiv& \frac{\p^2}{2Pz(1-z)} +\frac{m_{z}^2}{2zP}+\frac{m_{1-z}^2}{2(1-z)P} -\frac{m_1^2}{2P} \;,
\end{align}
where $m_{i}$ are the medium induced mass of the particles carrying momentum fraction $i$. Throughout this analysis we will use the leading order perturbative expressions
\begin{align}\label{eq:mDSqr}
    m^2_{\infty,g} = \frac{\mD^2}{2} = g^2T^2 \left( \frac{N_c}{6} + \frac{N_f}{12} \right)\;,\\
    m^2_{\infty,q} = g^2T^2 \frac{(N_c^2-1)}{8N_c^2}\;,
\end{align}
with $N_c=3$ and $N_f=3$.
The collisional broadening of the propagator can be expressed as the following collision integral 
\begin{align}\label{eq:ThreeBodyInt}
    &\Gamma_3(t) \circ G(t,\p;) =\nonumber\\
    & \frac{1}{C_R}\int_{\q}C_{\rm QCD}(t,\q) \Bigg\{ C_1 \Big[G(t,\p;) - G(t,\p-\q;)\Big]   \nonumber\\
    &+ C_z \Big[G(t,\p;) - G(t,\p+z\q;)\Big] \nonumber\\
    &+ C_{1-z} \Big[G(t,\p;) - G(t,\p+(1-z)\q;)\Big] \Bigg\} \;,
\end{align}
where the color factors are defined as 
\begin{align}
    C_1 =\frac{C^R_z+C^R_{1-z}-C^R_1}{2}\;, &\qquad 
    C_z =\frac{C^R_1+C^R_{1-z}-C^R_z}{2}\;, \\
    C_{1-z} =&\frac{C^R_1+C^R_z-C^R_{1-z}}{2}\;,
\end{align}
and $C^R_{z}$ is the Casimir of the particle with momentum fraction $z$, i.e. $C_R=C_A=N_c$ for gluons and $C_R=C_F= \frac{(N_c^2-1)}{2N_c}$ for quarks.
By following \cite{CaronHuot:2010bp}, one can use an integration by part to rewrite the following integral
\begin{align}
    &\int_{t_1}^\infty dt~ G(t,\q;t_1,\p) \nonumber \\
    = & \int_{t_1}^\infty dt~\left[ \frac{d}{dt} \left( \frac{e^{-i\delta E(\q)t}}{-i\delta E(\q)} \right) \right] e^{+i\delta E(\q)t} G(t,\q;t_1,\p) \;,
\end{align}
such that upon use of the evolution equation for the propagator in Eq.~(\ref{eq:Schrodinger}), the expression simplifies to
\begin{align}\label{eq:rearangement}
    &\int_{t_1}^\infty dt~ G(t,\q;t_1,\p) \nonumber \\
    =&  \frac{1}{-i \delta E(\q)} \left[ G(\infty,\q;t_1,\p)-G(t_1,\q;t_1,\p) \right]\nonumber \\
    & + \int_{t_1}^\infty dt~\frac{i}{\delta E(\q)}\Gamma_3(t)\circ  G(t,\q;t_1,\p) \;.
\end{align}
Now, as argued by \cite{CaronHuot:2010bp}, the terms in the second line of Eq.~(\ref{eq:rearangement}) do not contribute to the rate, as the first vanishes in the limit $t \to \infty$ due to rapid oscillations, while the second term merely correspond the initial condition in Eq.~(\ref{eq:InitialProp}) and thus cancels against the vacuum subtraction in Eq.~(\ref{eq:Formal}). By inserting Eq.~(\ref{eq:rearangement}) into Eq.~(\ref{eq:Formal}) re-arranging the order of integrations, and performing a time derivative w.r.t. $t$, the rate of medium-induced radiation can be compactly expressed in the form~\cite{CaronHuot:2010bp}
\begin{align}
    &\frac{d\Gamma^a_{bc}}{dz}(P,z,t)=\frac{g^2P^a_{bc}(z)}{ 4\pi P^2 z^2(1-z)^2} \nonumber\\
    &\quad{\rm Re}~\int_0^{t} dt_1~  \int_{\p,\q} ~ \frac{i\q.\p}{\delta E(\q)}\Gamma_3(t)\circ  G(t,\q;t_1,\p) \;. 
\end{align}
\subsubsection{Expressing the rate using wave function}
By introducing the wave function  
\begin{align}
    \vec{\psi}(\p,t,t_1) = \int_{\q} \frac{i \q}{\delta E(\q)} \Gamma_3(t) \circ G(t,\q;t_1,\p) \;,
\end{align}
we may further compactify the expression for the rate as follows 
\begin{align}
    &\frac{d\Gamma^a_{bc}}{dz}(P,z,t) \nonumber\\
    &=\frac{g^2P^a_{bc}(z)}{ 4\pi P^2 z^2(1-z)^2} {\rm Re}~\int_0^{t} dt_1~ \int_{\p} \p\cdot\vec{\psi}(\p,t,t_1 )\;.
\end{align}
Exploiting the linearity of the evolution equation ~(\ref{eq:Schrodinger}), one finds that the evolution equation for the wave function w.r.t. $t_1$ is given by
\begin{align}
    \left[ \partial_{t_1} - i\delta E(\p) - \Gamma_3(t_1) \circ \right] \vec{\psi}(\p,t,t_1 ) =&0\;,
\end{align}
which needs to be solved backward in time $t_1$ for $t>t_1>0$, with the initial condition 
\begin{align}
    \vec{\psi}(\p,t,t_1) =& \int_\q \frac{i\q}{\delta E(\q)} \Gamma_3\circ  (2\pi)^2 \delta^2(\p-\q)\;,\\
    =&  \Gamma_3(t)\circ  \frac{i\p}{\delta E(\p)} \;,
\end{align}
While the above re-arrangements can always be performed, we will in the following consider the radiative emission rates in a static QCD plasma, where $\Gamma_{3}(t)=\Gamma_{3}$ and $\vec{\psi}(\p,t,t_1 )= \vec{\psi}(\p,\Delta t)$ only depends on separation $\Delta t=t-t_1$. \\

We proceed to factor out the physical scales by defining the dimensionless variables
\begin{align}
    \Delta \tilde{t} = \frac{\mD^2}{2Pz(1-z)} \Delta t\;,\qquad
    \tilde{q} = \frac{q}{\mD}\;,~~
    \tilde{p} = \frac{p}{\mD}\;,
    \label{eq:DimensionlessVariablesIntro}
\end{align}
where $\mD^2/2Pz(1-z)$ is the inverse formation time of a splitting with small momentum transfer $\sim \mD$, and the energy becomes 
\begin{align}
    &\delta \tilde{E} (\tilde{\p}) = \frac{2Pz(1-z)}{\mD^2} \delta E(\p)\;,\\
    &= \tilde{\p}^2 +(1-z)\frac{m_{z}^2}{\mD^2}+z\frac{m_{1-z}^2}{\mD^2} -z(1-z)\frac{m_1^2}{\mD^2}\;.
\end{align}
By factoring out the parametric dependencies of the broadening kernel as $\tilde{C}(\tilde{\q}) = \frac{\mD^2}{C_R g^2T}C(\q)$, one finds
\begin{align}
    &\Gamma_3 = g^2 T ~\tilde{\Gamma}_3\;.
\end{align}
Now, expressing the wave function as  
\begin{align}
    \vec{\psi}(\p,\Delta t)  =&  g^2T~\frac{2Pz(1-z)}{\mD} \vec{\tilde{\psi}}(\tilde{\p},\Delta \tilde{t})\;.
\end{align}
the initial conditions can then be compactly expressed in terms of the dimensionless variables as
\begin{align}\label{eq:WaveFct}
    \vec{\tilde{\psi}}(\tilde\p,\Delta \tilde{t} = 0) =& \tilde{\Gamma}_3(t)\circ  \frac{i\tilde{\p}}{\delta \tilde{E}(\tilde{\p})} \;.
\end{align}
The evolution equation for the dimensionless wave function takes the form 
\begin{align}
    \left[  \partial_{\Delta \tilde{t}} + \delta\tilde{E}(\tilde{\p}) + \lambda\, \tilde{\Gamma}_3(t) \circ \right] \vec{\tilde{\psi}}(\tilde\p,\Delta \tilde{t})  =&0\;,
\end{align}
where $\lambda = g^2T \frac{2Pz(1-z)}{\mD^2}$ counts the number of small angle scatterings per formation time of a a splitting with small momentum transfer $\sim m_{D}$, and the splitting rate becomes
\begin{align}
    \frac{d\Gamma^a_{bc}}{dz}(P,z,\tilde{t})
    =&\frac{g^4 T P^a_{bc}(z)}{\pi } {\rm Re}~\int_0^{\tilde{t}} d{\Delta \tilde{t}}~ \int_{\tilde{\p}} \tilde{\p}\cdot \vec{\tilde{\psi}}(\tilde\p,\Delta \tilde{t}) \;.  
\end{align}
Still following earlier works~\cite{CaronHuot:2010bp}, the numerical determination of the rate can be further simplified by exploiting the isotropy of the wave-function in isotropic plasmas and introducing the transformation to the interaction picture. Defining
\begin{align}
    \tpsi_I(\tp,\Delta \tilde{t}) =& e^{i\delta\tilde{E}(\tp) \Delta \tilde{t}} \tilde{\p} \cdot \vec{\tpsi}( |\tp|,\Delta \tilde{t})\;,\\
    \vec{\tpsi}( |\tp|,\Delta \tilde{t}) =& e^{-i\delta\tilde{E}(\tp) \Delta \tilde{t}} \frac{\tilde{\p}}{\tp^2}  \tpsi_I(\tp,\Delta \tilde{t})\;,
\end{align}
one finds that $\tpsi_I(\tp,\Delta \tilde{t})$ follows the evolution equation
\begin{align}\label{eq:EvolInteraction}
    \left[  \partial_{\Delta \tilde{t}} + \lambda\, e^{i\delta\tilde{E}(\tp) \Delta \tilde{t}}  \tilde{\p}\cdot \tilde{\Gamma}_3 \circ e^{-i\delta\tilde{E}(\tp) \Delta \tilde{t}} \frac{\tilde{\p}}{\tp^2} \right] \tpsi_I(\tp,\Delta \tilde{t})  =&0\;,
\end{align}
with the initial condition 
\begin{align}\label{eq:InitialInteraction}
    \tpsi_I(\tp,\Delta \tilde{t}=0) = \tilde{\p} \cdot \tilde{\Gamma}_3\circ  \frac{i\tilde{\p}}{\delta \tilde{E}(\tilde{\p})} \;,
\end{align}
and the splitting rate can be compactly expressed as 
\begin{align}\label{eq:RateEqInteraction}
    \frac{d\Gamma^a_{bc}}{dz}
    =&\frac{g^4 T P^a_{bc}(z)}{\pi } {\rm Re}~\int_0^{\tilde{t}} d{\Delta \tilde{t}}~ \int_{\tilde{\p}}  e^{-i\delta\tilde{E}(\tp) \Delta \tilde{t}} \tpsi_I(\tp,\Delta \tilde{t}) \;,  
\end{align}
which is the form of the equation that we use for our numerical determination of the medium induced splitting rate for fixed values of $P$ and $z$.

We use a logarithmic grid to discretize the momentum $\tilde{p}$ and employ a standard numerical integration from the GNU scientific library \cite{Galassi_gnuscientific} to obtain the initial wave function $\tpsi_I(\tp,\Delta \tilde{t} =0)$ from Eq.~(\ref{eq:InitialInteraction}) at each point. Subsequently, the discretized wave functions $\tpsi_I(\tp,\Delta \tilde{t})$ are evolved using an Euler scheme, where we use a spline interpolation to interpolate the discrete wave function when numerically integrating the collision integral in Eq.~(\ref{eq:EvolInteraction}). 
Eventually, the two-dimensionally tabulated values of the wave function $\tpsi_I(\tp,\Delta \tilde{t})$ are made continuous using a two-dimensional spline and integrated numerically with the CUBA library~\cite{Hahn:2004fe}  to obtain the rate in Eq.~(\ref{eq:RateEqInteraction}).
We note that for the next-to-leading perturbative, as well as for the non-perturbative momentum broadening kernel, the $1/\qt^3$ IR behavior can lead to instabilities when evolving the coupled set of evolution equations. However, this problem can be resolved by separating the soft and hard contributions to momentum broadening, and treating the soft contributions in an expansion of the momentum transfer $\qt$ to perform the integrations analytically, as discussed in detail in Appendix \ref{ap:FiniteMedium}.

Even though the numerical solution for the rate can be obtained at all scales for a highly energetic parton, one can get away with using approximations in certain regimes which simplifies the calculation drastically. Numerous approximation have been developed in the literature \cite{Baier:1996sk} \cite{Zakharov:1996fv} \cite{Mehtar-Tani:2019tvy,Mehtar-Tani:2019ygg,Barata:2020sav} \cite{Andres:2020vxs}, during the following sections we will review the latest developments together with some traditional approximations, which we will compare to the full rate in Figure.~\ref{fig:FiniteMediumVSApprox}.

\subsection{Opacity expansion} 
Simplification to the rate occur, when the medium is short
and the hard particle does not frequently interact with the medium. In this regime, the rate can be computed perturbatively in an expansion in the number of interactions $N$ with the medium. This expansion is also known as the Gyulassy, Levai and Vitev (GLV) approximation\footnote{We note here that in contrast to the traditional GLV approximation we will not neglect thermal masses, and we will not take the soft gluon approximation.} \cite{Gyulassy:1999zd,Gyulassy:2000er}. It is easier to compute the expansion in the interaction picture introduced earlier, the wave function for the first order ($N=1$) is directly the initial condition defined in Eq.~(\ref{eq:InitialInteraction}) as we already take one scattering in the definition of the wave function 
\begin{align}
    \tpsi^{(1)}_I(\tp) = \tilde{\p} \cdot \tilde{\Gamma}_3\circ  \frac{i\tilde{\p}}{\delta \tilde{E}(\tilde{\p})} \;.
\end{align} 

By inserting the wave function into the definition of the rate in Eq.~(\ref{eq:RateEqInteraction}), one obtains
\begin{align}
    &\left. \frac{d\Gamma^a_{bc}}{dz} \right|_{N=1}(P,z,\tilde{t})
    \nonumber\\
    &=\frac{g^4 T P^a_{bc}(z)}{\pi } {\rm Re}\int_0^{\tilde{t}} d{\Delta \tilde{t}} \int_{\tilde{\p}}  e^{-i\delta\tilde{E}(\tp) \Delta \tilde{t}} \tilde{\p} \cdot \tilde{\Gamma}_3\circ  \frac{i\tilde{\p}}{\delta \tilde{E}(\tilde{\p})} \;.
\end{align}
The time integration can be done analytically and one finds  \cite{Gyulassy:1999zd,Gyulassy:2000er}
\begin{align}\label{eq:FiniteOpacity}
    &\left. \frac{d\Gamma^a_{bc}}{dz} \right|_{N=1}(P,z,\tilde{t})\nonumber\\
    &=\frac{g^4 T P^a_{bc}(z)}{\pi } \int_{\tilde{\p}}  \frac{1-\cos\left(\delta\tilde{E}(\tp) \tilde{t} \right)}{\delta\tilde{E}(\tp)} \tilde{\p} \cdot  \tilde{\Gamma}_3\circ  \frac{i\tilde{\p}}{\delta \tilde{E}(\tilde{\p})} \;.
\end{align}

\subsection{Resummed opacity expansion} 
Besides the straight opacity expansion, the authors of \cite{Andres:2020kfg} developed a resummation that tries to capture additional re-scatterings with the medium. During this section we will present this procedure, starting with the second order ($N=2$) correction to the wave function which obeys the following evolution equation 
\begin{align}
    \partial_{\Delta \tilde{t}} \tpsi^{(2)}_I(\tp,s) = 
    - \lambda\, e^{i\delta\tilde{E}(\tp) s}  \tilde{\p}\cdot \tilde{\Gamma}_3 \circ e^{-i\delta\tilde{E}(\tp) s} \frac{\tilde{\p}}{\tp^2} \tpsi^{(1)}_I(\tp) \;,
\end{align}
with initial condition  $ \tpsi^{(2)}_I(\tp,\Delta \tilde{t}=0) =0$. Integrating with respect to time, one finds 
\begin{align}\label{eq:SecondOrderWaveFct}
    &\tpsi^{(2)}_I(\tp,\Delta \tilde{t}) = \nonumber\\
    &- \lambda\, \int_{0}^{\Delta \tilde{t}} ds ~e^{i\delta\tilde{E}(\tp) s}  \tilde{\p}\cdot \tilde{\Gamma}_3 \circ e^{-i\delta\tilde{E}(\tp) s} \frac{\tilde{\p}}{\tp^2} \tpsi^{(1)}_I(\tp) \;.
\end{align}
Explicitly, the correction is given by 
\begin{align}
    &\tpsi^{(2)}_I(\tp,\Delta \tilde{t}) =-\lambda\, \int_{0}^{\Delta \tilde{t}} ds ~e^{i\delta\tilde{E}(\tilde{\p}) s}  \tilde{\p}\cdot \int_{\tilde{\q}} \nonumber\\
    &\left[ C_1 \tilde{C}(\tilde{\q}) + \frac{C_z}{z^2} \tilde{C}\left(\frac{\tilde{\q}}{z}\right) + \frac{C_{1-z}}{(1-z)^2} \tilde{C}\left(\frac{\tilde{\q}}{1-z}\right)\right] \nonumber\\
    &\left[e^{-i\delta\tilde{E}(\tilde{\p}) s} \frac{\tilde{\p}}{\tp^2} \tpsi^{(1)}_I(\tp) - e^{-i\delta\tilde{E}(\tilde{\p}-\tilde{\q}) s} \frac{\tilde{\p} - \tilde{\q}}{|\tilde{\p} - \tilde{\q}|^2} \tpsi^{(1)}_I(\tp-\tq)\right],
\end{align}
where we have utilized a change of variable in the $\tilde{\q}$ integral to combine the different terms in the collision integral in Eq.~(\ref{eq:ThreeBodyInt}). Now following~\cite{Andres:2020kfg} and considering the difference
\begin{align}\label{eq:Difference}
    &\tpsi^{(1)}_I(\tp) - e^{i(\delta\tilde{E}(\tilde{\p})-\delta\tilde{E}(\tilde{\p}-\tilde{\q})) s} \frac{ \tilde{\p}^2 - \tilde{\p}\cdot\tilde{\q}}{|\tilde{\p} - \tilde{\q}|^2} \tpsi^{(1)}_I(\tp-\tq) \;,
\end{align}
one concludes, that for small momentum transfer ($\tq\ll1$), the two terms cancel each other, while for larger momentum transfer ($\tq\gg1$) the phase factor oscillates rapidly and the second term does not contribute significantly to the integral. By introducing a cut-off scale $\mu^2$ and dropping the second term in Eq.~(\ref{eq:Difference}), one can then approximately keep track of the contributions $\tilde{\Sigma}(\mu^2) = \int_{ \tilde{\q}^2 > \mu^2 } \tilde{C} ( \tilde{\q})$ with large momentum transfer, yielding
\begin{align}
    \tpsi^{(2)}_I(\tp,\Delta \tilde{t}) =-\lambda\, \int_{0}^{\Delta \tilde{t}} ds ~ \tpsi^{(1)}_I(\tp)  \tilde{\Sigma}_3(\mu^2,z) \;,
\end{align}
where $\tilde{\Sigma}_3(\mu^2,z)=\Big[ C_1 \tilde{\Sigma}(\mu^2) + C_z \tilde{\Sigma}(\mu^2/z^2) + C_{1-z} \tilde{\Sigma}(\mu^2/(1-z)^2)\Big]$. Inserting Eqns.~(\ref{eq:InitialInteraction}) and (\ref{eq:SecondOrderWaveFct}) into (\ref{eq:RateEqInteraction}),
the expansion of the splitting rate is now given by
\begin{align}
    &\left. \frac{d\Gamma^a_{bc}}{dz} \right|_{N=X}(P,z,\tilde{t}) 
    \nonumber\\
    &= \frac{g^4 T P^a_{bc}(z)}{\pi } {\rm Re}\int_0^{\tilde{t}} d{\Delta \tilde{t}} \int_{\tilde{\p}}  e^{-i\delta\tilde{E}(\tp) \Delta \tilde{t}} \tpsi^{(1)}_I(\tp)  \nonumber\\
    &+ \frac{g^4 T P^a_{bc}(z)}{\pi } {\rm Re}\int_0^{\tilde{t}} d{\Delta \tilde{t}} \int_{0}^{\Delta \tilde{t}} ds ~ \nonumber\\
    &\times\int_{\tilde{\p}}  e^{-i\delta\tilde{E}(\tp) \Delta \tilde{t}}   \tpsi^{(1)}_I(\tp) \left( -\lambda  \tilde{\Sigma}_3(\mu^2,z) \right) + \cdots \;.
\end{align}
After performing the time integral ($ds$), one finds
\begin{align}
    &\left. \frac{d\Gamma^a_{bc}}{dz} \right|_{N=X}(P,z,\tilde{t})
    \nonumber\\
    &= \frac{g^4 T P^a_{bc}(z)}{\pi } {\rm Re}\int_0^{\tilde{t}} d{\Delta \tilde{t}} \int_{\tilde{\p}}  e^{-i\delta\tilde{E}(\tp) \Delta \tilde{t}} \tpsi^{(1)}_I(\tp)  \nonumber\\
    &+ \frac{g^4 T P^a_{bc}(z)}{\pi } {\rm Re}\int_0^{\tilde{t}} d{\Delta \tilde{t}}  \nonumber\\
    &\times \int_{\tilde{\p}}  e^{-i\delta\tilde{E}(\tp) \Delta \tilde{t}}   \tpsi^{(1)}_I(\tp) \left( -\lambda  \tilde{\Sigma}_3(\mu^2,z) \Delta \tilde{t} \right) + \cdots \;.
\end{align}
One notices that subsequent terms with additional time integration will exponentiate, yielding the final result 
\begin{align}\label{eq:FiniteOpacityX}
    &\left. \frac{d\Gamma^a_{bc}}{dz} \right|_{N=X}(P,z,\tilde{t})\nonumber\\
    &= \frac{g^4 T P^a_{bc}(z)}{\pi } {\rm Re}\int_0^{\tilde{t}} d{\Delta \tilde{t}} \int_{\tilde{\p}}  e^{-(i\delta\tilde{E}(\tp)  +\lambda  \tilde{\Sigma}_3(\tp^2,z))\Delta \tilde{t}} \tpsi^{(1)}_I(\tp)   \;,
\end{align}
where following \cite{Andres:2020kfg}, we employed $\mu^2 = \tp^2$ for the cutoff scale. 

\subsection{Harmonic Oscillator expansion}

When the typical energy evolved in the radiation is much larger than the medium temperature $(Pz(1-z) \gg T)$, the formation time is large so that multiple soft scatterings have to be resummed. By treating the multiple soft scatterings in diffusion approximation, the evolution equation for the Green's function can be re-cast into the form of a harmonic oscillator type equation which can be solved analytically \cite{Arnold:2008iy}.
Rather than using this approximation only, we will make use of recent calculations which go beyond this simple harmonic oscillator limit by treating the hard scatterings as a perturbative correction on top of the resummed infinitely many soft scatterings \cite{Mehtar-Tani:2019tvy,Mehtar-Tani:2019ygg,Barata:2020sav}. Here, we will only compute the first correction, i.e. a single hard scattering in addition to many soft scatterings. 

Starting with the short-distance behavior defined in Eq.~(\ref{UV_limit}), one can introduce a scale $Q^2$ to evaluate the logarithm and separate it as follows 
\begin{align}
    \bar{C}(\bbp) =& \frac{1}{C_R}C(\bbp)\\
    =& \frac{\gs^4 T^3}{16\pi}\mathcal{N} \bbp^2 \ln \left( \frac{4Q^2}{\xi \mDsq } \right) + \frac{\gs^4 T^3}{16\pi}\mathcal{N} \bbp^2 \ln \left( \frac{1}{Q^2 \bbp^2} \right) \label{eq:Separation}
\end{align}
where  $\xi= 4\frac{\gs^4 T^2 }{\mDsq} e^{-4\pi\frac{  \hat{q}_0}{C_R \gs^4 T^3 \mathcal{N}}}\simeq 0.1702$ for $T=500$MeV as denoted in Tab.~\ref{tab:full_Cbp} . Based on this separation, the Harmonic oscillator (HO) kernel is now defined as the first part of Eq.~(\ref{eq:Separation}), i.e.
\begin{align}
    \bar{C}^{HO}(\bbp) =& \frac{\gs^4 T^3}{16\pi}\mathcal{N} \bbp^2 \ln \left( \frac{4Q^2}{\xi \mDsq } \right)\;,
\end{align}
which is used to calculate the Green's function subject to multiple soft-scatterings, while the remainder is treated perturbatively.
Instead of using only the short-distance limit, i.e. $\bar{C}^{\rm pert} (\bbp) = \frac{\gs^4 T^3}{16\pi}\mathcal{N} \bbp^2 \ln \left( \frac{1}{Q^2 \bbp^2} \right)$, it is better to define the correction to the kernel as the difference 
\begin{equation}
    \bar{C}^{\rm pert} (\bbp) = \bar{C}(\bbp) - \bar{C}^{HO}(\bbp)\;,
\end{equation}
where we use the full definition of $\bar{C}(\bbp)$, i.e. the numerical spline.
The radiation spectrum will also be separate to the sum of the HO and the first correction
\begin{align}\label{eq:FiniteHO}
    \frac{dI^{NLO}}{dz}(P,z,t) = \frac{dI^{HO}}{dz}(P,z,t)  + \frac{dI^{(1)}}{dz}(P,z,t) \;,
\end{align}
where the correction is computed using a first order opacity expansion with the kernel $C^{\rm pert} (\bbp)$ around the harmonic oscillator solution.
Following \cite{Mehtar-Tani:2019tvy,Mehtar-Tani:2019ygg,Barata:2020sav,Barata:2021wuf}, the scale $Q^2$ is the typical momentum of the radiated quanta defined self-consistently by using 
\begin{align}
    Q^2 (P,z) =& \sqrt{ Pz(1-z) \hat{q}_{\rm eff}(Q^2)}\;,\\
    \hat{q}_{\rm eff}(Q^2) =& \frac{\gs^4 T^3}{4\pi}\mathcal{N}  \left[ C_1+ C_z z^2 + C_{1-z} (1-z)^2 \right] \nonumber\\
    &\times\ln \left( \frac{4Q^2}{\xi \mDsq } \right)\;,
\end{align}
where $\hat{q}_{\rm eff}(Q^2)$ is the coefficient of the three-body interaction term $\Gamma_3$, obtained by plugging $C^{HO}(\bbp)$ in Eq.~(\ref{eq:ThreeBodyInt}).

\subsubsection{Leading Order}
Using $C^{HO}(\bbp)$ the rate equations can be solved analytically \cite{Zakharov:1996fv,Zakharov:1997uu,Baier:1996kr}, historically the result was obtained in terms of the spectrum
\begin{equation}
    \frac{dI^{HO}}{dz}(P,z,t) = \frac{g^2P^a_{bc}(z)}{4\pi^2}\ln|\cos\Omega t|\;,
\end{equation}
where we define the frequency
\begin{equation}
    \Omega = \frac{1-i}{2} \sqrt{\frac{\hat{q}_{\rm eff}(Q^2)}{Pz(1-z)}}\;.
\end{equation}
By applying a time derivative \cite{CaronHuot:2010bp}, one obtains the leading order harmonic oscillator rate
\begin{equation}
    \frac{d\Gamma^{HO}}{dz}(P,z,t) = -\frac{g^2P^a_{bc}(z)}{4\pi^2}{\rm Re~}\Omega\tan\Omega t\;.
\end{equation}

\subsubsection{Next to Leading order}
While the leading order HO term can be seen as a resummation of multiple soft scatterings with the medium, the next-to-leading order correction introduces the effect of one `hard' scattering with the medium\footnote{Note that in the original papers~\cite{Mehtar-Tani:2019tvy,Mehtar-Tani:2019ygg,Barata:2020sav}, the NLO Harmonic Oscillator approximation has been named `Improved opacity expansion', since it involves an expansion in the number of hard scatterings while keeping infinitely many soft scatterings.}.
One obtain the correction by making use of the separation in Eq.~(\ref{eq:FiniteHO}), which translates to a separation of the propagators 
\begin{align}
    G(t,\bp;t_1,\p) = G^{HO}(t,\bp;t_1,\bm{y}) + G^{(1)}(t,\bp;t_1,\bm{y})\;.
\end{align}
By inserting the full propagator into the evolution Eq.~(\ref{eq:Schrodinger}), and using the fact that the propagator $G^{HO}(t,\q;t_1,\p)$ is solution to the equation 
\begin{align}
    &\left[ i \partial_t + \frac{\partial_{\bp}^2}{2Pz(1-z)} + M_{\rm eff} + i \Gamma^{HO}_3(\bp) \right] G^{HO}(\bp,t;\bm{y},t_1) \nonumber\\
    &= i\delta(t-t_1) \delta^{(2)}(\bp-\bm{y})\;.
\end{align}
One finds the evolution equation of the next-to-leading order propagator $G^{(1)}(t,\bp;t_1,\bm{y})$
\begin{align}
    &\left[ i \partial_t + \frac{\partial_{\bp}^2}{2Pz(1-z)} + M_{\rm eff} + i \Gamma^{HO}_3(\bp) \right] G^{(1)}(\bp,t;\bm{y},t_1) \nonumber\\
    &= -i\Gamma^{\rm pert}_3(\bp) G^{HO}(\bp,t;\bm{y},t_1)\;,
\end{align}
where one neglects next-to-next-to-leading order terms $\propto \Gamma^{\rm pert}_3(\bp) G^{(1)}(\bp,t;\bm{y},t_1)$.
Strikingly, the evolution equation can be solved analytically to obtain the spectrum \cite{Mehtar-Tani:2019tvy,Mehtar-Tani:2019ygg,Barata:2020sav}
\begin{align}
    &\frac{dI^{(1)}}{dz}(P,z,t) \nonumber\\
    =& \frac{g^2 P^a_{bc}(z)}{4\pi^2} {\rm Re} \int_0^t ds~ \int_0^\infty \frac{2 d u}{u} \left[ C_1 \bar{C}^{\rm pert}(u)+C_z \bar{C}^{\rm pert}(zu) \right.\nonumber\\
    &\left.+C_{1-z} \bar{C}^{\rm pert}((1-z)u)\right] e^{k^2(s)u^2} \;,\\
    =& \frac{g^2 P^a_{bc}(z)}{4\pi^2} {\rm Re} \int_0^t ds~ \int \frac{2 d u}{u} \bar{C}^{\rm pert}(u)\left[ C_1e^{k^2(s)u^2} +C_z e^{\frac{k^2(s)}{z^2} u^2} \right.\nonumber\\
    &\left.+C_{1-z} e^{\frac{k^2(s)}{(1-z)^2}u^2}\right]  \;,
\end{align}
where we define
\begin{equation}
    k^2(s)= \frac{iPz(1-z)\Omega}{2} \left[ \cot\Omega s - \tan\Omega(t-s) \right]\;.
\end{equation}
When presenting numerical results for the NLO harmonic oscillator approximation, we compute the integrated spectrum in Eq.~(\ref{eq:FiniteHO}) and subsequently perform a numerical derivative w.r.t. $t$ to obtain the rate shown in Fig.~\ref{fig:FiniteMediumVSApprox}.

\section{Numerical Results}\label{sec:Results}

\begin{figure}
    \includegraphics[width=0.45\textwidth]{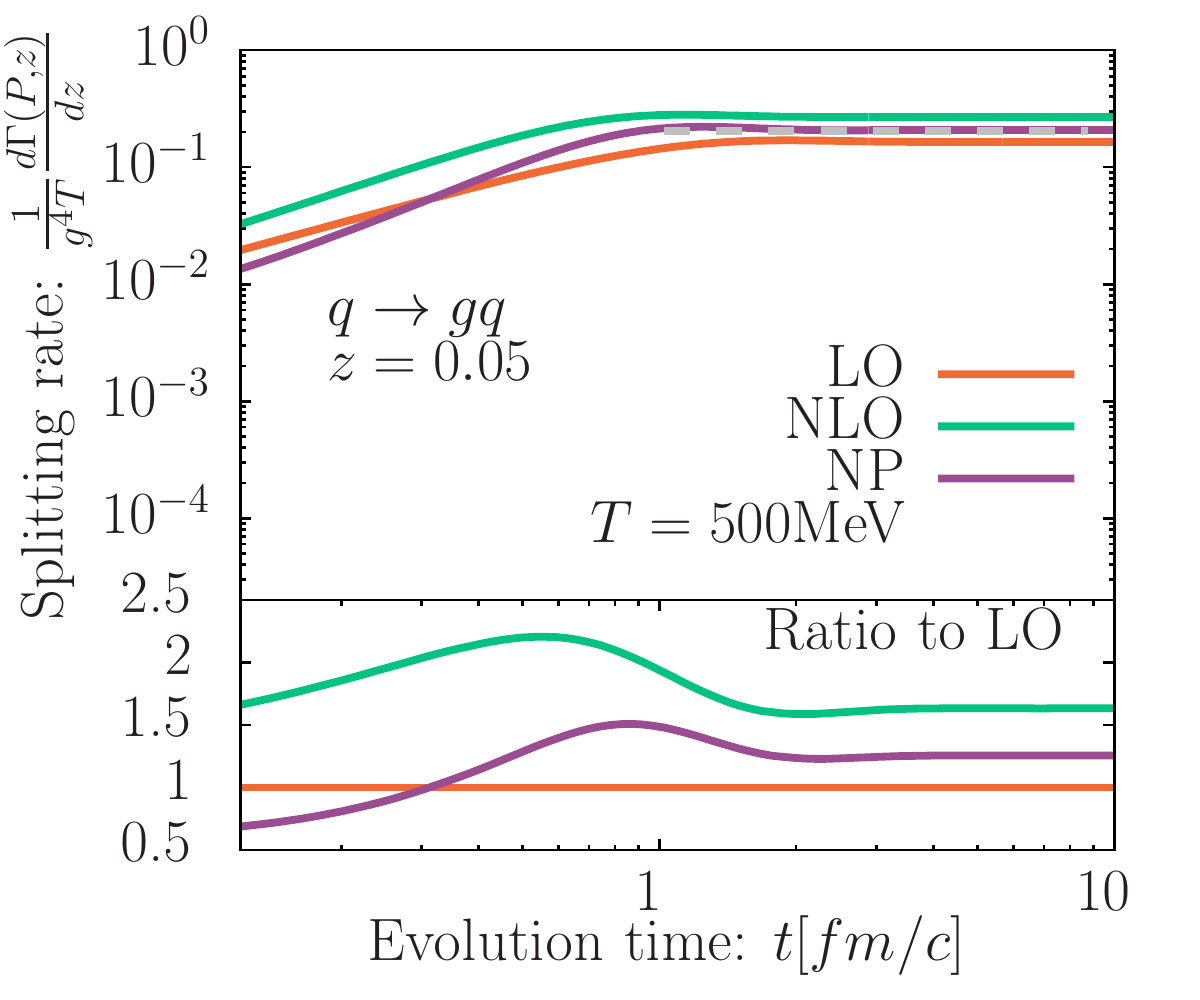}
    \includegraphics[width=0.45\textwidth]{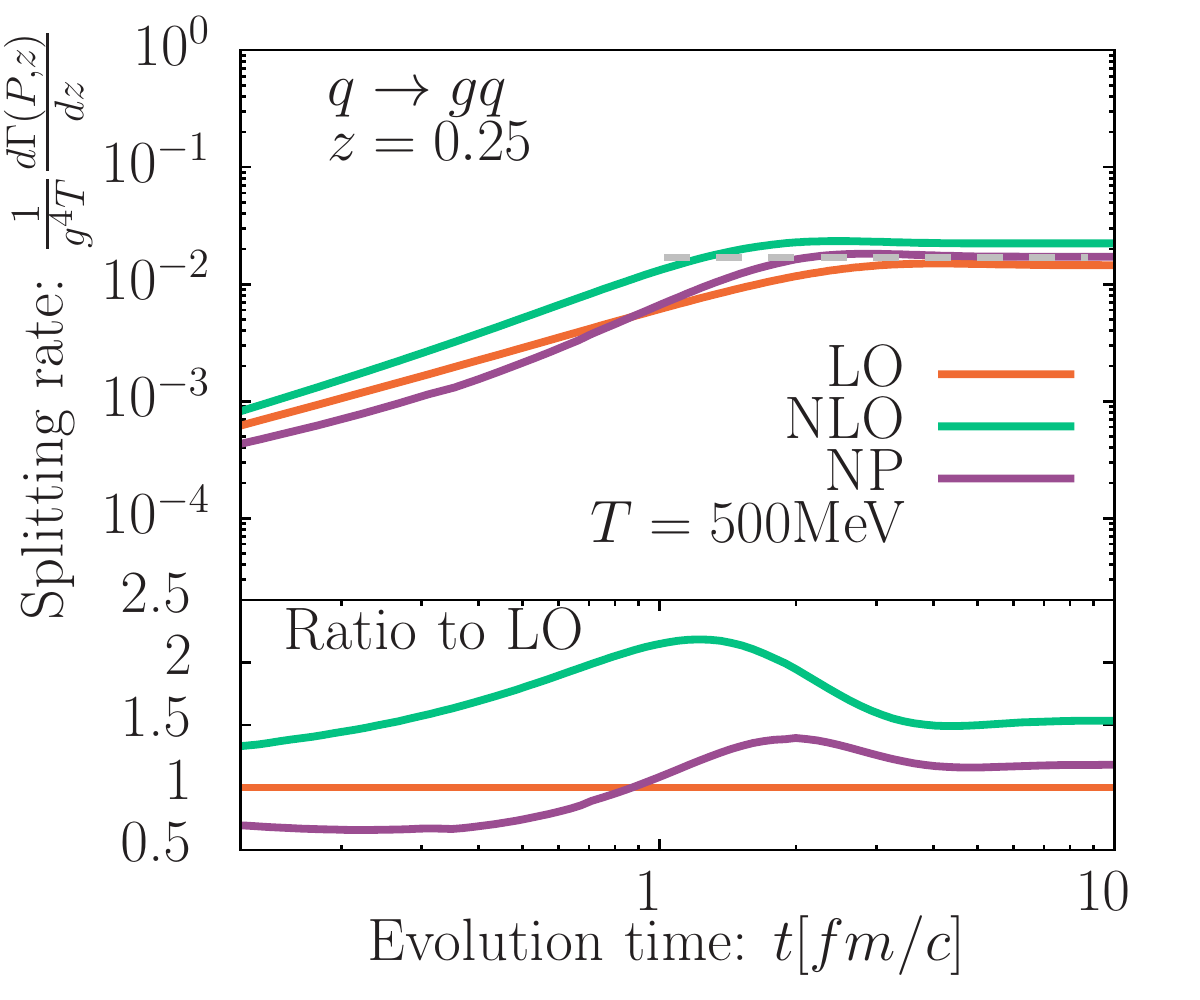}
    \includegraphics[width=0.45\textwidth]{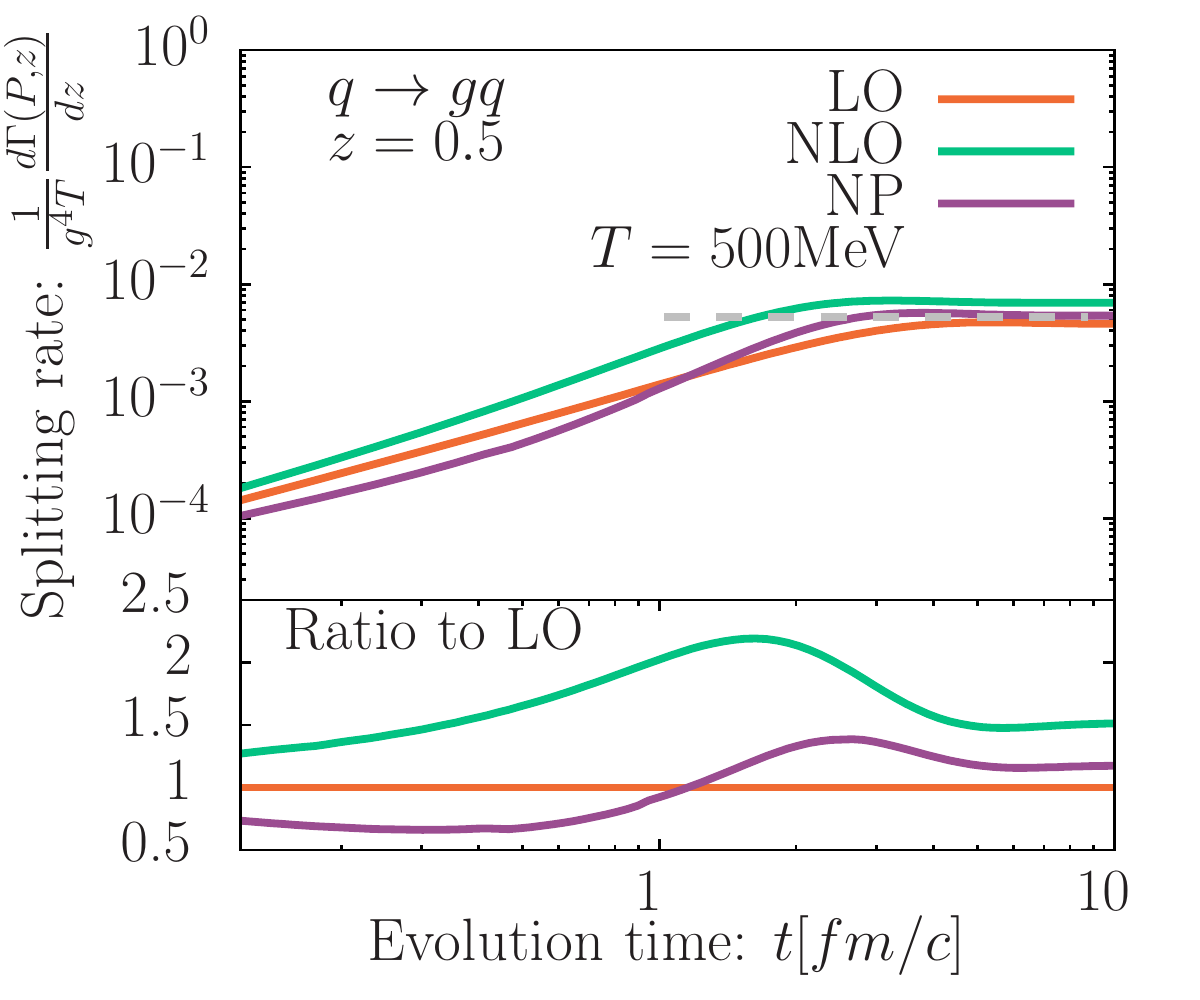}
    \caption{Splitting rate for the medium-induced emission of a gluon from a parent quark with energy $P=300T$ in an equilibrium plasma with temperature $T=500$MeV as a function of the evolution time $t$. Each panel represent a different gluon momentum fraction $z=0.05,0.25,0.5$ from top to bottom. Different curves in each panel show the results for the different, leading order (LO), next-to-leading order (NLO) and non-perturbative (NP) momentum broadening kernels in Fig.~\ref{fig:InverseKernel}. Dashed lines correspond to the (AMY) splitting rates~\cite{Arnold:2003zc} in an infinite medium  \cite{Moore:2021jwe}. The lower panel of each plot displays the ratio to the LO results.  }
    \label{fig:FiniteMediumVSKernels}
\end{figure}

\begin{figure}
    \includegraphics[width=0.45\textwidth]{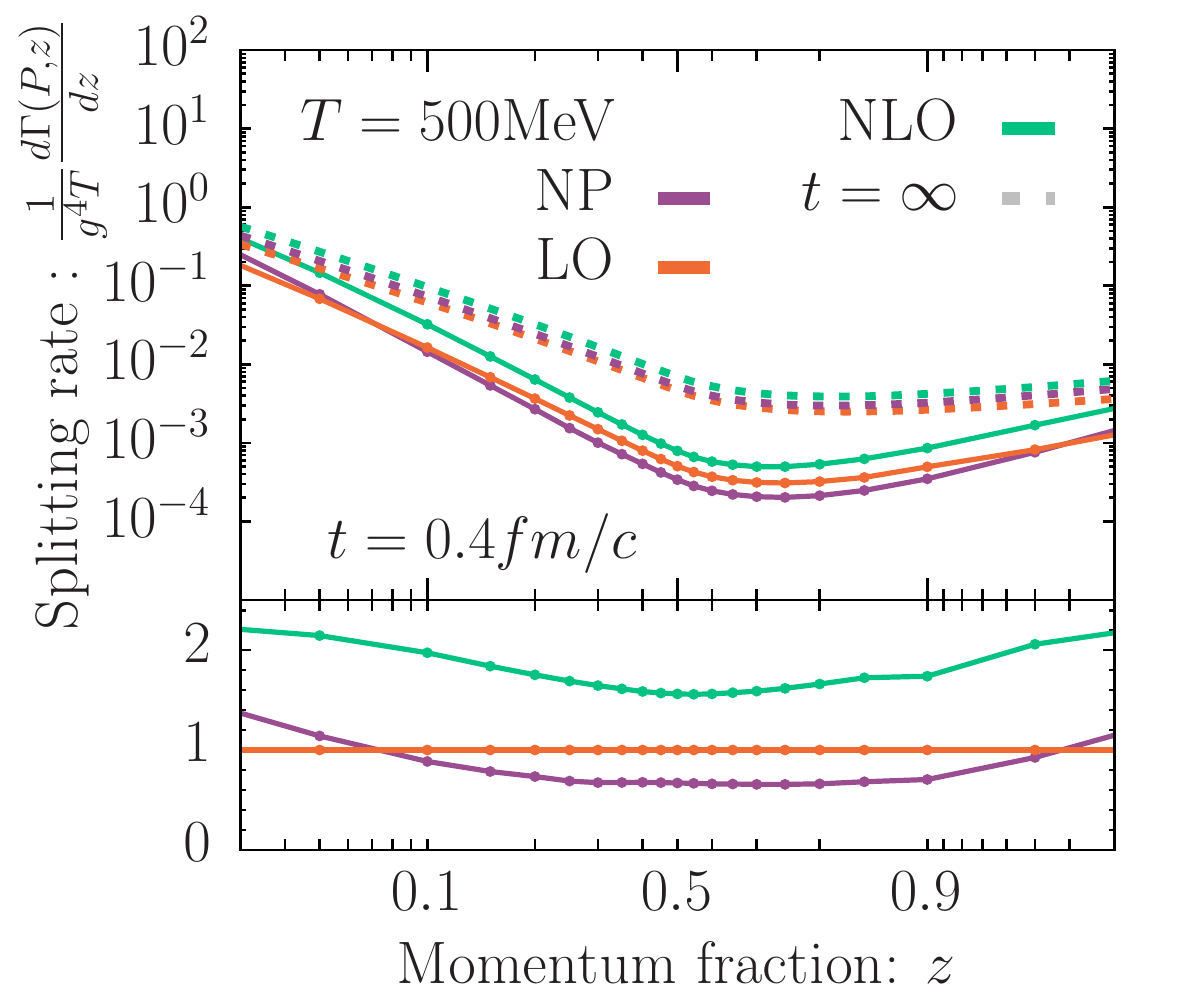}
    \includegraphics[width=0.45\textwidth]{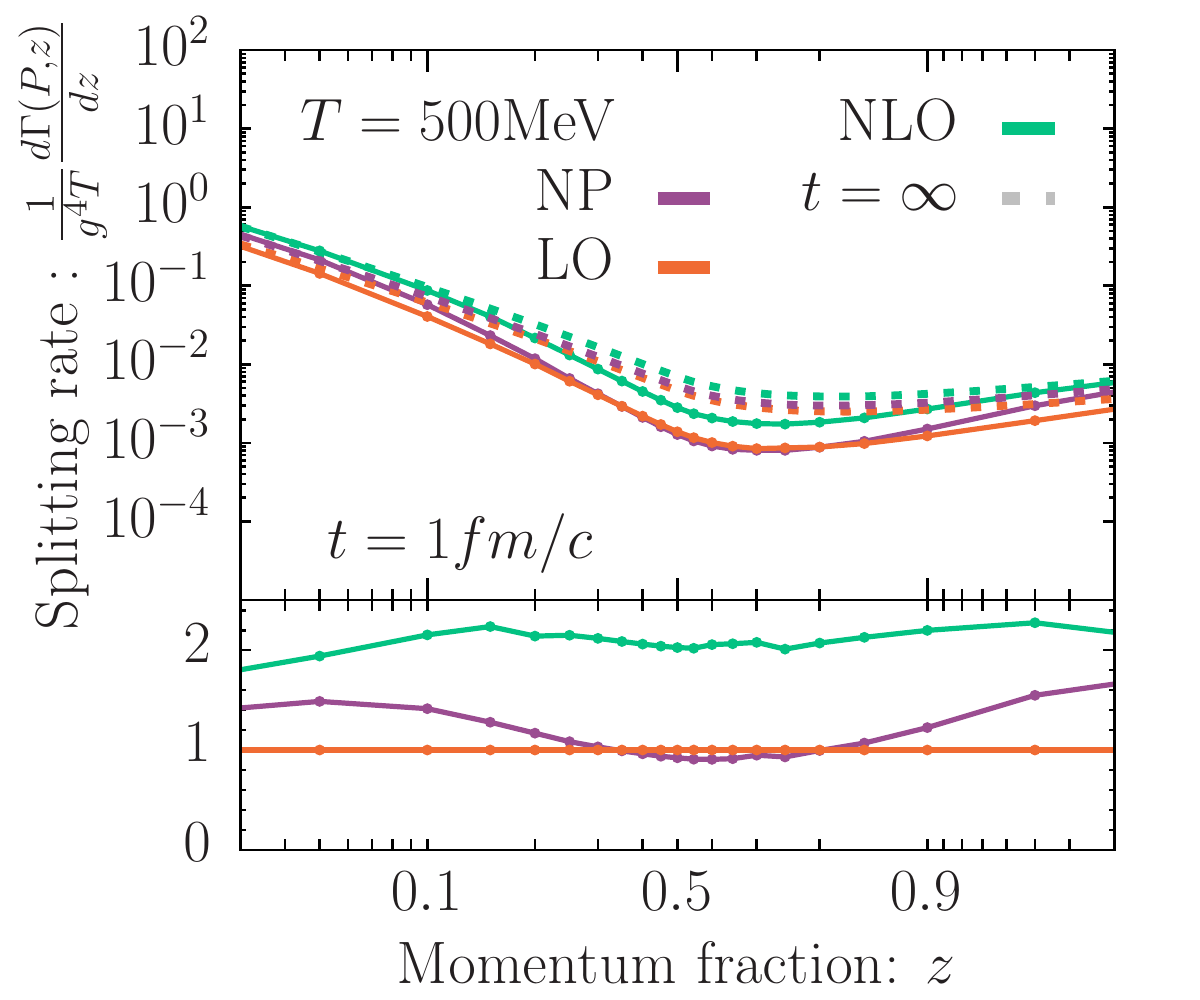}
    \includegraphics[width=0.45\textwidth]{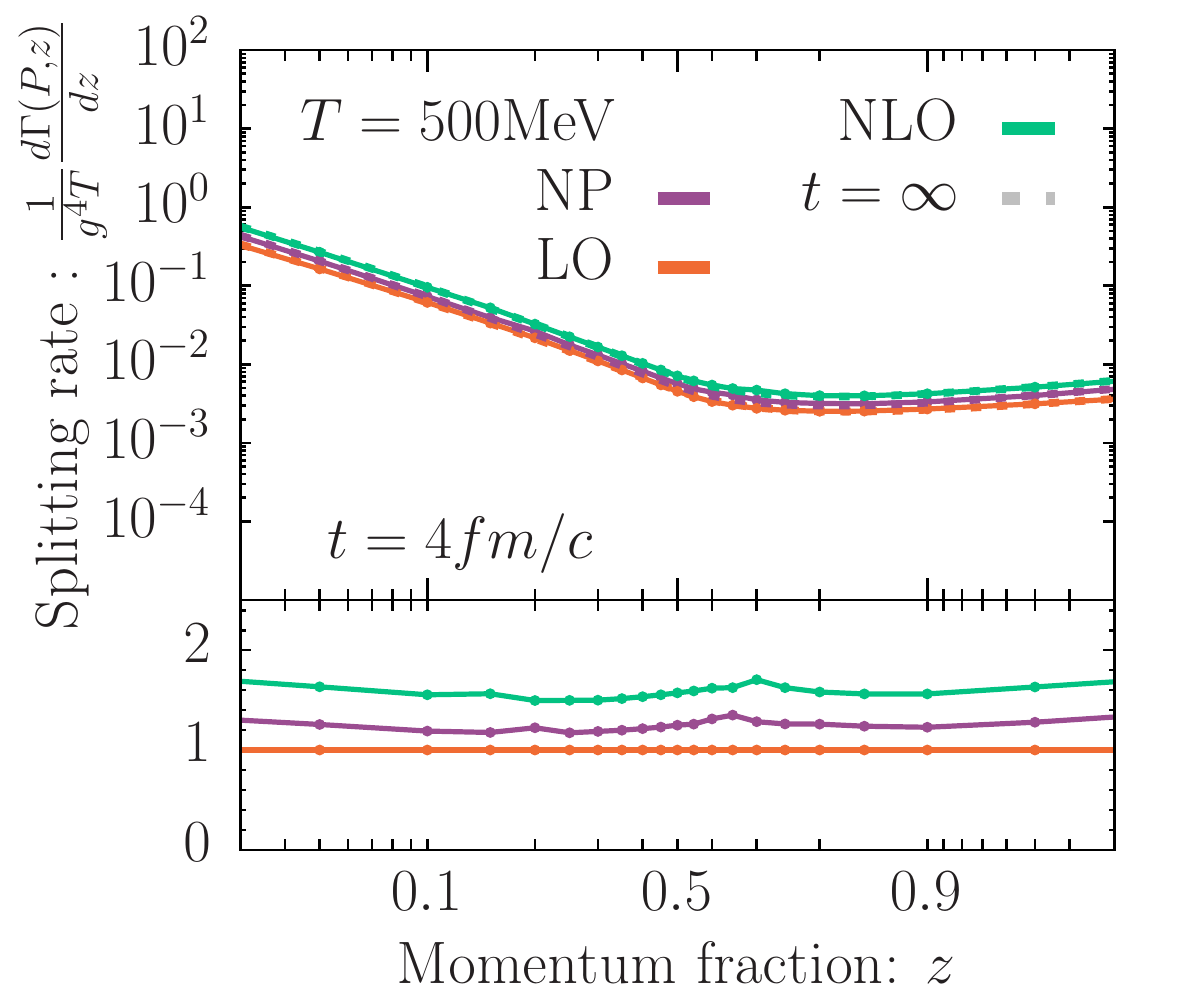}
    \caption{Splitting rate for the medium-induced emission of a gluon from a parent quark with energy $P=300T$ in an equilibrium medium with temperature $T=500$MeV as a function of momentum fraction of the radiated gluon $z$. Different panels show the rate $d\Gamma/dz$ at fixed times $t=0.4,1,4 fm/c$ from top to bottom.  Different curves in each panel show the results for the different, leading order (LO), next-to-leading order (NLO) and non-perturbative (NP) momentum broadening kernels in Fig.~\ref{fig:InverseKernel}. Dashed lines  $(t=\infty)$ correspond to the (AMY) splitting rates~\cite{Arnold:2003zc} in an infinite medium  \cite{Moore:2021jwe}. The lower panel of each plot shows the ratio to the splitting rate for the LO momentum broadening kernel.}
    \label{fig:FiniteMediumFctOfz}
\end{figure}

\begin{figure}
    \includegraphics[width=0.45\textwidth]{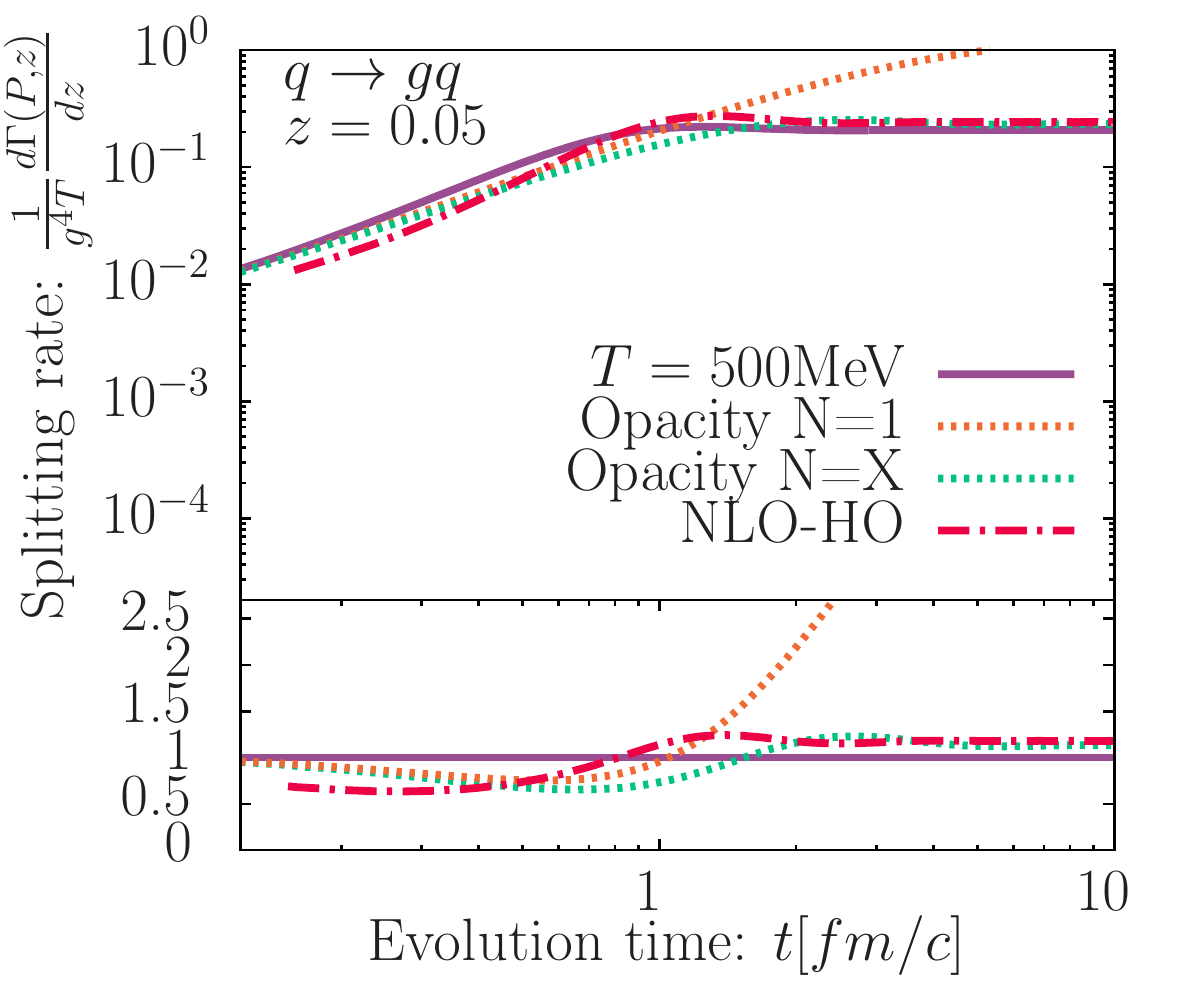}
    \includegraphics[width=0.45\textwidth]{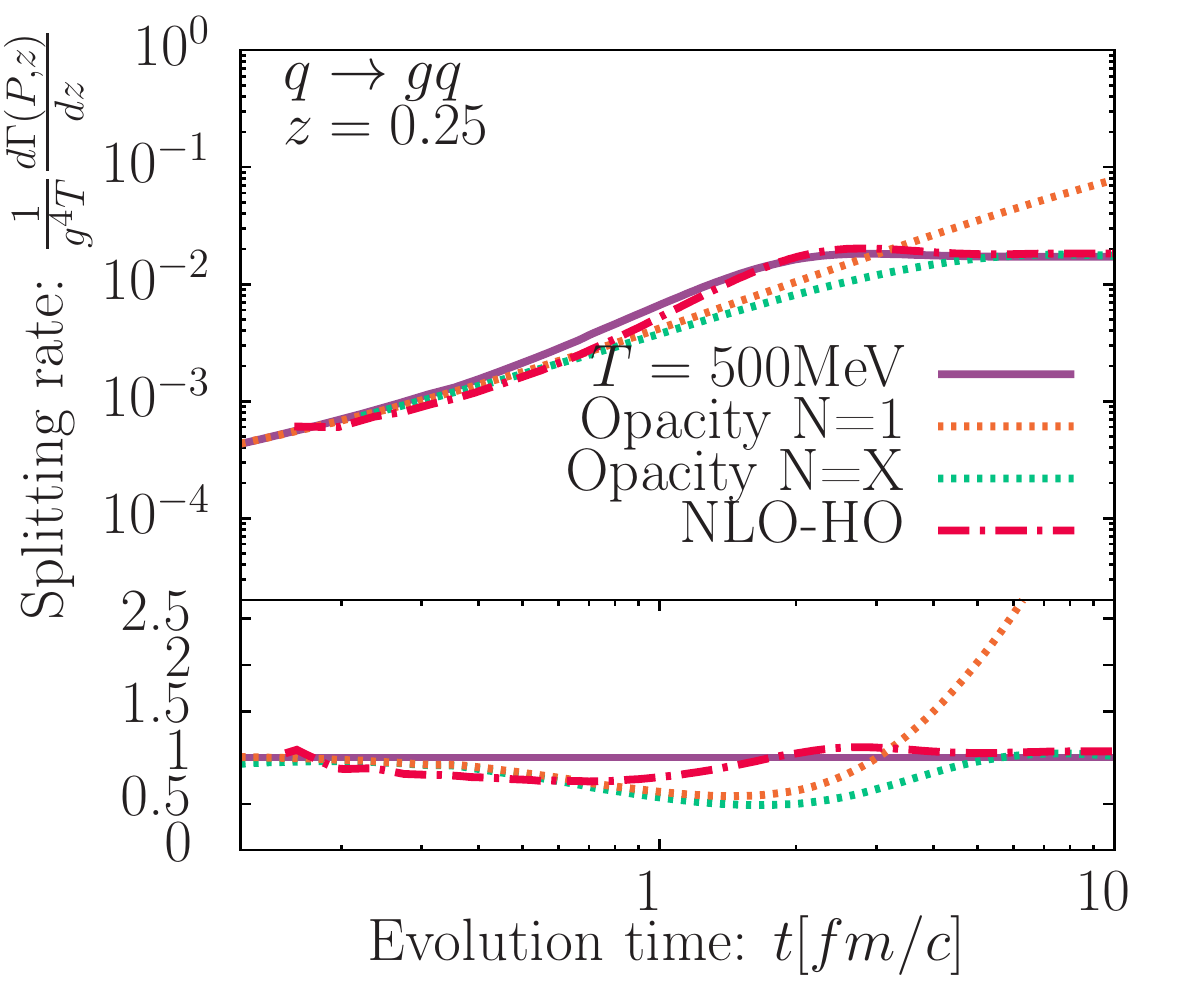}
    \includegraphics[width=0.45\textwidth]{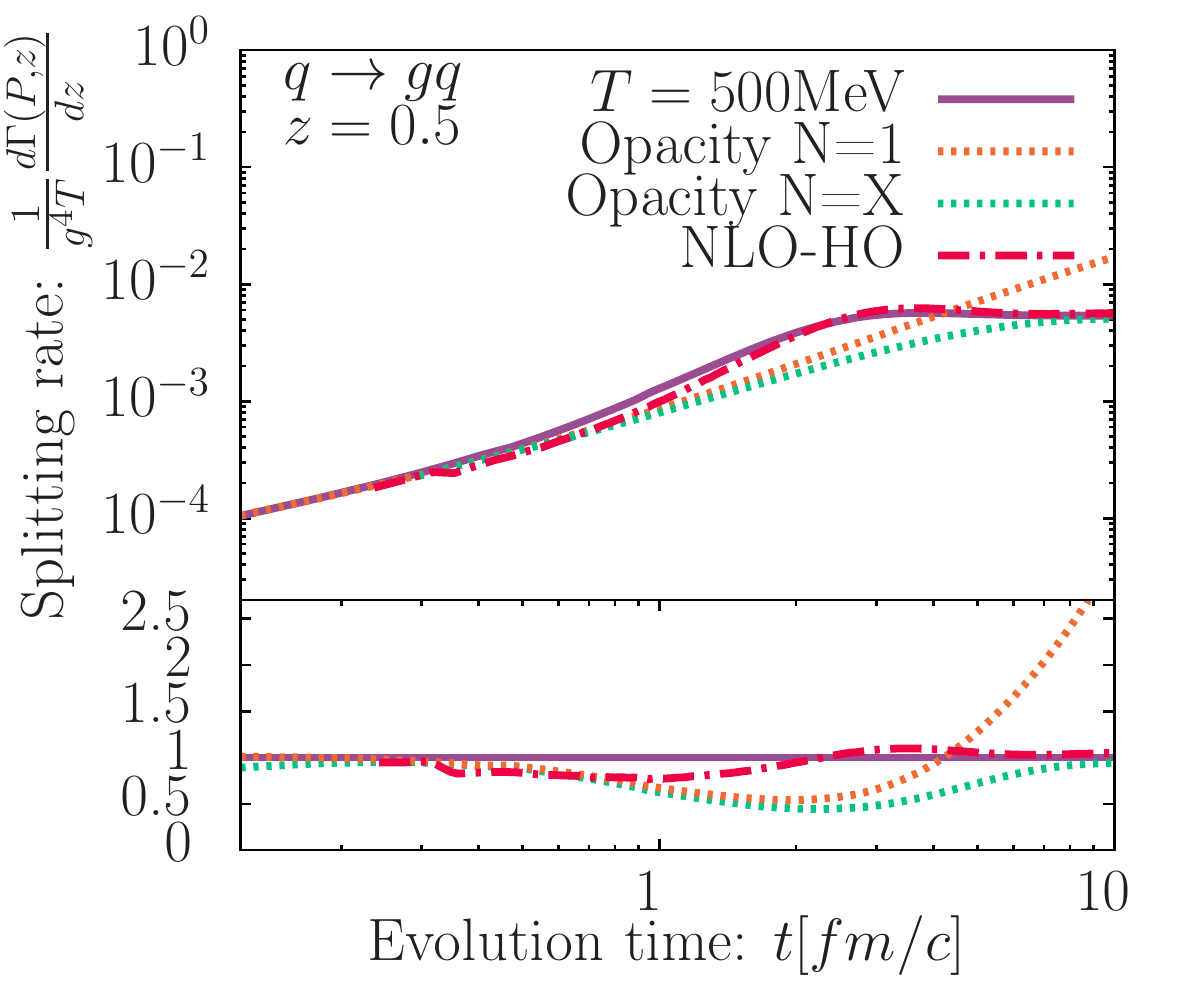}
    \caption{Splitting rate for the medium-induced emission of a gluon from a parent quark with energy $P=300T$ as a function of the evolution time $t$. Each panel represent a different gluon momentum fraction $z=0.05,0.25,0.5$ from top to bottom. We compare different approximations of the in-medium splitting rate, namely the opacity expansion at $N=1$ Eq.(\ref{eq:FiniteOpacity}), the resummed opacity rate of Eq.~(\ref{eq:FiniteOpacityX}) ($N=X$) and the NLO expansion around the Harmonic Oscillator Eq.~(\ref{eq:FiniteHO}) (NLO-HO) to the full result ($T=500{\rm MeV}$). Note that all results are obtained with the non-perturbative collision kernel. The lower panel of each plot displays the ratio to the full rate.}
    \label{fig:FiniteMediumVSApprox}
\end{figure}

\begin{figure}
    \includegraphics[width=0.45\textwidth]{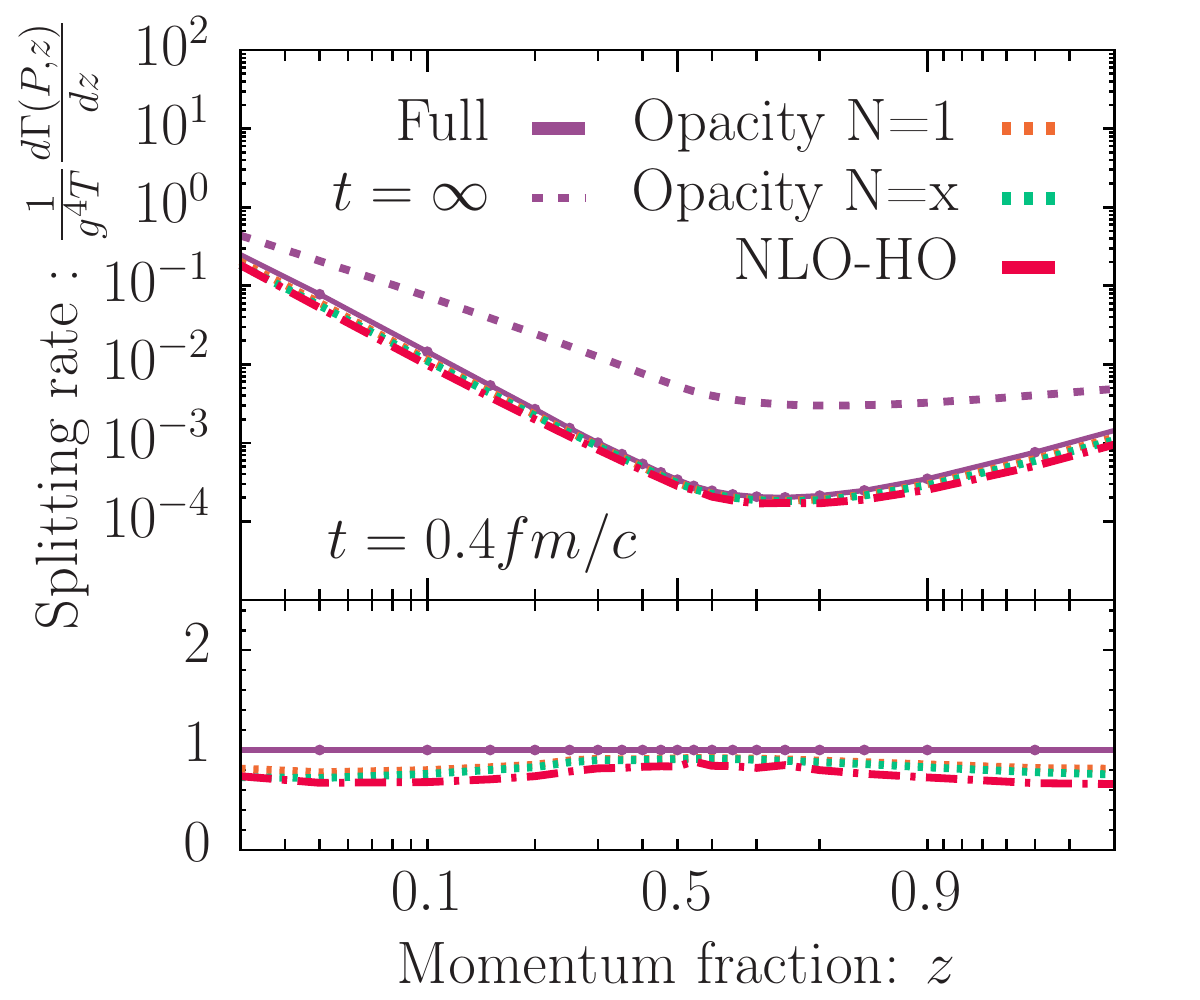}
    \includegraphics[width=0.45\textwidth]{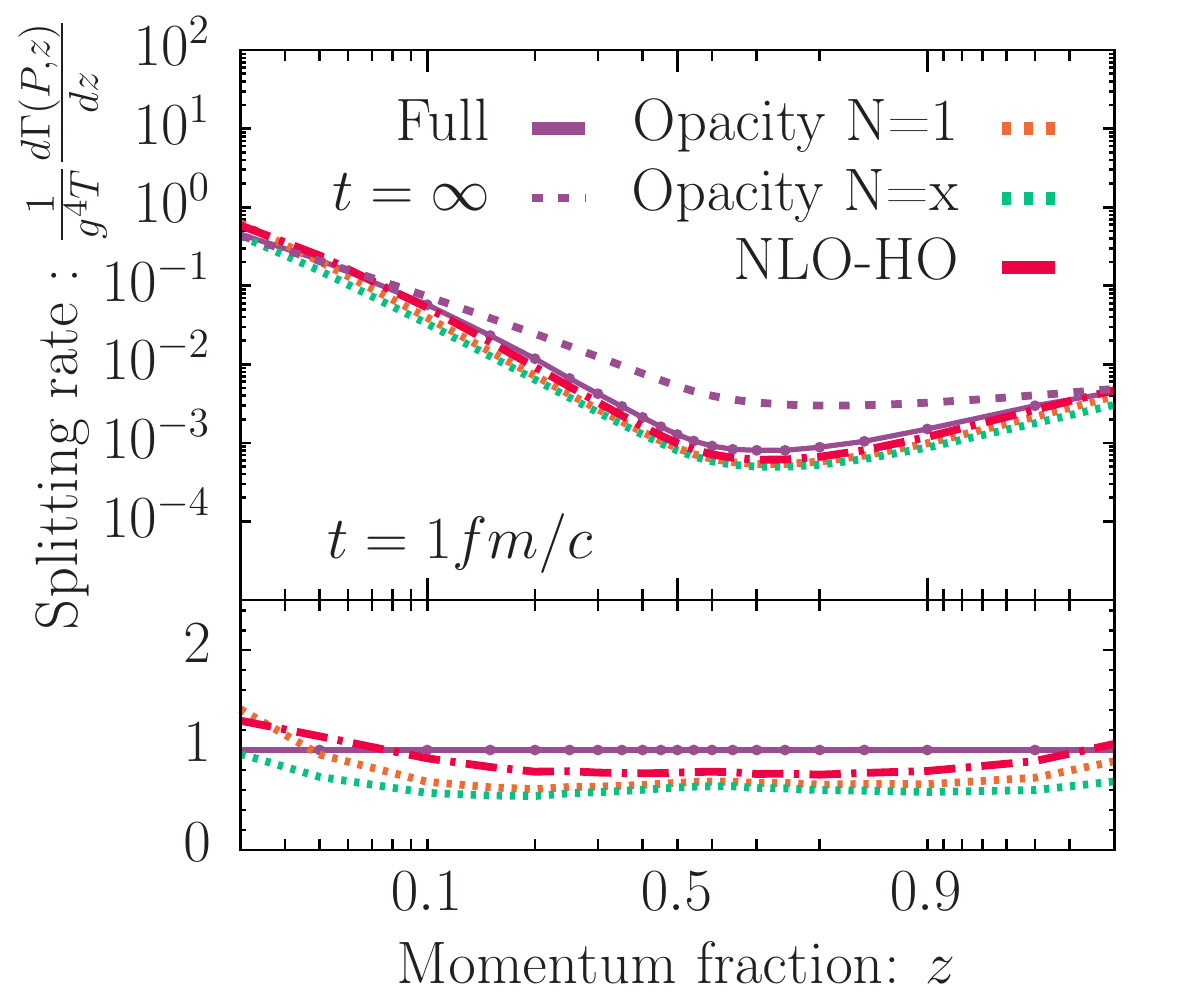}
    \includegraphics[width=0.45\textwidth]{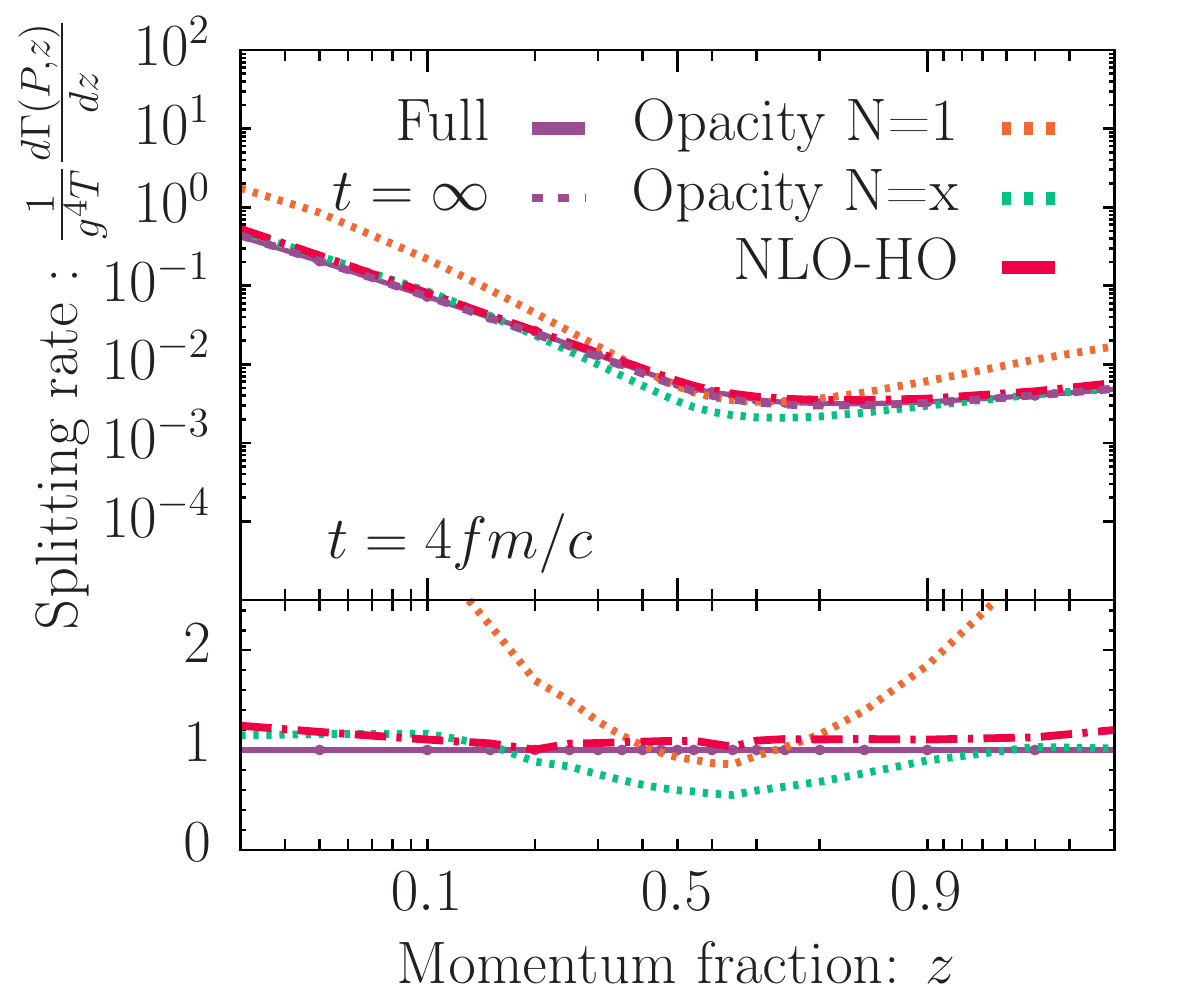}
    \caption{Splitting rate for the medium-induced emission of a gluon from a parent quark with energy $P=300T$ as a function of momentum fraction of the radiated gluon $z$. Different panels show the rate $d\Gamma/dz$ at fixed times $t=0.4,1,4 fm/c$.  We compare different approximations of the in-medium splitting rate, namely the opacity expansion at $N=1$ Eq.(\ref{eq:FiniteOpacity}), the resummed opacity rate of Eq.~(\ref{eq:FiniteOpacityX}) ($N=X$) and the NLO expansion around the Harmonic Oscillator Eq.~(\ref{eq:FiniteHO}) (NLO-HO) to the full result ($T=500{\rm MeV}$). Note that all results are obtained with the non-perturbative collision kernel. The lower panel of each plot shows the ratio to the full splitting rate.}
    \label{fig:FiniteMediumVSApproxFctOfz}
\end{figure}

We now turn to the discussion of numerical results for the in-medium radiation rate. We numerically obtain the rate for the different (LO,NLO,NP) broadening kernels $C(\qt)$ as described in detail in App.~\ref{ap:FiniteMedium}; the software to calculate the rates including the tabulation of the broadening  kernel is publicly available on GitHub~\cite{GitHub}. We will illustrate our results at the example of the radiation of a gluon by a parent quark of energy $P=300T$ in an equilibrium medium with constant temperature $T=500 {\rm MeV}$ below, and refer to Appendix~\ref{app:extra} for additional results regarding the energy $(P)$ and temperature $(T)$ dependence. We present our results for the rate $d\Gamma/dz$ in Figure~\ref{fig:FiniteMediumVSKernels} as a function of time $t$ for three different momentum fractions $z=0.05,0.25,0.5$ and in Figure~\ref{fig:FiniteMediumFctOfz} as a function of momentum fraction $z$ for four different times $t=0.15,0.4,1,4 {\rm fm}/c$. Different curves in each panel of Figs.~\ref{fig:FiniteMediumVSKernels} and \ref{fig:FiniteMediumFctOfz} show the rates obtained using the non-perturbatively (NP) determined $C(\qt)$, along with the results for the leading order (LO) and next-to-leading order (NLO) perturbative collision kernel (c.f. Sec.~\ref{sec:Broadening}). Insets at the bottom of each graphic display the ratio to the LO results, which are frequently employed in phenomenological studies of jet quenching.

With regards to time dependence in Fig.~\ref{fig:FiniteMediumVSKernels}, one finds that the splitting rates exhibit a linear behavior at early times and quickly saturate at later times where the splitting rate converges to the rate for an infinite medium. We indicate the infinite medium (AMY) rate by a gray dashed line, which can be determined entirely in impact parameter space(c.f.~\cite{Moore:2021jwe}), and thus provides an important validation of the numerical procedure. When comparing the results obtained for the different collision kernels, we observe that the non-perturbative result starts lower than the LO rates before it settles above the LO and below the NLO.  We believe that this behavior can be attributed to the fact that at early times, radiative emissions occurs primarily due to a single hard scattering for which the non-perturbative kernel $C(\qt \gg m_D)$ falls below the LO kernel. Conversely, at later times radiative emission also occur due multiple soft scatterings $\qt \sim m_D$ for which the non-perturbative $C(\qt \lesssim m_D)$ behaves more similar to NLO perturbative kernel.

Similar effects can be observed when considering the $z$ dependence of the rate in Fig.~\ref{fig:FiniteMediumFctOfz}, while keeping in mind that the formation time of the radiation behaves as $t_{f} \sim \frac{2 P z(1-z)}{\qt^2}$ where $\qt^2$ is the transverse momentum acquired due to potentially multiple scatterings over the course of the formation time. 
While at early times, the non-perturbatively determined rate $d\Gamma/dz$ is suppressed compared to the LO rate for all momentum fractions $z$, it starts to rise above the LO results as the rates for soft $(z \ll1)$ and hard $(z \sim 1)$ branchings approach the infinite medium limit $t \gg t_{f}$. Nevertheless, since for quasi-democratic $(z\sim1/2)$ splittings the formation time $t_{f}$ remains large, finite size effects still lead to a significant suppression of the rate of quasi-democratic $(z\sim1/2)$ splittings compared to the infinite medium rates ($t=\infty$).

Notably, we find that in both Figs.~\ref{fig:FiniteMediumVSKernels} and \ref{fig:FiniteMediumFctOfz} the result for the non-perturbative kernel does not depart from a band of $\pm50\%$ around LO, while the NLO result can becomes over $2\times$ larger than the LO result.

We also computed the various approximations to the splitting rates discussed in Sec.~\ref{sec:Formalism}.
In Figures \ref{fig:FiniteMediumVSApprox} and \ref{fig:FiniteMediumVSApproxFctOfz} we compare the full in-medium rates to the first order opacity expansion ($N=1$), the resummed opacity expansion ($N=X$), and the next-to-leading order expansion around the HO (NLO-HO) approximation. We emphasize that in all cases we employ the same non-perturbative broadening kernel $C_{\rm QCD}(\qt)$ at $T=500$MeV, such that any differences are solely due to underlying approximations in the calculation of the rate. Different panels in Fig.~\ref{fig:FiniteMediumVSApprox} show the results for $d\Gamma/dz$ as a function of time for three gluon momentum fractions $z=0.05,0.25,0.5$; the bottom insets in each panel represent the ratio of the respective approximation to the full in-medium splitting rate. 
We observe that, as expected, the early time linear behavior is captured by the opacity expansion since the parton does not have sufficient time to re-interact with the medium. However, soon after the leading order $(N=1)$ opacity expansion starts to over-estimate the rate, while the resummed $(N=X)$ opacity expansion is able to reproduce the rate rather well even at late times especially for soft splittings ($z(1-z)\ll 0.25$). Similarly, the NLO expansion around the HO also performs fairly well at all times, especially if one considers quasi-democratic splittings ($z\sim 1/2$). 

With regards to the $z$ dependence shown in Fig.~\ref{fig:FiniteMediumVSApproxFctOfz}, we find that, as pointed out above, the $N=X$ opacity dependence works particularly well at small/large momentum fractions $z$, while the NLO expansion around the HO is typically most accurate for quasi-democratic splittings ($z\sim 1/2$). Nevertheless, the overall $z$ dependence in Fig.~\ref{fig:FiniteMediumVSApproxFctOfz} is rather well reproduced by both approaches, and the deviations from the full rate behave fairly uniformly as a function of $z$ as can be inferred from the ratios in the insets. Evidently, the leading order opacity expansion is only applicable for times much smaller than the formation time, and fails rather badly on large time scales.

Beyond the cases shown in Figs.~\ref{fig:FiniteMediumVSApprox},\ref{fig:FiniteMediumVSApproxFctOfz},\ref{fig:FiniteMediumVSKernels},\ref{fig:FiniteMediumFctOfz} we have also investigated the behavior for different momenta $P$ of the emitter, as detailed in Appendix~\ref{app:extra}. Generally, we find that the opacity expansion works well at early times $t \ll t_{f}$, where $t_{f} \sim \sqrt{\frac{2z(1-z)P}{\hat{q}}}$ is the  typical formation time for the emission due to multiple soft scatterings. Since in this regime, multiple soft scatterings with $\qt \lesssim m_D$ can not generate sufficient transverse momentum, the radiative emission is primarily due to a single (rare) hard scattering with $\qt \gg m_D$. As for  $\qt \gg m_D$ the non-perturbative broadening kernel behaves similar to the LO perturbative kernel, the rates obtained in this regime are also similar.  When $t \gg t_{f}$ radiative emissions occur primarily due to multiple soft scatterings over the course of one formation time, except for the Bethe-Heitler regime at very low energies $2 P z (1-z) \lesssim T$, where the formation time becomes shorter than the mean free path between soft scatterings, such that a single soft $\qt \lesssim m_D$ scattering is responsible for the emission, and the rates determined for the non-perturbative broadening kernel behave similar to the NLO perturbative kernel, which exhibits a similar IR behavior. While the Bethe-Heitler regime can again be described in terms of an opacity expansion, emissions with $2 P z (1-z) \lesssim T$ and $t \gg t_{f}$ are due to multiple scatterings and suffer from Landau-Pomeranchuk-Migdal (LPM) suppression~\cite{Landau:1953um,Migdal:1956tc}. We find that in this regime, the rates determined for the non-perturbative broadening kernel typically lie between the LO and NLO perturbative determinations, and are best described by the NLO-HO approximation, which accounts for the effects of multiple soft scatterings along with a single hard scattering.

By comparing the effect of various approximations in Figs.~\ref{fig:FiniteMediumVSApprox} and \ref{fig:FiniteMediumVSApproxFctOfz}, with 
the impact of the different LO, NLO and NP collision kernels in Figs.~\ref{fig:FiniteMediumVSKernels} and \ref{fig:FiniteMediumFctOfz}, we generally find that the different approximations to the splitting rates perform rather well within their respective range of validity, whereas the choice of the broadening kernel $C(\qt)$ is clearly more impactful for the calculation of the in-medium splitting rates.

\section{Conclusion}\label{sec:Conclusion}
Building on the determination of the collisional broadening kernel $C_{\rm QCD}(\bp)$ in ~\cite{Moore:2021jwe}, we performed a Fourier transform of $C_{\rm QCD}(\bp)$, to determine the non-perturbative broadening kernel $C_{\rm QCD}(\bqp)$ in momentum space in order to compute radiative emissions rates in a QCD medium of finite extent. 

We presented results for the in-medium splitting rates obtained with the non-perturbative collision kernel and compared them to the results obtained with  leading and next-to-leading order perturbative collision kernels, as well as with different approximations of the in-medium splitting rates, which are commonly employed in the literature. While approximations to the splitting rate calculation are quite effective in reproducing the rate within their respective range of validity, differences between the LO kernel, which is usually used in phenomenological studies of jet quenching, and the non-perturbative kernel can easily be on the order of $30\%$. We conclude that, while for sophisticated numerical simulations one can reconstruct the full rate to obtain precise results, for (semi-)analytical calculations a combination of the resummed opacity and NLO-HO rates is likely sufficient, as theoretical improvements mostly rely on the precise knowledge of the collisional broadening kernel. 

With regards to the phenomenological applications of our work, we note that the collisional broadening kernel and in-medium splitting rates obtained in this paper can be incorporated into a study of jet quenching either using a kinetic approach \cite{Mehtar-Tani:2018zba,Adhya:2019qse,schlichting_medium-induced_2021} or with a MonteCarlo simulations \cite{Caucal:2018ofz,Chen:2017zte,Putschke:2019yrg,Schenke:2009gb}. Similarly, one could also utilize the same broadening kernel to include non-perturbative contributions to the elastic scatterings.  We finally note that a recent study using the same EQCD setting obtained non-perturbative contributions to the thermal masses \cite{Moore:2020wvy} and it would be interesting to investigate their impact on the rate calculation specifically and jet quenching in general.

\section*{Acknowledgement}
We thank Guy D. Moore and Niels Schlusser for insightful discussions, collaboration on our previous work~\cite{Moore:2021jwe} and their help in performing the Fourier transform of the momentum broadening kernel. This work is
supported in part by the Deutsche Forschungsgemeinschaft (DFG, German Research Foundation)
through the CRC-TR 211 ’Strong-interaction matter under extreme conditions’– project
number 315477589 – TRR 211
and the German Bundesministerium f\"{u}r Bildung und Forschung (BMBF) through Grant No. 05P21PBCAA. The authors also gratefully acknowledge computing time
provided by the Paderborn Center for Parallel Computing (PC2).

\appendix 
\section{Hankel transformation}\label{ap:Hankel}
Below we provide details of the procedure followed to obtain the Hankel transform of the momentum broadening kernel.

\subsection{Numerical implementation of the Hankel transformation}
To perform the numerical integration in Eq.~(\ref{eq:HankelTransformDeriv}), we split the integral using the zeros $\{x_i\}$ of the Bessel function $(J_1(x_i)=0)$ as follows
\begin{align}
    \Delta C^{\rm NP}(\qp) =& \frac{2\pi}{\qp} \sum_{i=0}^{\infty}\int_{x_i/\qp}^{x_{i+1}/\qp} \! \d b_\perp~  b_\perp J_1(b_\perp\,\qp) \nonumber\\
    &\times\frac{d}{db_\perp}\Delta C^{\rm NP}(b_\perp)\,.
\end{align}
Let us define the series
\begin{align}
    A_n = \int_{x_i/\qp}^{x_{i+1}/\qp} \! \d b_\perp~  b_\perp J_1(b_\perp\,\qp) \frac{d}{db_\perp}\Delta C^{\rm NP}(b_\perp)\;.
\end{align}
The Hankel transform is then given by the sum $A=\sum_{n=0}^{\infty} A_n$, however, this sum is slowly convergent. The convergence can be accelerated using a method known as Shanks transformation \cite{Shanks1955}, where one defines the series
\begin{align}
    A = & \lim_{n\to\infty } S(A_{n}) \;,\\= & \lim_{n\to\infty } A_{n+1} - \frac{(A_{n+1}-A_n)^2}{(A_{n+1}-A_n) - (A_n - A_{n-1})}\;.
\end{align}
The result is then obtained by truncating the sum when we obtain convergence up to a small tolerance threshold, i.e.~ when $\left| \frac{S(A_{n+1})}{S(A_n)} -1\right| \leq 10^{-8}$. 

\subsection{Transformation of the short-distance behavior}
At short-distances the broadening kernel follows a similar behavior to the LO kernel, which is the short-distance limit of 
\begin{align}
    &C(\bbp) = \frac{ \CR g^4T^3\mathcal{N}}{8\pi \mD^2}[ \gamma_e + \log(\mD \bbp/2) + K_0(\mD\bbp)] \;.
\end{align}
In momentum space this broadening kernel is given by \cite{Ghiglieri:2013gia}
\begin{align}
    C(\qp) = \frac{ \CR g^4T^3\mathcal{N}}{\mD^2} \frac{\mD^2}{ \qp^2(\qp^2 + \mD^2)}\;.
\end{align}
Since we are only interested in the leading UV behavior, provided by the $b^2\log(b)$ term, the non-perturbative broadening kernel will follow the same behavior as the LO kernel
\begin{align}
    C^{\rm UV}(\qp) = \frac{\CR \gs^4T^3 \mathcal{N}}{8 \pi} \frac{8\pi}{\qp^4}\;.
\end{align}
\subsection{Transformation of the long-distance behavior}
We proceed to transform the long-distance behavior given in Eq.~(\ref{IR_limit}).
Let us first consider the Hankel transformation of the linear function 
\begin{align}
    C(\bbp) = A + B \bbp\;.
\end{align}
The constant term leads to a delta function $\delta^{(2)}(\qp)$ in momentum space which can be discarded, and the linear term will lead to
\begin{align}
    C(\qp) = B\frac{2\pi}{\qp^3}\;.
\end{align}
Note that in order to verify this identity, it is actually easier to compute the following inverse transform 
\begin{align}
    &\int \frac{d^2\qp}{(2\pi)^2 } \frac{2\pi}{\qp^3} (1-e^{i\qp\cdot\bp}) \nonumber\\
    &= \frac{1}{2\pi} \int_{-\infty}^{\infty} dx \int_{-\infty}^{\infty} dy~\frac{1}{(x^2+y^2)^{3/2}} (1-e^{i x\bbp})\;, \nonumber\\
    &= \frac{1}{2\pi} \int_{-\infty}^{\infty} dx~ \frac{2}{x^2} (1-\cos{x\bbp}) = \bbp\;.
\end{align}

\section{Numerical implementation}\label{ap:FiniteMedium}
Below we provide some additional details on the numerical calculation of the splitting rate for finite media following the approach of \cite{CaronHuot:2010bp}. We employ a forward Euler scheme, to evolve the wave function from $\Delta \tilde{t}=0$ to $\Delta \tilde{t} = \tilde{t}$ according to the differential equation~(\ref{eq:EvolInteraction}) and use our results to perform the integral in Eq.~(\ref{eq:RateEqInteraction}).
\subsection{Separating the soft scale}
\begin{figure}
    \centering
    \includegraphics[width=0.45\textwidth]{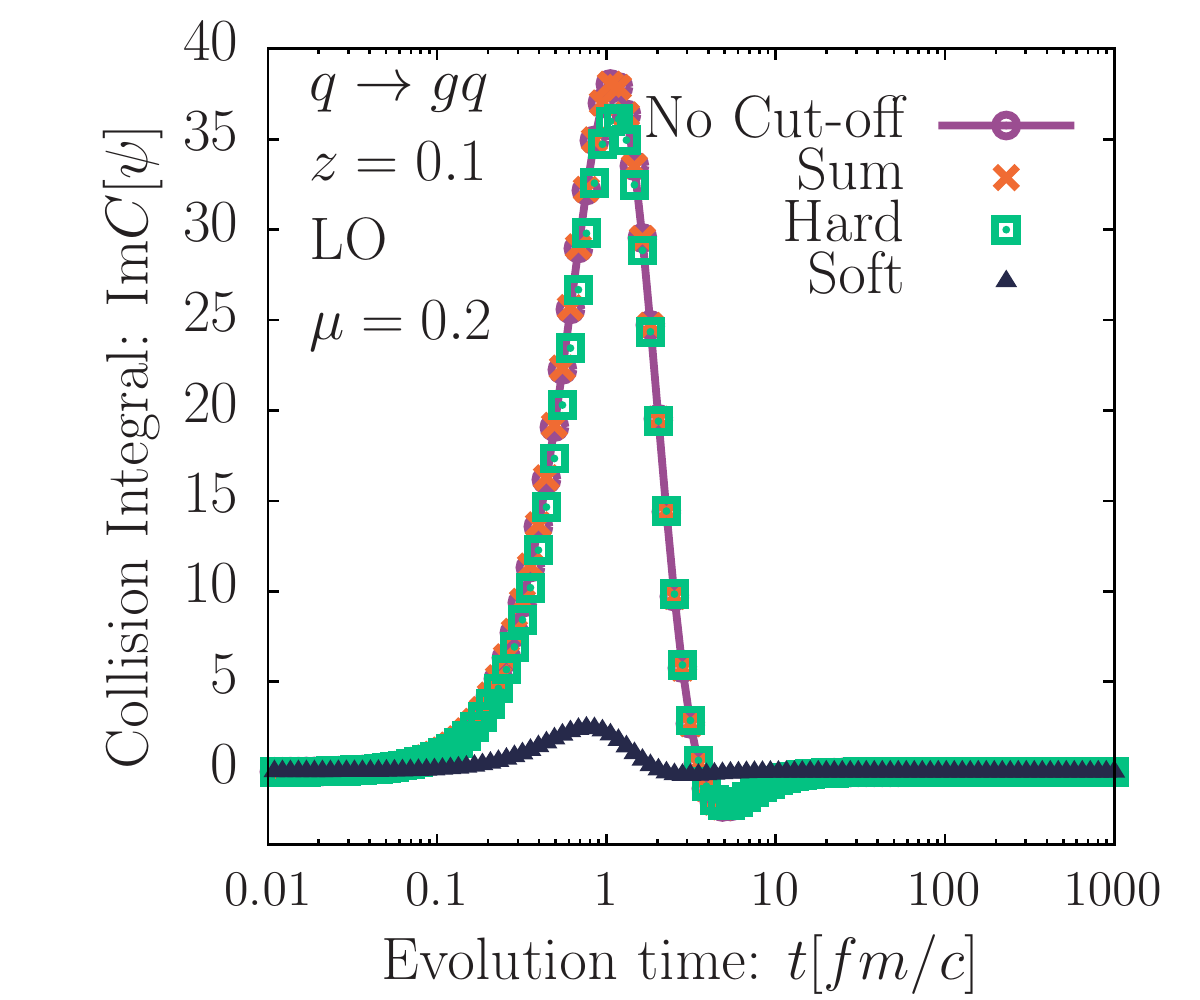}
    \includegraphics[width=0.45\textwidth]{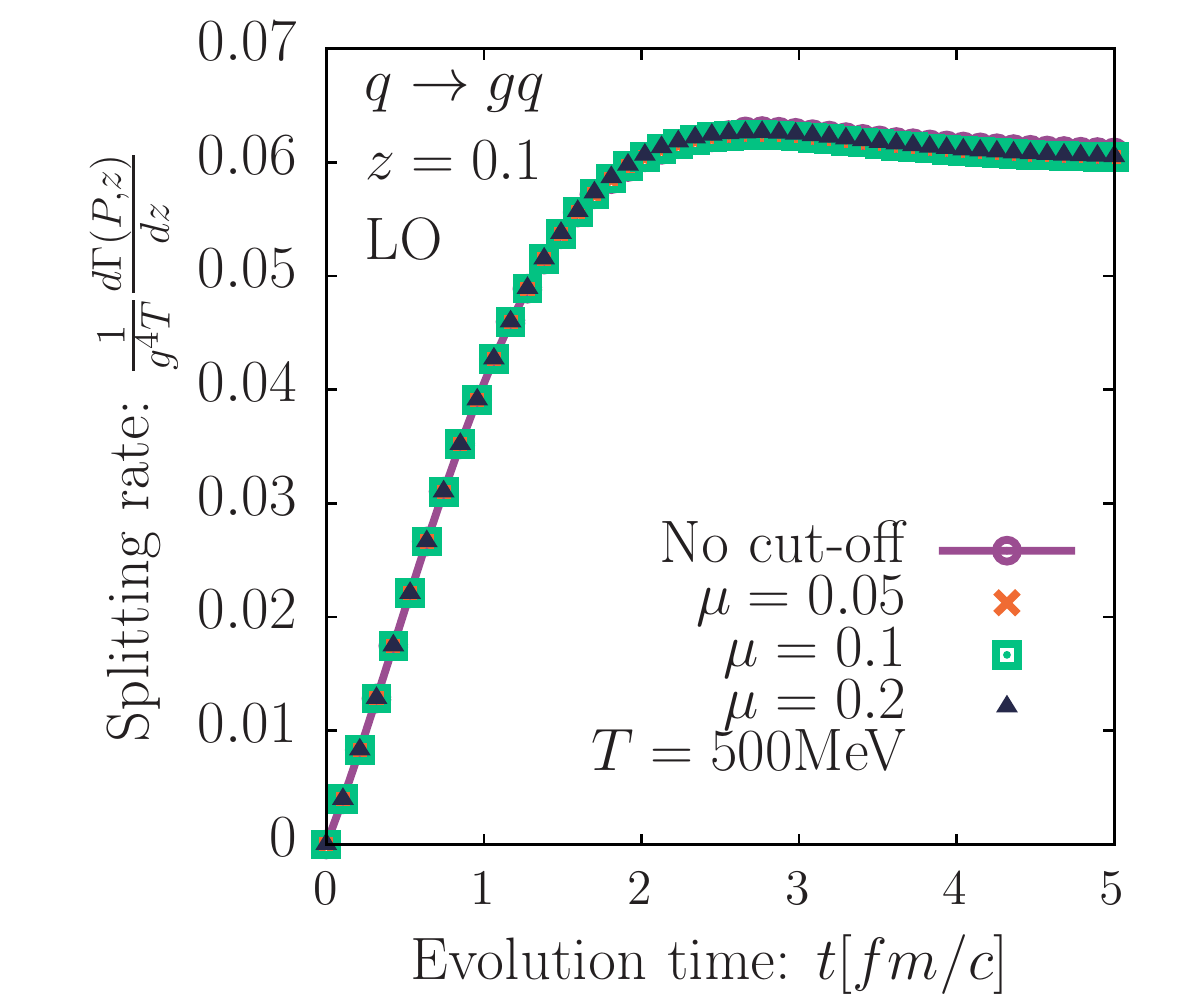}
    \caption{ Validation of the separation into soft and hard components in the calculation for a quark with momentum $P=300T$ radiating a gluon with momentum fraction $z=0.1$. (top) Comparison of the imaginary part of the initial collision integral using the LO perturbative broadening kernel. Different curves separately show the hard and soft contributions to the collision integral for $\mu=0.2$; the sum is compared to the full leading order collision integral without separating the soft-scales. (bottom) Evolution of the in-medium splitting rate with varying cut-off $\mu=0.05,0.1,0.2$. The in-medium splitting rate using the LO perturbative broadening kernel do not show any significant dependence on the choice of the cut-off scale.}
    \label{fig:CutOffDependence1}
\end{figure}
\begin{figure}
    \centering
    \includegraphics[width=0.45\textwidth]{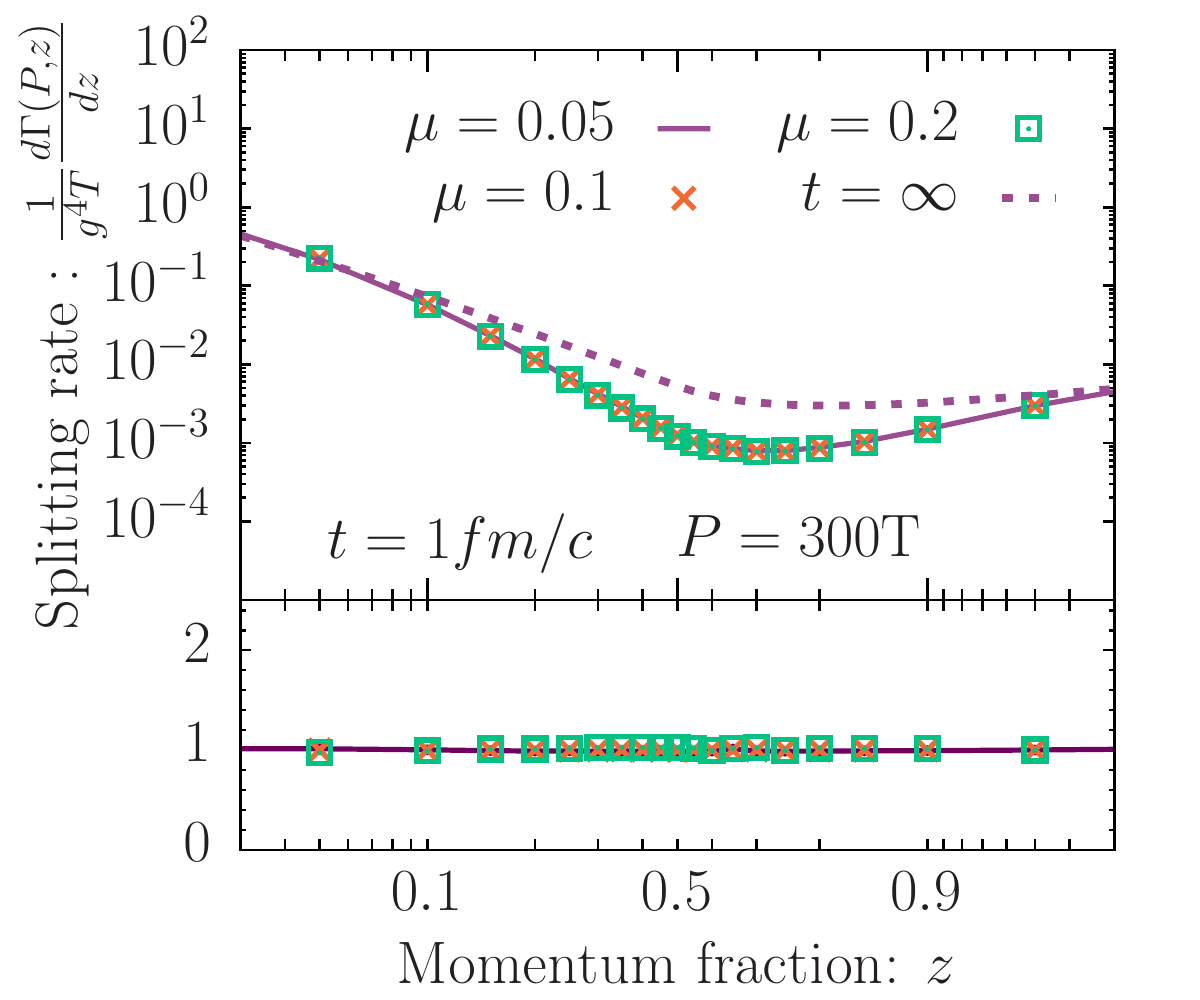}
    \includegraphics[width=0.45\textwidth]{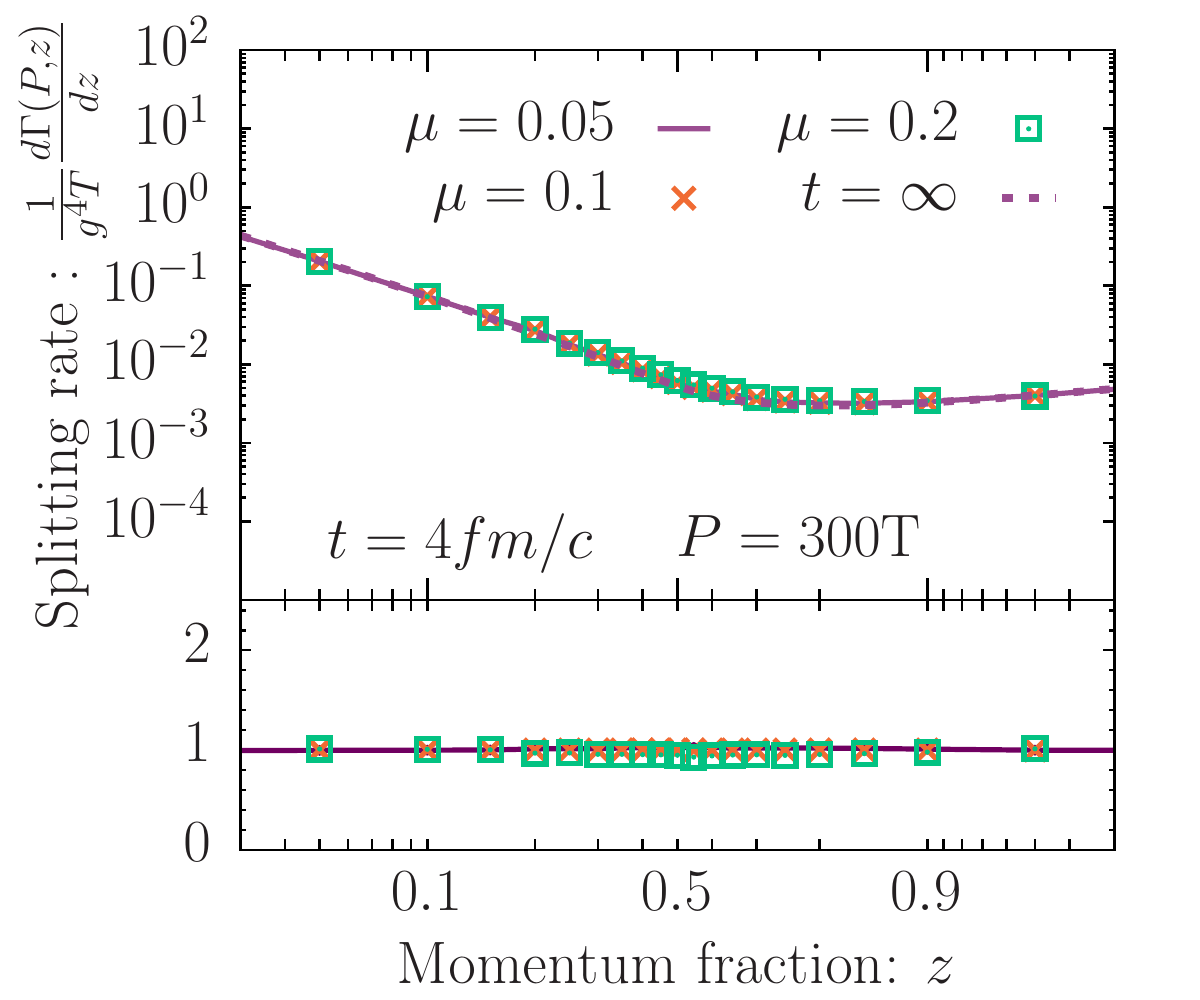}
    \includegraphics[width=0.45\textwidth]{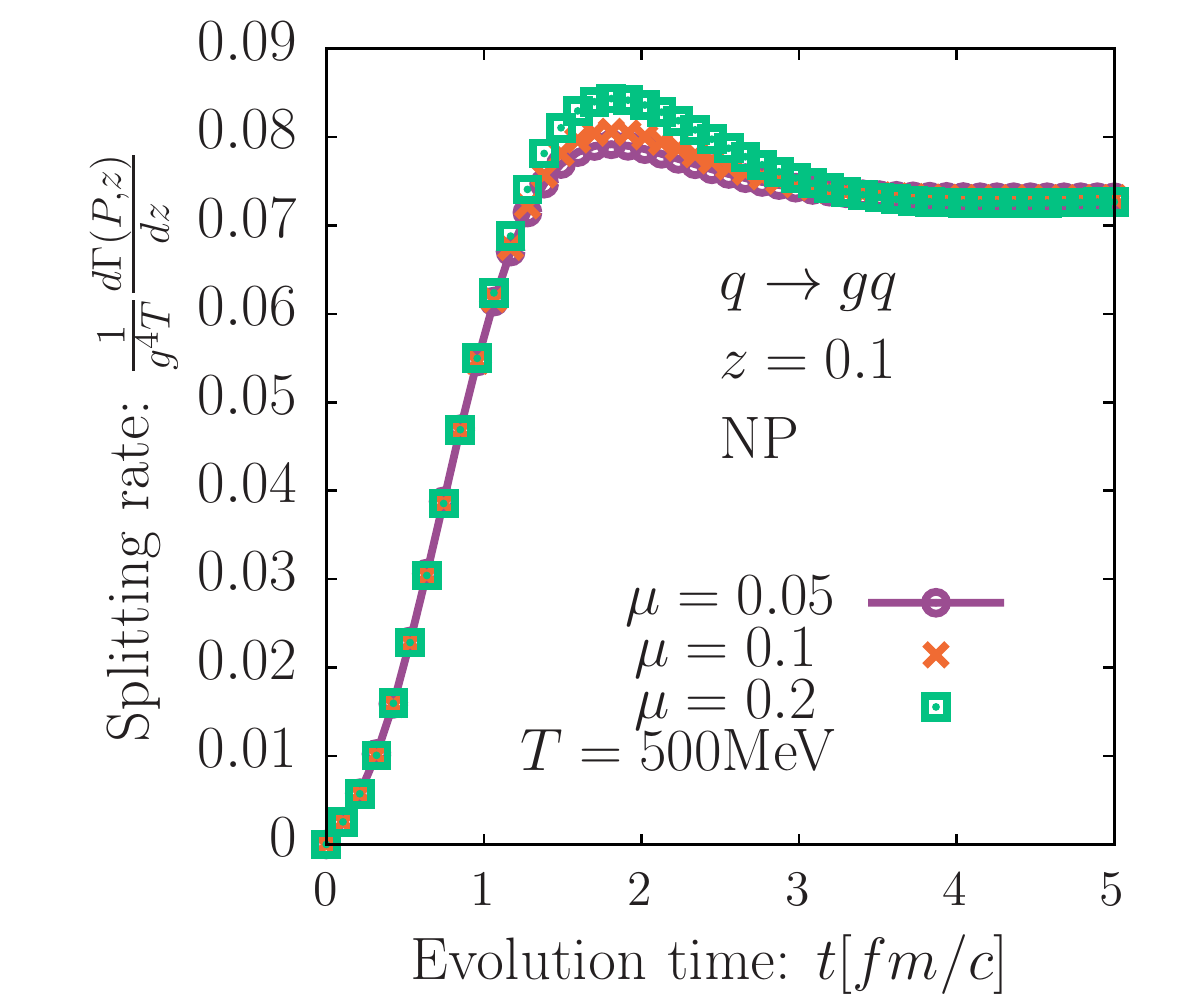}
    \caption{ Dependence of the in-medium splitting rate on the soft cut-off $\mu=0.05,0.1,0.2$ for a quark with momentum $P=300T$ radiating a gluon with momentum fraction $z$. Upper panels show the spectrum as a functions of $z$ at $t=1 {\rm fm}/c$ (top) and $t=4 {\rm fm}/c$ (center).  The bottom panel shows the time evolution of the rate for $z=0.1$. All results are obtained for the non-perturbative broadening kernel.}
    \label{fig:CutOffDependence2}
\end{figure}
When solving the evolution equation using the NLO and non-perturbative broadening kernels, we find that the $1/q^3$ behavior at small momentum leads to numerical instabilities. In order to stabilize this evolution, we will consider the soft interactions in the collision integral separately.
Starting with Eq.~(\ref{eq:ThreeBodyInt}), we rewrite the collision integral using the interaction picture and combine the different momentum integrals using variable change to find
\begin{align}
    &\tilde{\Gamma}_3 \circ \tpsi(\tp)\nonumber\\
    &=e^{i\delta\tilde{E}(\tp) \Delta \tilde{t}}  \tilde{\p}\cdot  \nonumber\\
    &\int_{\tilde{\q}}\left[ C_1 \tilde{C}(\tilde{\q}) + \frac{C_z}{z^2} \tilde{C}\left(\frac{\tilde{\q}}{z}\right) + \frac{C_{1-z}}{(1-z)^2} \tilde{C}\left(\frac{\tilde{\q}}{1-z}\right)\right] \nonumber\\
    &\left[e^{-i\delta\tilde{E}(\tp) \Delta \tilde{t}} \frac{\tilde{\p}}{\tp^2} \tpsi_I(\tp) - e^{-i\delta\tilde{E}(|\tilde{\p} - \tilde{\q}|) \Delta \tilde{t}} \frac{\tilde{\p} - \tilde{\q}}{|\tilde{\p} - \tilde{\q}|^2} \tpsi_I(|\tilde{\p} - \tilde{\q}|)\right],
\end{align}
By introducing an intermediate cut-off $\mu$ in the momentum exchange $\tilde{q}$, the collision integral is separated to hard and soft interactions
\begin{align}
    C[\tpsi_I] = C_{\rm hard}[\tpsi_I] + C_{\rm soft}[\tpsi_I]\;.
\end{align}
The soft interaction can be treated in a diffusion approximation using an expansion in momentum exchange $\tilde{q}$.
We specifically expand the following term from Eq.~(\ref{eq:EvolInteraction}) of the collision integral 
\begin{align}
    &\frac{\tp^2-\ttp\cdot\ttq}{|\ttp-\ttq|^2}\tpsi_I(|\tp-\tq|)
    = \tpsi_I(\tp) + \frac{\tq}{\tp}\cos\theta \left[  \tpsi_I(\tp)  -\tp \tpsi'_I(\tp) \right] \nn\\
    &+ \frac{q^2}{2 \tp^2} \Big[ 2\cos2\theta \tpsi_I(\tp) \nn\\
    &+ \tp( 1-3 \cos^2\theta) \tpsi'_I(\tp) + \tp^2\cos\theta^2 \tpsi''_I(\tp) \Big] \;,
\end{align}

where $\theta$ is the angle between $\ttp$ and $\ttq$. Plugging the expansion to the collision integral and performing the angular integral, we find
\begin{align}
    &C_{\rm soft}[\tpsi_I] =\tpsi_I(\tp)\left(I^{(0)}_1(\tp, \Delta \tilde{t}) - I_2(\tp, \Delta \tilde{t}) - I_1^{(3)}(\tp, \Delta \tilde{t}) \right) \nn\\
    &+ \frac{p}{2} \tpsi_I'(\tp) \left(  2I_2(\tp, \Delta \tilde{t}) -I_2^{(3)}(\tp, \Delta \tilde{t}) \right) - \frac{p^2}{2} \tpsi_I''(\tp) I_3(\tp, \Delta \tilde{t})\;,
\end{align}
where $I_i$ are the following integral moments
\begin{align}
\label{eq:B5}
    I_1^{(0)}(\tp, \Delta \tilde{t}) =& \frac{1}{2\pi}\int_0^{\mu} d\tq~ \mathcal{C}(\tq,z) \left[ 1- e^{-i \Delta \tilde{t} \tq^2} \right. \nonumber\\
    &\left. \times\mathcal{J}_0\left(2\Delta \tilde{t} \tp\tq \right) \right] \;,\\
    I_1^{(3)}(\tp, \Delta \tilde{t}) =&\frac{-1}{2\pi} \int_0^{\mu} d\tq~\mathcal{C}(\tq,z) 
    \frac{\tq^2}{\tp^2}  e^{-i \Delta \tilde{t} \tq^2} \nonumber\\
    & \times\mathcal{J}_2\left( 2\Delta \tilde{t} \tp\tq \right) \;,\\
    I_2(\tp, \Delta \tilde{t}) =&\frac{1}{2\pi} \int_0^{\mu} d\tq~\mathcal{C}(\tq,z) 
    \frac{\tq}{p}  i e^{-i \Delta \tilde{t} \tq^2} \nonumber\\
    & \times\mathcal{J}_1\left( 2\Delta \tilde{t} \tp\tq \right) \;,
\end{align}
\begin{align}
    I_2^{(3)}(\tp, \Delta \tilde{t}) =&\frac{1}{2\pi} \int_0^{\mu} d\tq~\mathcal{C}(\tq,z) 
    \frac{\tq^2}{\tp^2}  e^{-i \Delta \tilde{t} \tq^2}  \left[ \frac{3}{2\Delta\tilde{ t} \tp\tq}\right. \nonumber\\
    &\left. \times\mathcal{J}_1\left( 2\Delta \tilde{t} \tp\tq \right)  - 2\mathcal{J}_0\left( 2\Delta \tilde{t} \tp\tq \right)\right] \;,\\
    I_3(\tp, \Delta \tilde{t}) =&\frac{1}{2\pi} \int_0^{\mu} d\tq~\mathcal{C}(\tq,z) 
    \frac{\tq^2}{\tp^2}  e^{-i \Delta \tilde{t} \tq^2}  \left[ \frac{1}{2\Delta\tilde{ t} \tp\tq}\right. \nonumber\\ \label{eq:B10}
    &\left. \times\mathcal{J}_1\left( 2\Delta \tilde{t} \tp\tq \right)  - \mathcal{J}_2\left( 2\Delta \tilde{t} \tp\tq \right)\right] \;,
\end{align}
where $\mathcal{J}_i(x)$ are the Bessel functions of the first kind and 
\begin{align}
    \mathcal{C}(\tq,z)=   \tq \left[ C_1 \tilde{C}(\tq) + \frac{C_z}{z^2} \tilde{C}\left(\frac{\tq}{z}\right) + \frac{C_{1-z}}{(1-z)^2} \tilde{C}\left(\frac{\tq}{1-z}\right)\right]
\end{align}
We perform the integrals in Eqns.~(\ref{eq:B5}-\ref{eq:B10}) numerically and tabulate them for a fixed time step $\Delta\tilde{t}$. Combining the soft component with the hard component, then makes up the full collision integral in Eq.~\ref{eq:EvolInteraction}, which can be used to evolve the wave function. While the hard component is easily evolved using an Euler explicit scheme, we employ an implicit scheme for the soft component to deal with instabilities.

We have explicitly validated the procedure at the hand of the LO kernel, for which the rate can be evaluated with and without introducing the cut-off scale $\mu$. In the top panel of Fig.~\ref{fig:CutOffDependence1}, we show how hard and soft contributions combine to yield the full collision integral in Eq.~\ref{eq:EvolInteraction}. Note that, to produce this figure we used a rather large value of $\mu=0.2$ to render the soft contributions visible on the plot. The bottom panel of Fig.~\ref{fig:CutOffDependence1} shows an examplary results for the effect of the choice of the cut-off scale $\mu= 0.05,0.1,0.2$ for the time evolution of the in-medium splitting computed using the LO kernel. Excellent agreement of the curves shows that for the LO kernel there is almost no dependence on the cut-off scale $\mu$. 
 
Next we investigate the sensitivity to the cut-off scale $\mu$ for the non-perturbative (NP) broadening kernel. We present in Fig.~\ref{fig:CutOffDependence2} the dependence on the cutoff of the in-medium splitting rate computed using the non-perturbative broadening kernel. We observe how for sufficiently small cut-off scales $\mu\ll1$ the change of the cut-off scale has almost no effect on the resulting in-medium splittings rates. Only for larger values of the cut-off scale $\mu=0.2$ one starts to notice deviations around the time when the medium-induced begins to saturate. We note for completeness that the results shown in Sec.~\ref{sec:Results} are obtained for a value $\mu=0.05$.

\section{Some additional results on the temperature ($T$) and momentum $(P)$ dependence} 
\label{app:extra}
\begin{figure}
    \includegraphics[width=0.45\textwidth]{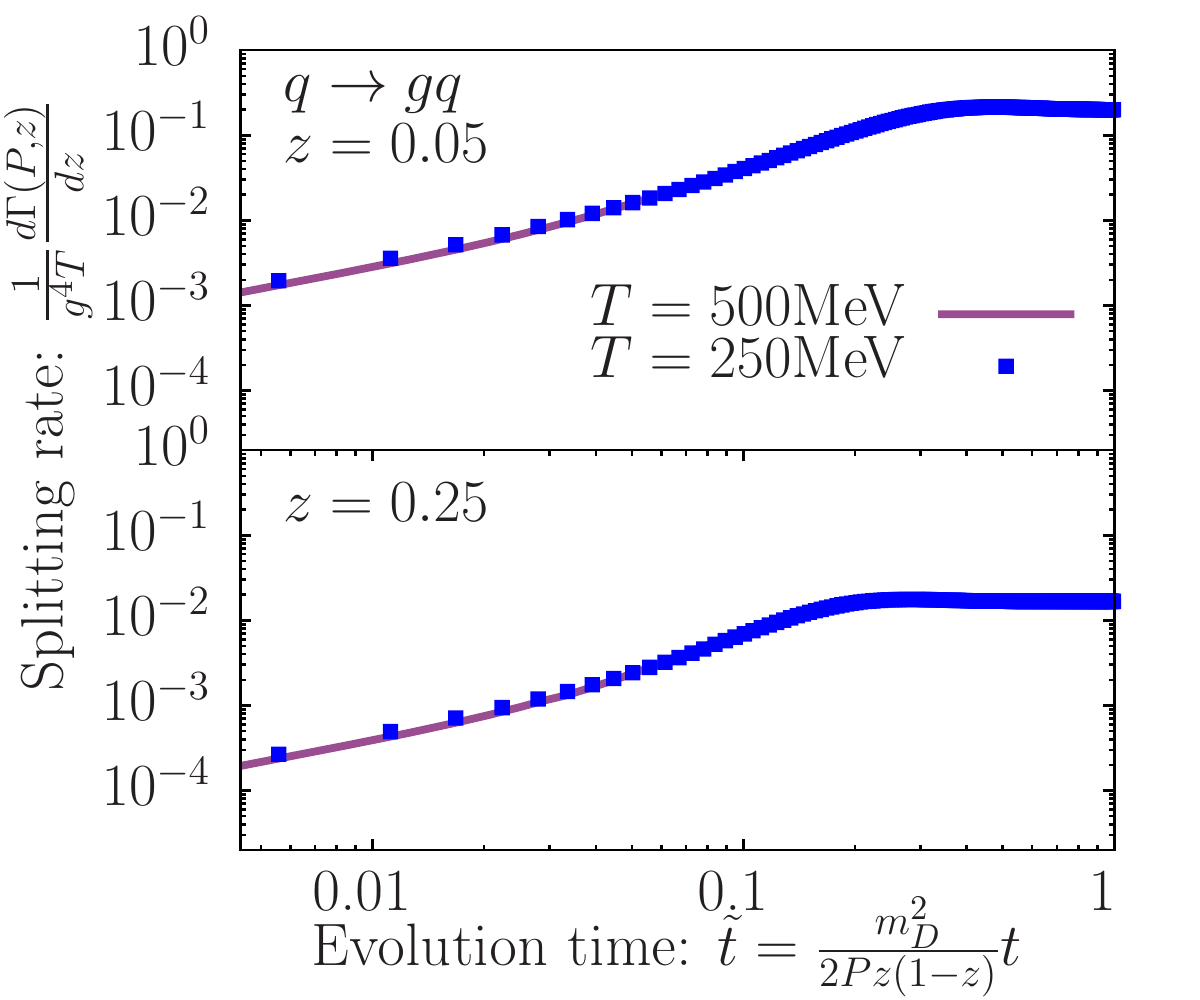}
    \includegraphics[width=0.45\textwidth]{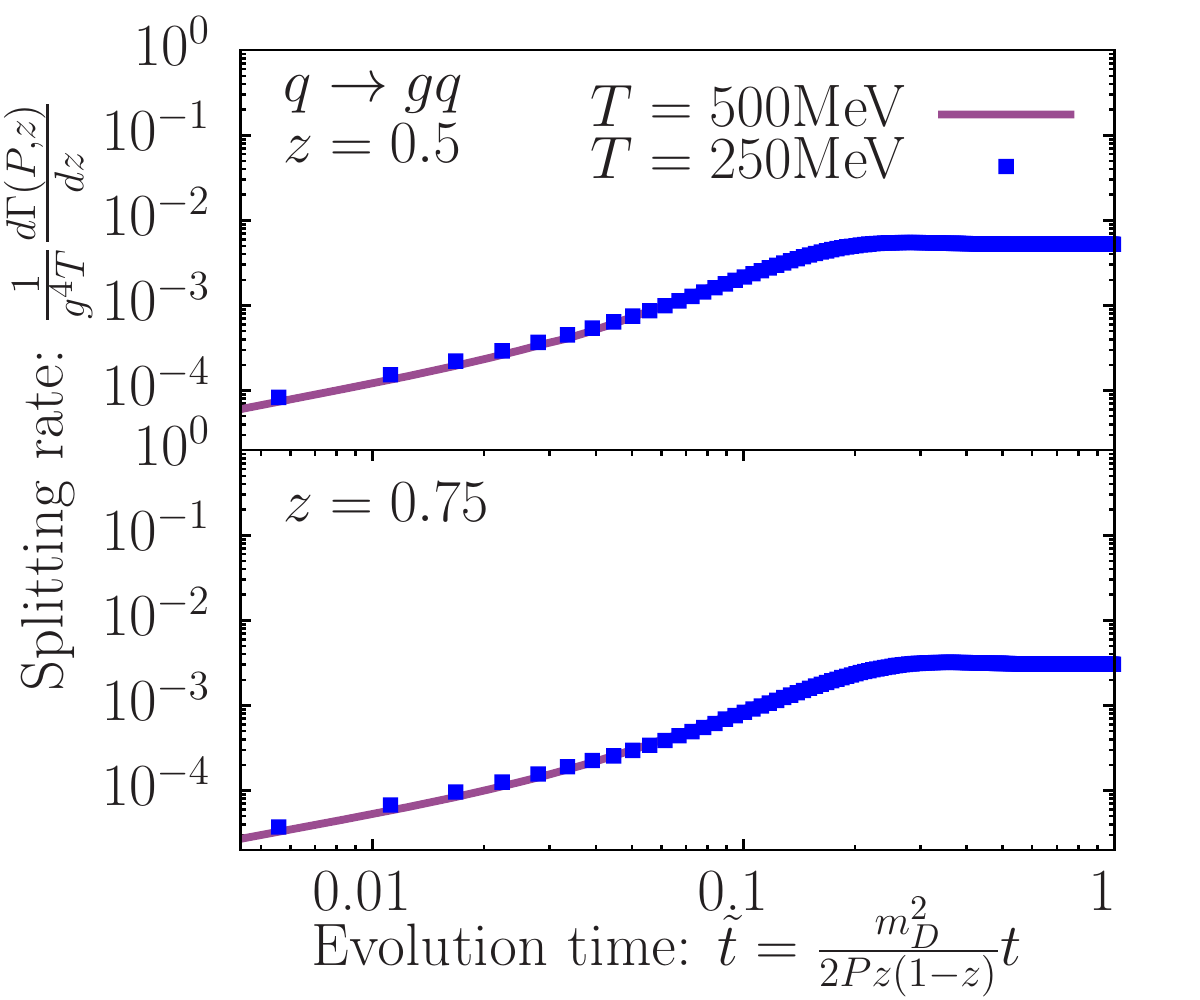}
    \caption{Evolution of the splitting rate for the medium-induced emission of a gluon from a parent quark with energy $P=300T$ in an equilibrium plasma with temperatures $T=250,500$MeV as a function of the scaled evolution time $\tilde{t}=\frac{m_D^2}{2Pz(1-z)}t$. Each panel represent a different gluon momentum fraction $z=0.05,0.25,0.5,0.75$ from top to bottom.}
    \label{fig:TemperatureDependence}
\end{figure}

\begin{figure*}
    \includegraphics[height=0.345\textwidth,trim=0 0 0.8cm 0, clip]{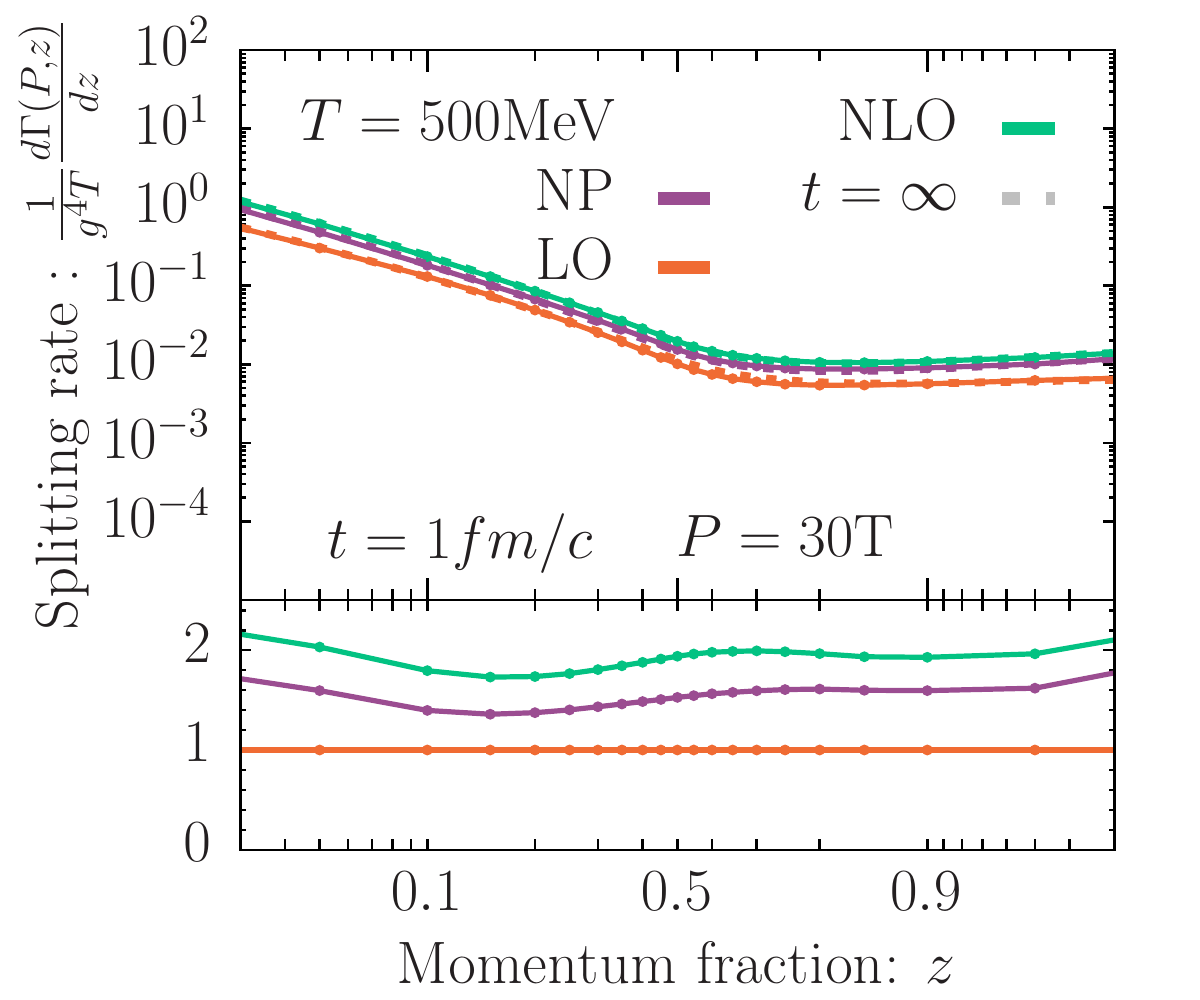}\includegraphics[height=0.345\textwidth,trim=2.4cm 0 0.8cm 0, clip]{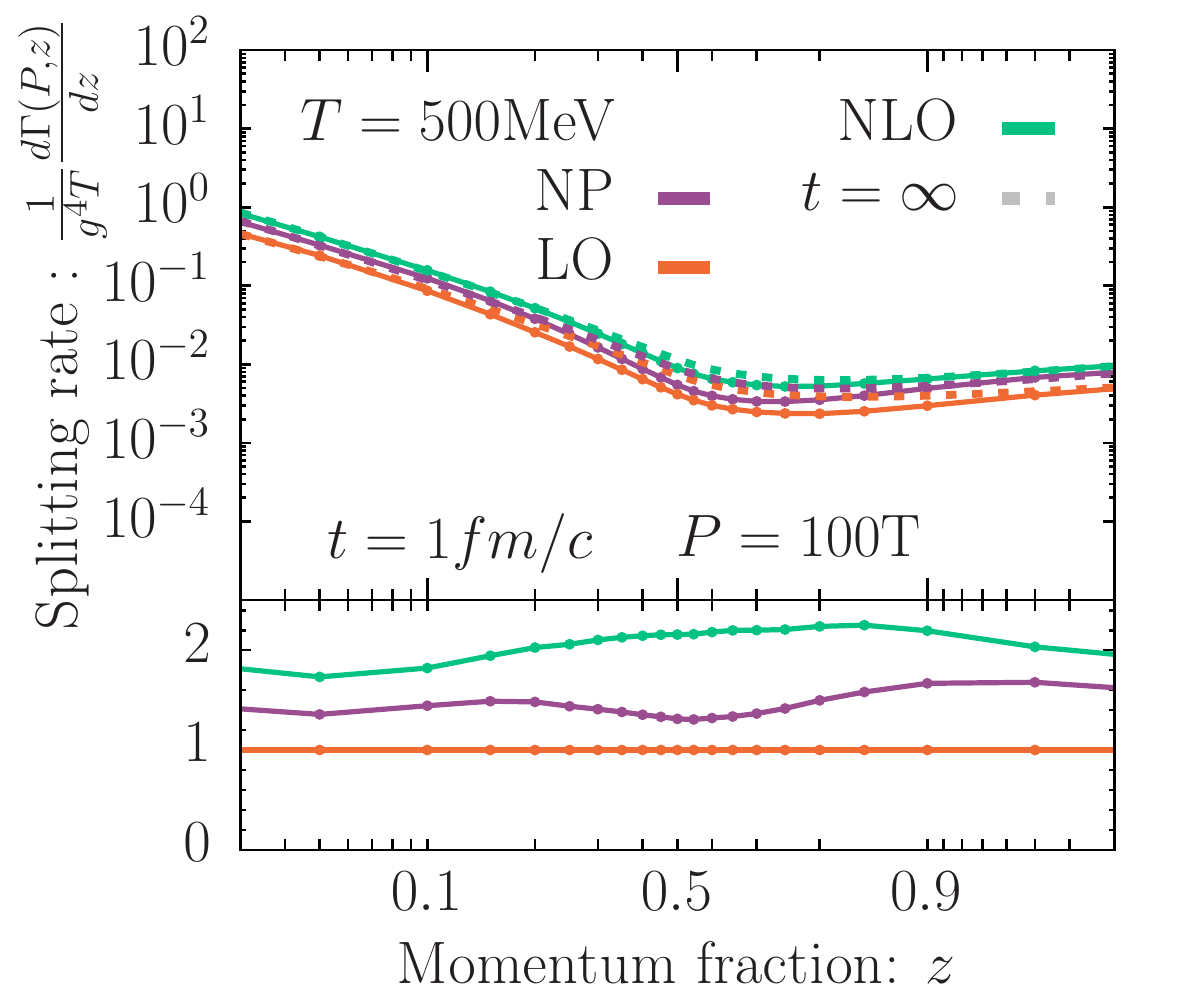}\includegraphics[height=0.345\textwidth,trim=2.4cm 0 0.8cm 0, clip]{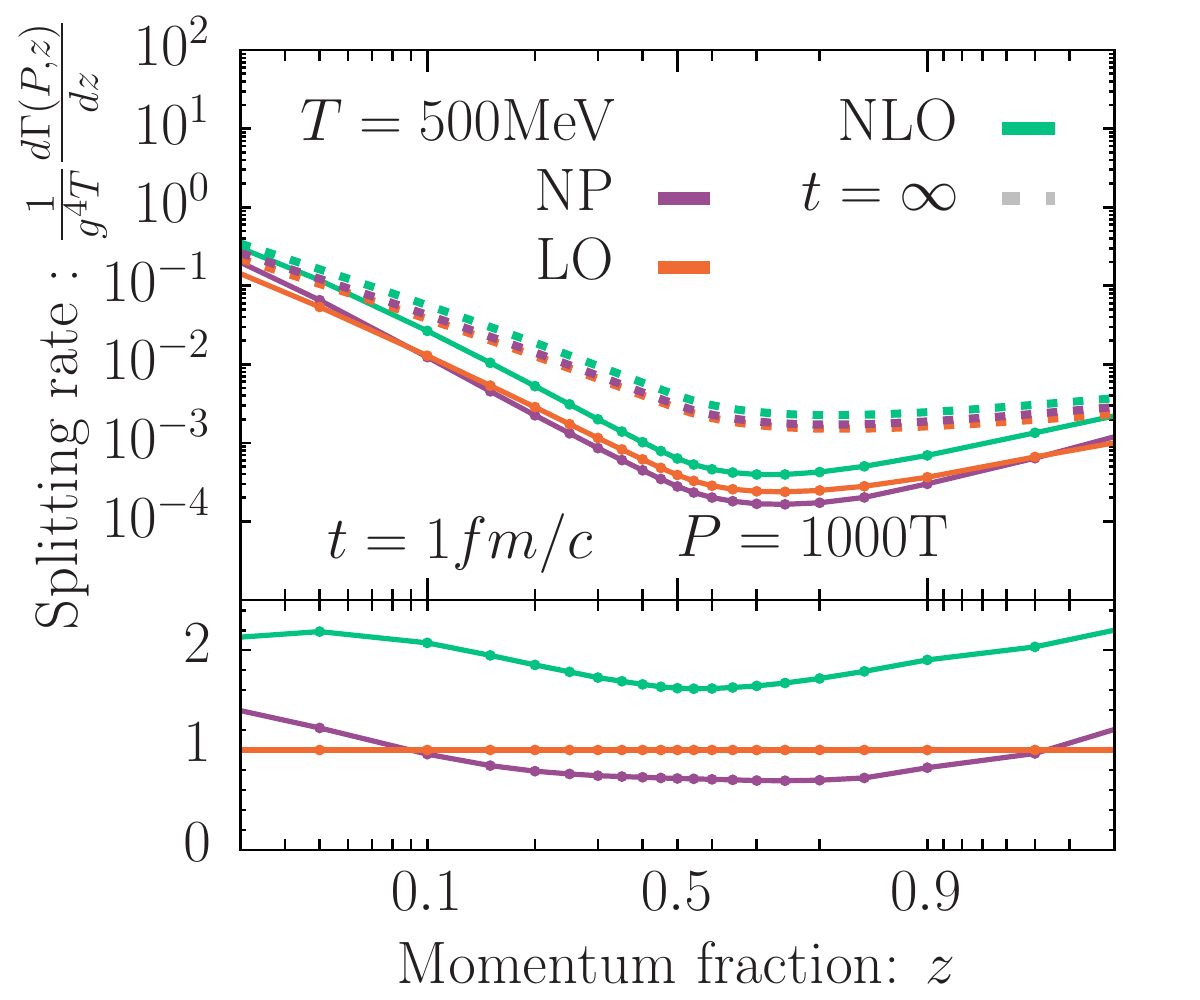}
    \includegraphics[height=0.345\textwidth,trim=0 0 0.8cm 0, clip]{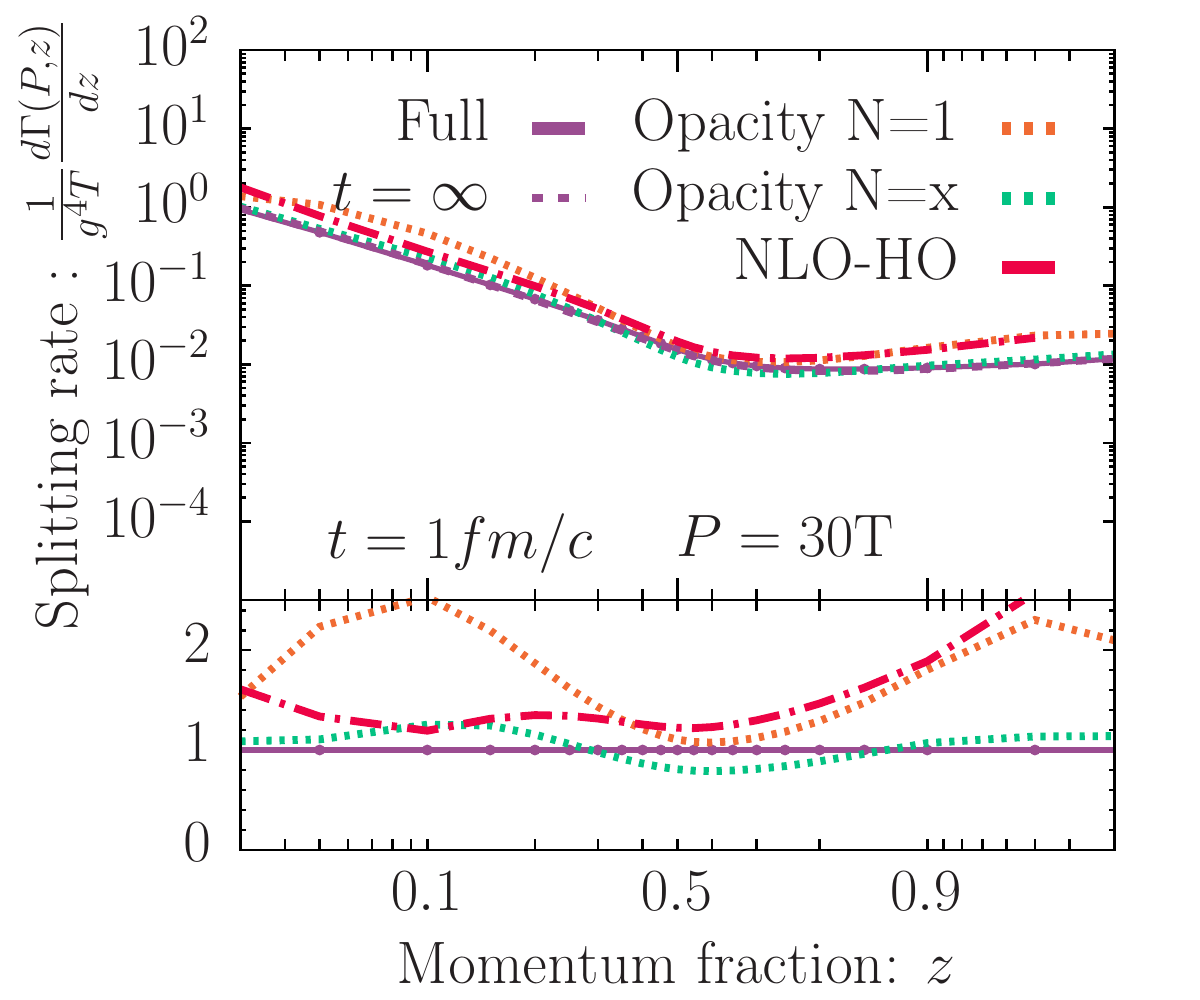}\includegraphics[height=0.345\textwidth,trim=2.4cm 0 0.8cm 0, clip]{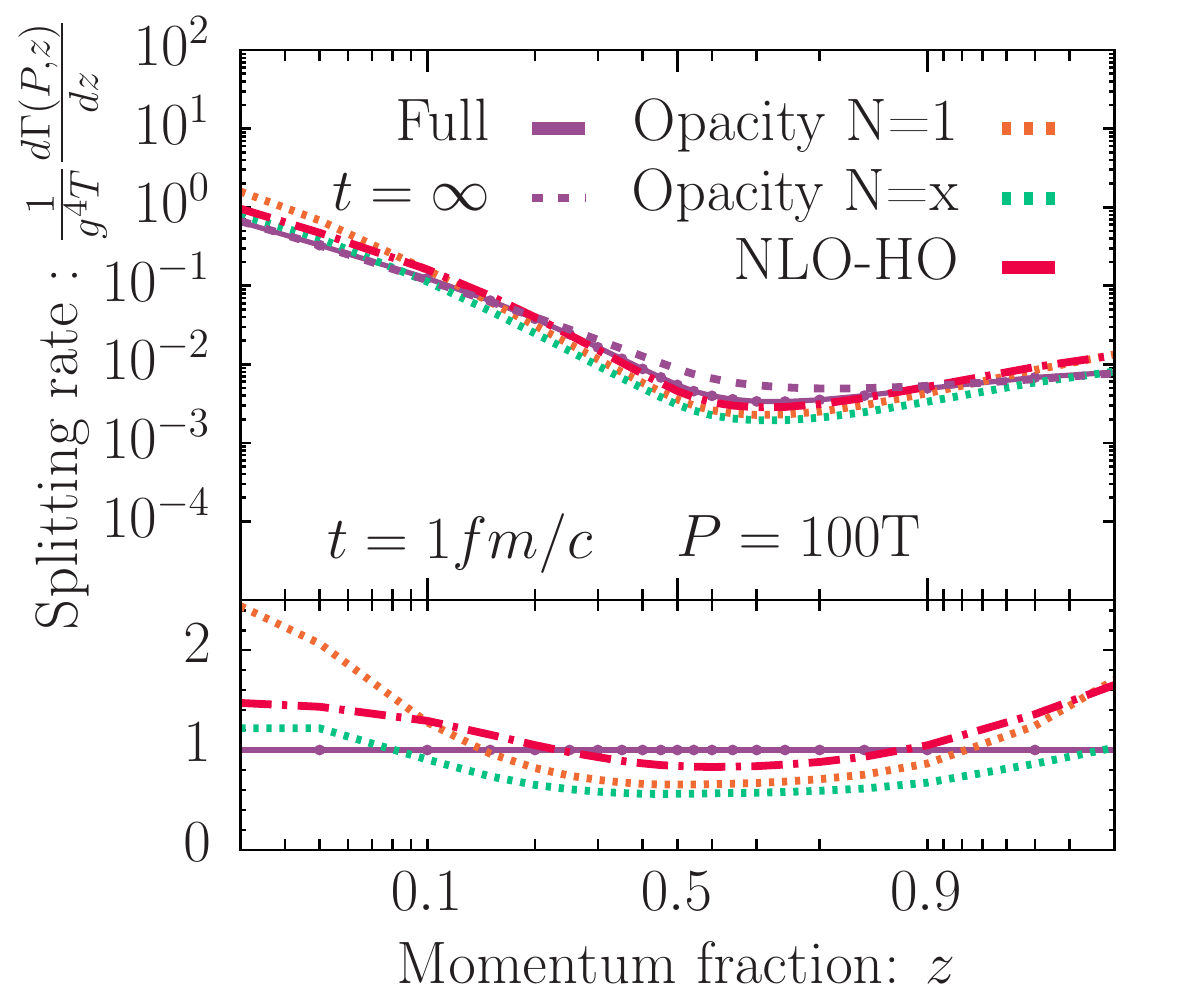}\includegraphics[height=0.345\textwidth,trim=2.4cm 0 0.8cm 0, clip]{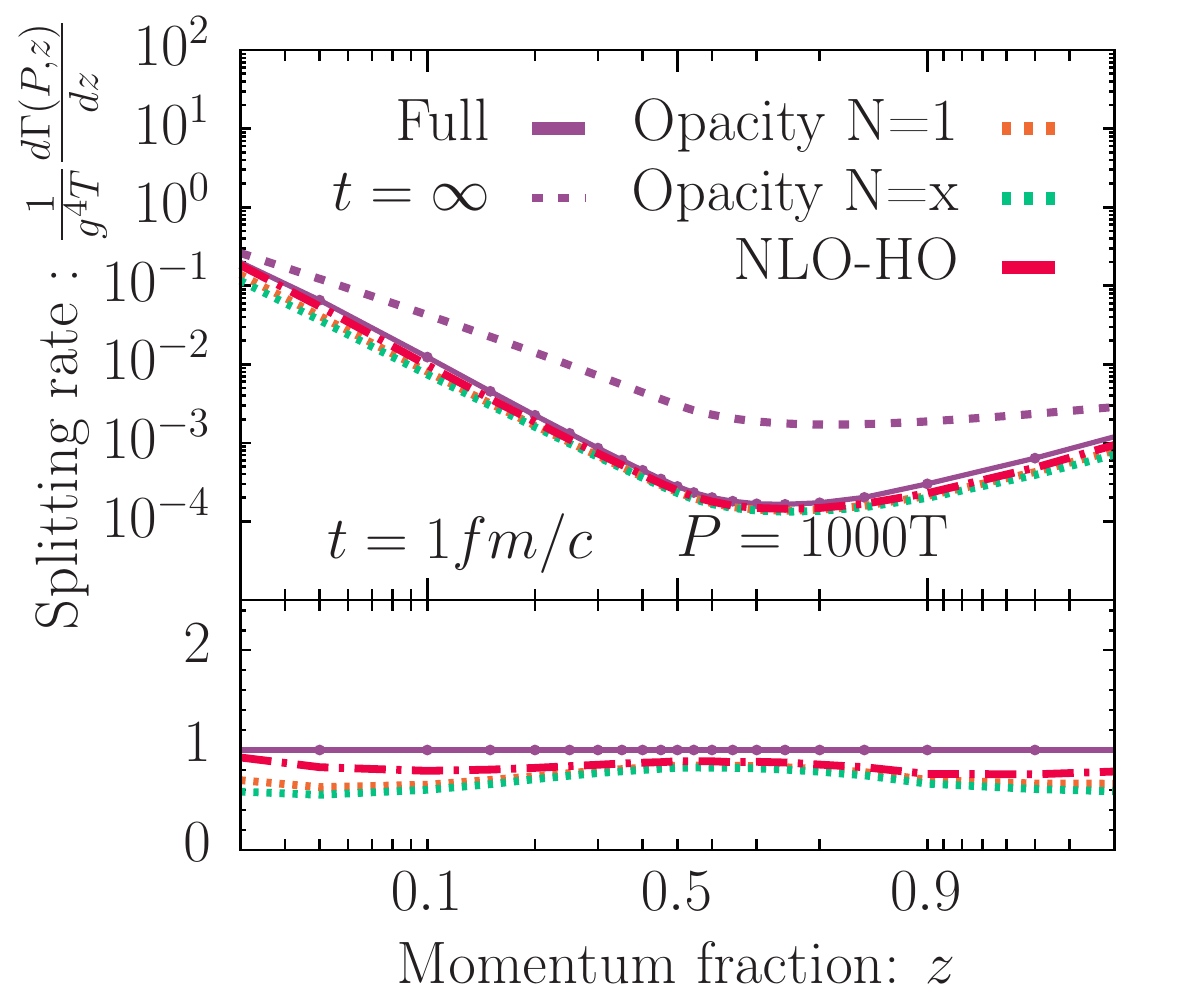}
    \caption{
        Splitting rate for the medium-induced emission of a gluon from a parent quark as a function of momentum fraction of the radiated gluon $z$. Different panels show the rate $d\Gamma/dz$ at fixed times $t=1fm/c$ for different values of the parent quark energy $P=30,100,1000T$ from left to right. We compare the rates obtained using different broadening kernels in the top row and by employing different approximations to the rate calculation in the bottom row. Dashed lines correspond to the (AMY) splitting rates~\cite{Arnold:2003zc} in an infinite medium  \cite{Moore:2021jwe}
    }
    \label{fig:RateEarlyTime}
\end{figure*}

\begin{figure*}
    \includegraphics[height=0.345\textwidth,trim=0 0 0.8cm 0, clip]{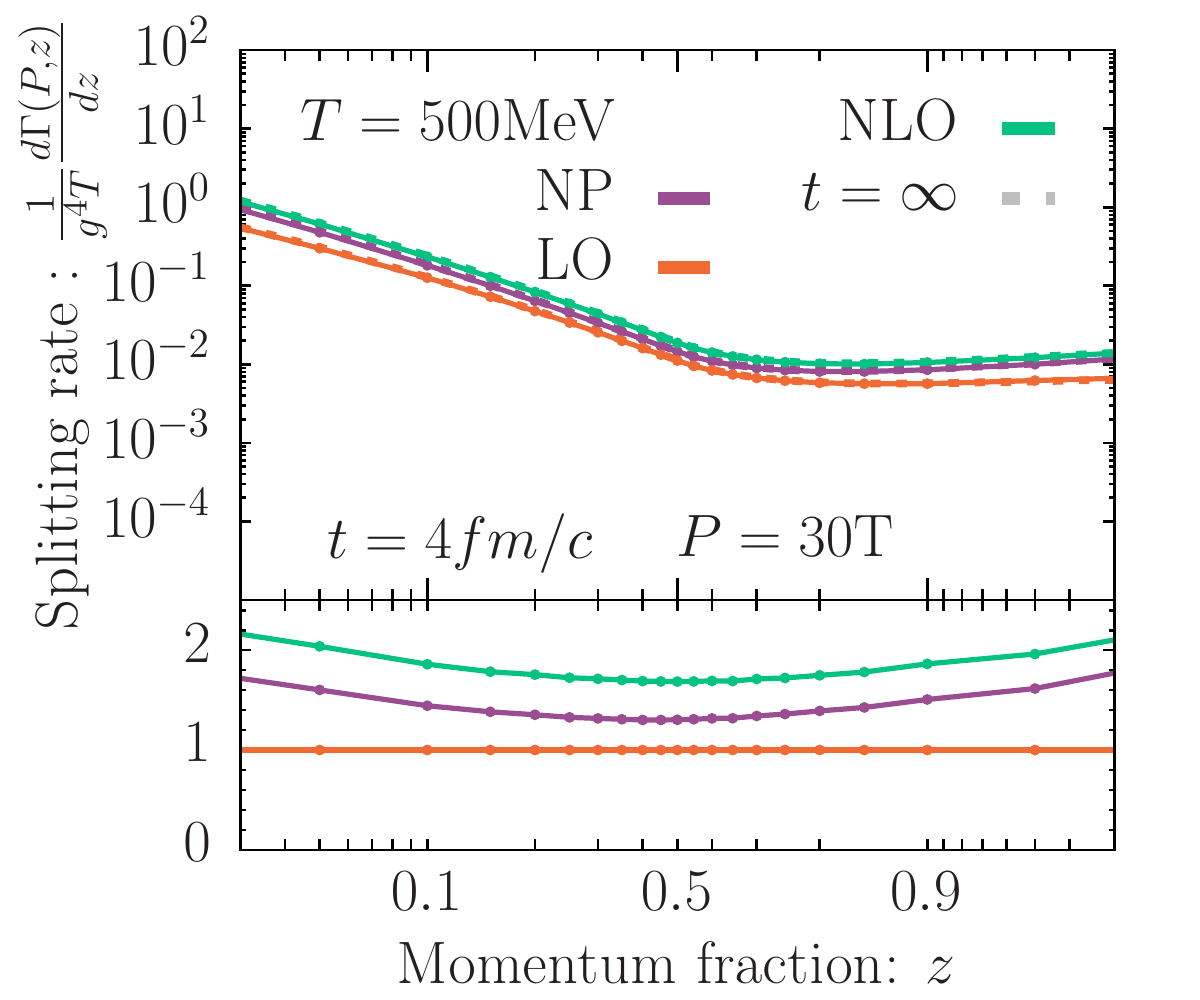}\includegraphics[height=0.345\textwidth,trim=2.4cm 0 0.8cm 0, clip]{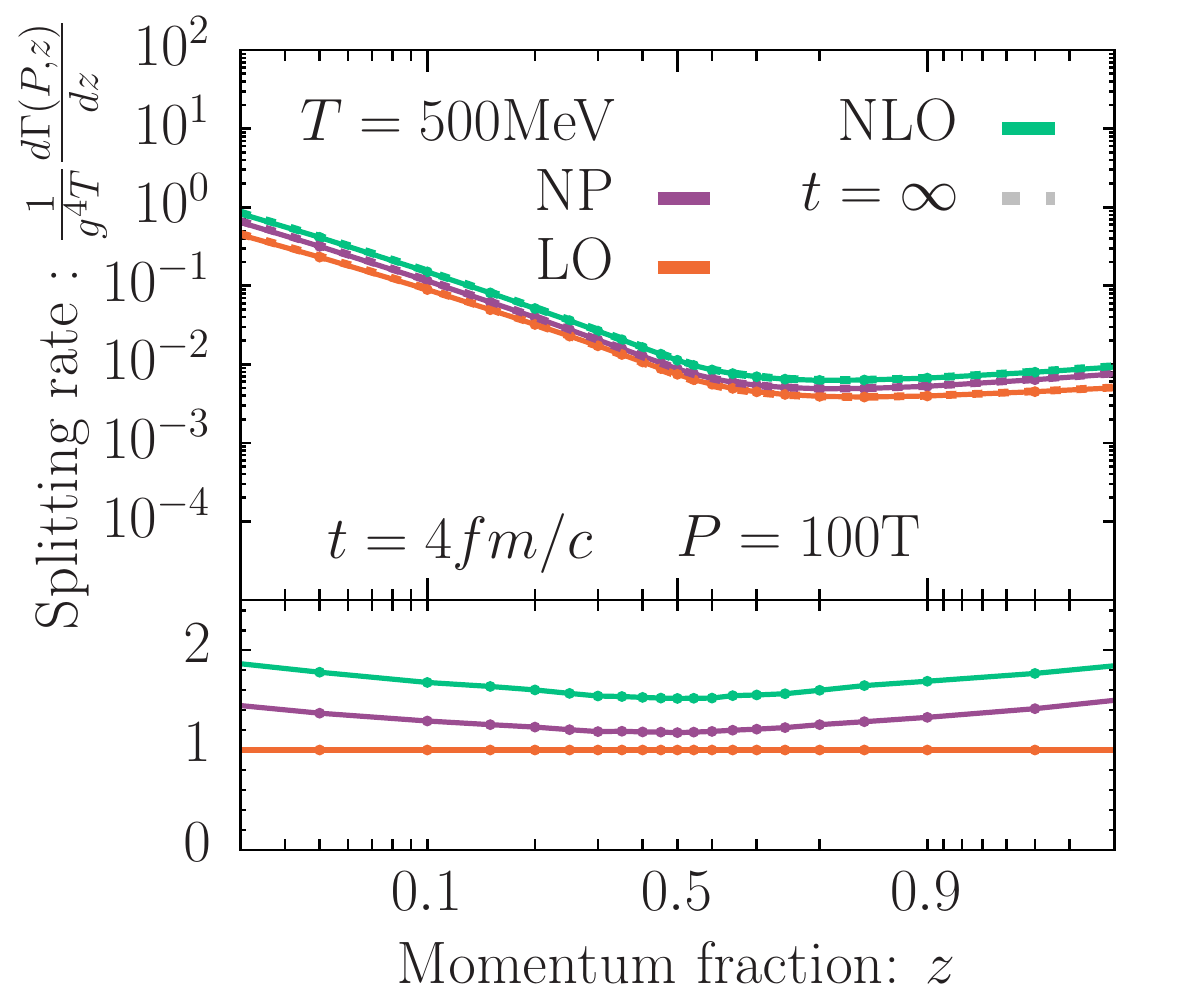}\includegraphics[height=0.345\textwidth,trim=2.4cm 0 0.8cm 0, clip]{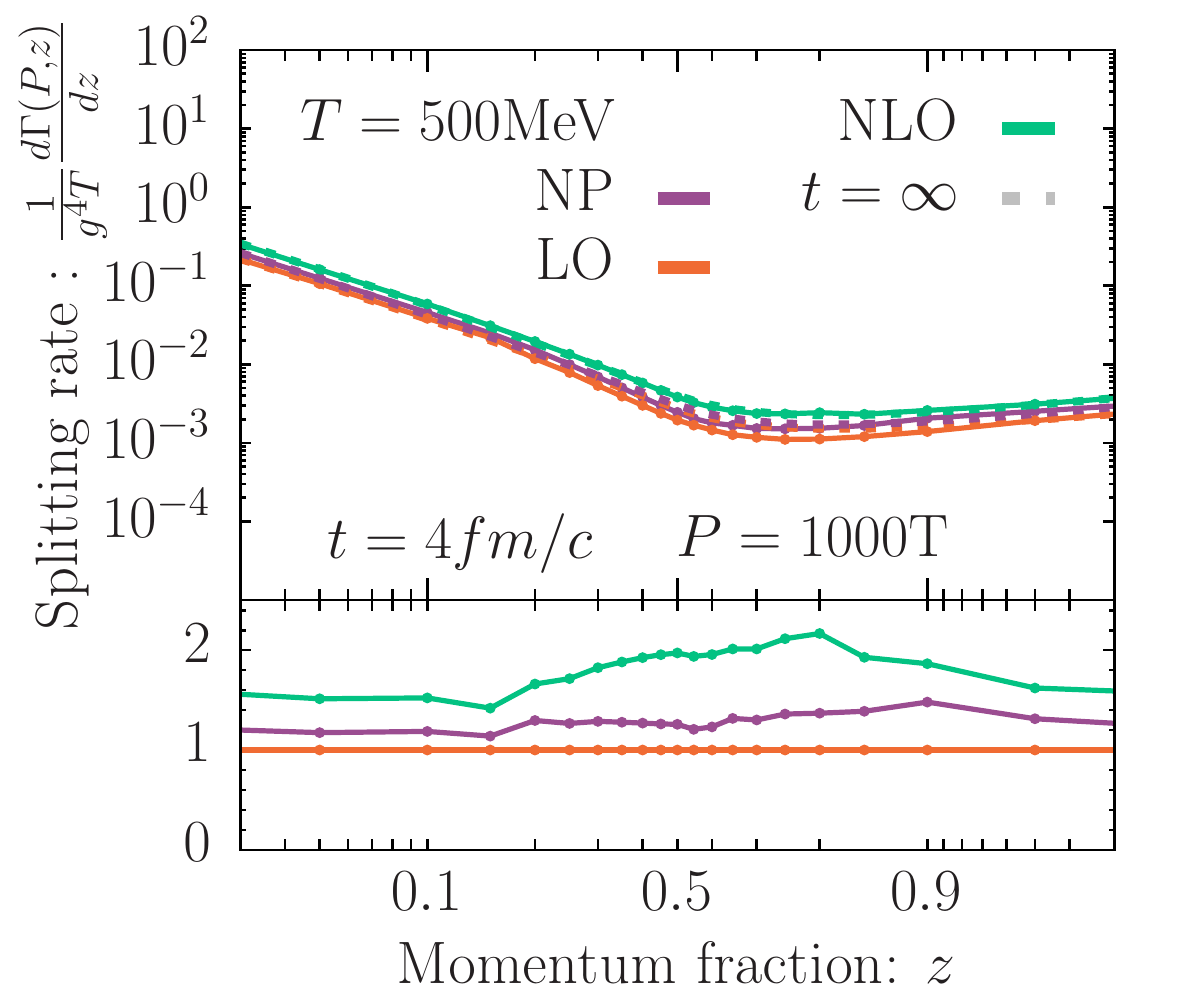}
    \includegraphics[height=0.345\textwidth,trim=0 0 0.8cm 0, clip]{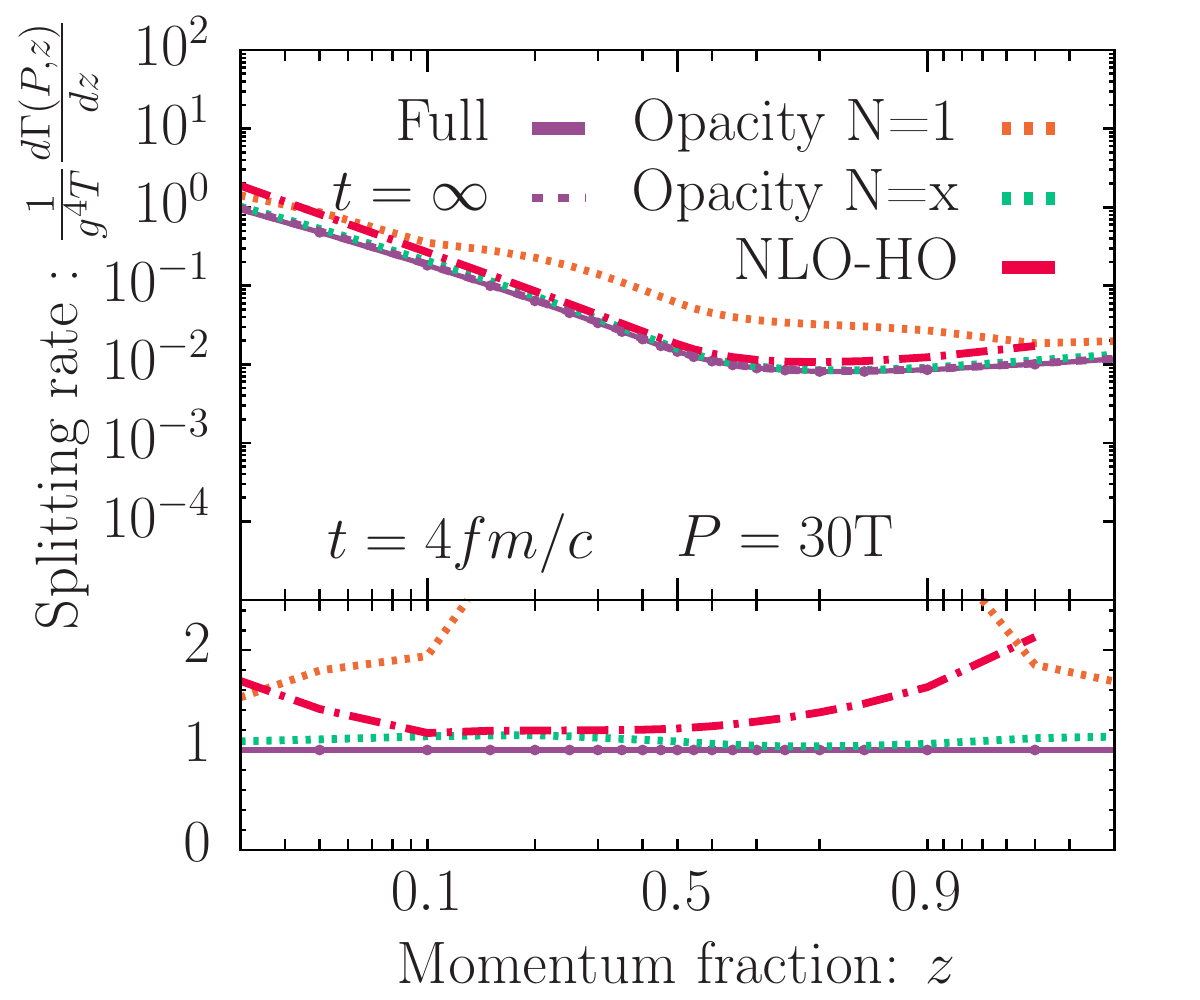}\includegraphics[height=0.345\textwidth,trim=2.4cm 0 0.8cm 0, clip]{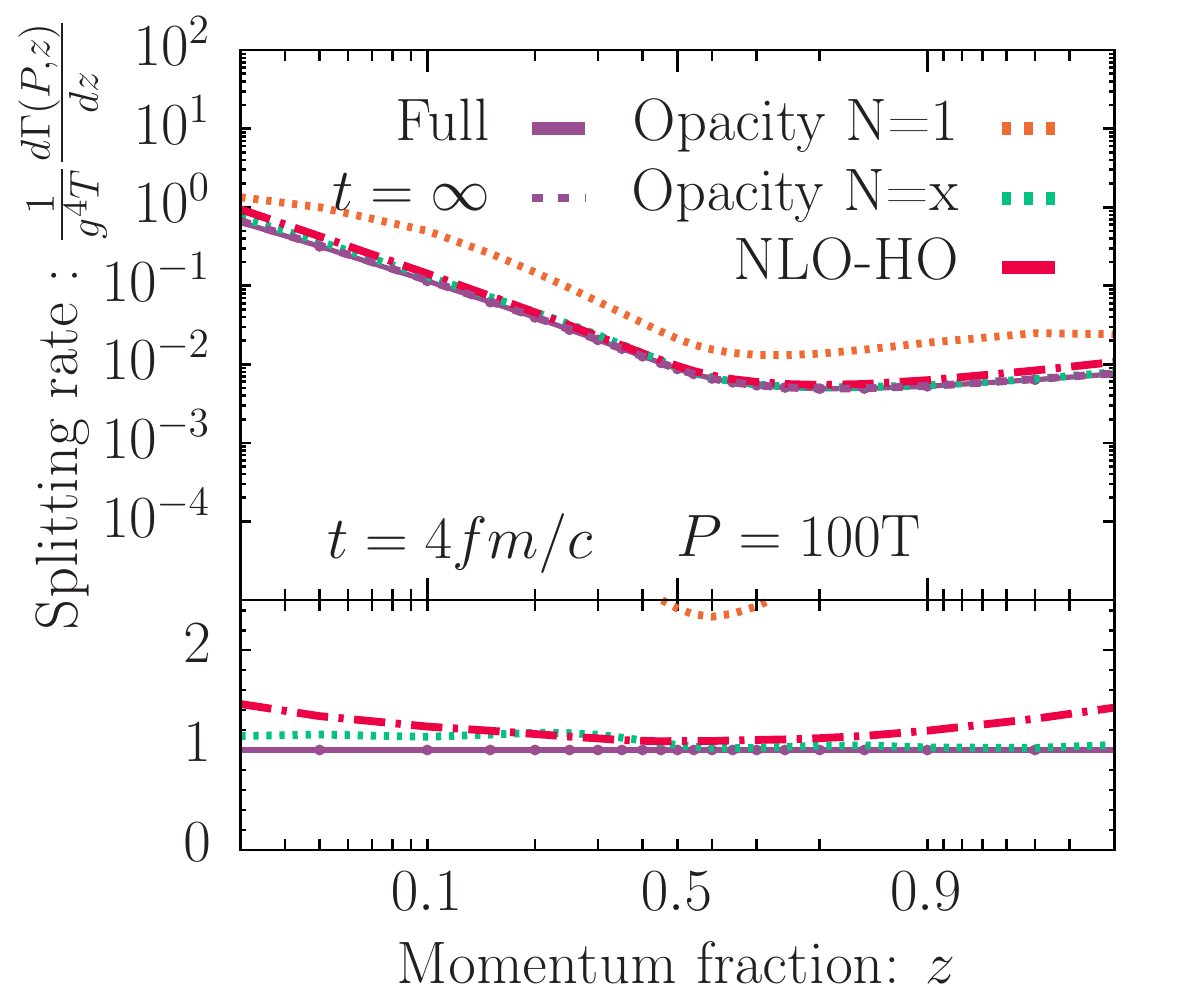}\includegraphics[height=0.345\textwidth,trim=2.4cm 0 0.8cm 0, clip]{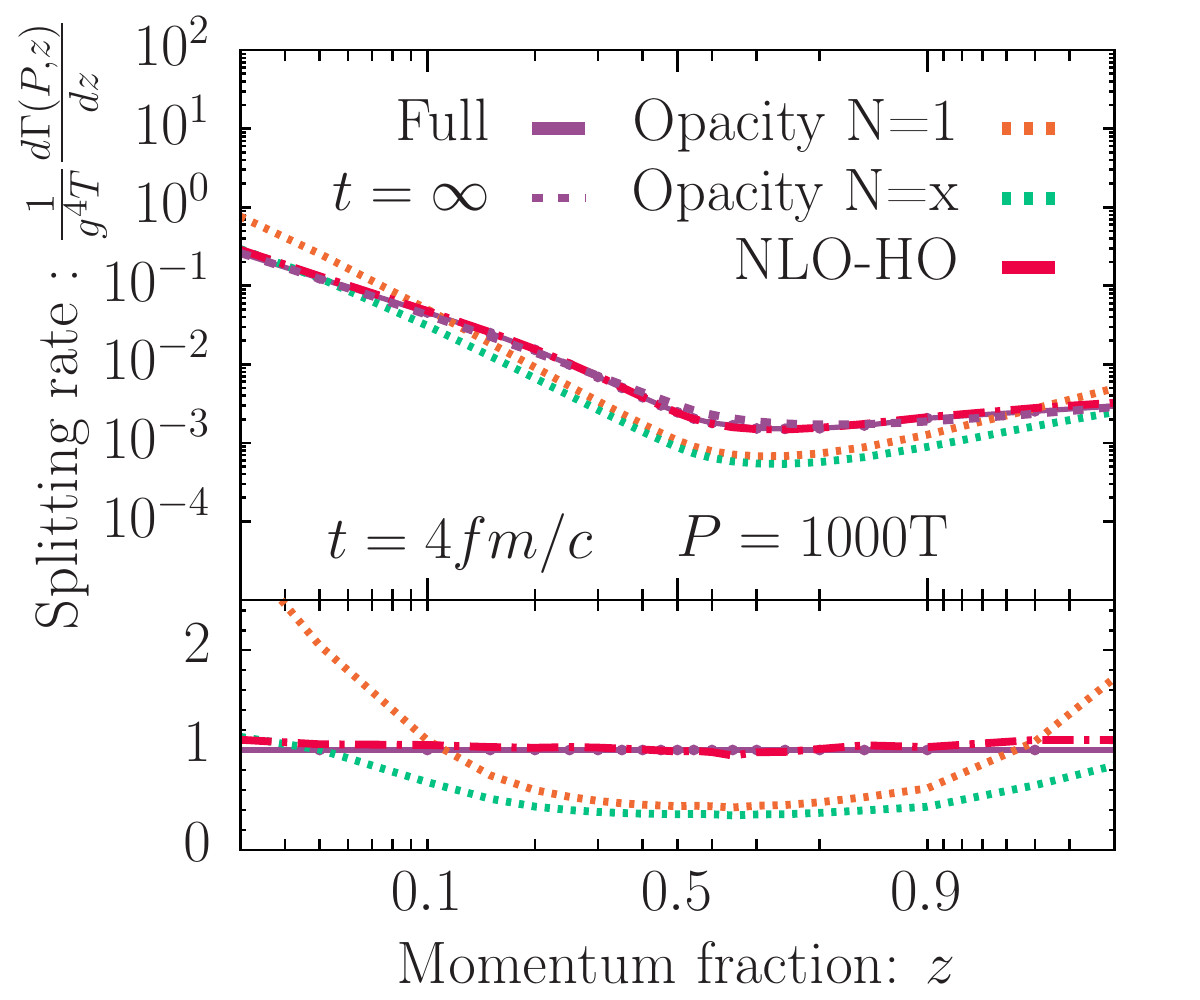}
    \caption{
        Splitting rate for the medium-induced emission of a gluon from a parent quark as a function of momentum fraction of the radiated gluon $z$. Different panels show the rate $d\Gamma/dz$ at fixed times $t=4fm/c$ for different values of the parent quark energy $P=30,100,1000T$ from left to right. We compare the rates obtained using different broadening kernels in the top row and by employing different approximation to the rate calculation in the bottom row. Dashed lines correspond to the (AMY) splitting rates~\cite{Arnold:2003zc} in an infinite medium  \cite{Moore:2021jwe}
    }
    \label{fig:RateLateTime}
\end{figure*}
Generally the in-medium splitting rate depends on the temperature $T$ of the plasma, the momentum $P$ of the emitter, the momentum fraction $z$ of the splitting and the time $t$ of evolution inside the QGP medium. Below we collect some additional results on the temperature ($T$) and momentum $(P)$ dependence of the in-medium splitting rates that corroborate our conclusions of the main text.

With regards to the temperature dependence, we present in Fig.~\ref{fig:TemperatureDependence} our results for the splitting rate of a quark with momentum $P=300T$ radiating a gluon with momentum fraction $z=0.05,0.25,0.5,0.75$ , for the two different temperature $T=250,500$MeV for which the non-perturbative broadening kernel has been determined. By expressing the rate $\Gamma$ in units of $g^4T$ as a function of the scaled evolution time $\tilde{t} = \frac{m_D^2}{2Pz(1-z)}t$ introduced in Eq.~(\ref{eq:DimensionlessVariablesIntro}), one finds that the result for the two different temperatures in Fig.~\ref{fig:TemperatureDependence} are in very good agreement with each other, indicating that the dominant temperature dependence can be accounted for by this simple scaling.

With regards to the momentum dependence, we explore different possibilities in Figs.~\ref{fig:RateEarlyTime}-\ref{fig:RateLateTime} by considering the splitting of a hard quark with different momenta $P=30,100,1000T$
\footnote{We note that for lower energies $P=30,100T$, we employed a smaller cutoff $\mu=0.005$ to ensure numerical stability. }
from left to right into a quark and a gluon with momentum fraction $z$ at two fixed times $t=1$fm/c and $t=4$fm/c respectively.
In the top panels we display the comparison between the splitting rate obtained using the LO and NLO perturbative broadening kernels as well as the non-perturbative  broadening kernel similarly to Fig.~\ref{fig:FiniteMediumFctOfz}, while the bottom panels show comparison between the different approximation to the splitting rate calculation as shown in Fig.~\ref{fig:FiniteMediumVSApprox}. We find that for low typical momentum $2Pz(1-z)\sim T$ the splitting rate obtained using the non-perturbative kernel displays a similar behavior to the one obtained using the NLO broadening kernel, which can be expected as the rate of soft splittings is dominated by small momentum transfer where NLO and NP kernels display similar behavior. Conversely, for high typical momentum $2Pz(1-z)\gg T$, which are more sensitive to large transverse momentum transfer, the non-perturbative splitting rate is closer to the LO order result, as the LO and NP kernels are in better agreement with each other for high $q_\bot$ . We can again confirm our conclusion from the main text,  that for a broad range of momenta $P$, splitting fraction $z$ and times $t$, the uncertainties in the elastic broadening kernel typically translate into larger uncertainties in the splitting rate as opposed to the different approximations to the splitting rate, which can reproduce the in-medium splitting rate rather well within their respective range of validity.

\bibliography{main.bib}
\end{document}